\documentclass[letterpaper,twocolumn,10pt]{article}
\usepackage{usenix}

\usepackage{amsmath,amssymb}
\usepackage{algorithm}
\usepackage{algorithmic}
\usepackage{graphicx}
\usepackage{booktabs}
\usepackage{multirow}
\usepackage{makecell}
\usepackage{url}
\usepackage{xcolor}
\usepackage{xspace}
\usepackage{subcaption} 
\usepackage{enumitem}

\usepackage{xurl}       
\hypersetup{hidelinks}
\renewcommand{\textcolor}[2]{\begingroup\color{black}#2\endgroup} 

\newcommand{\sys}{\textsc{ActProbe}\xspace}

\newcommand{\eg}{\textit{e.g.}}

\begin{document}

\date{}

\title{\Large \bf Detecting and Localizing Segment-Level Poisoning in Multi-Source \\ LLM-Agent Inputs}

\author{
{\rm Xue Tan$^{*}$, Changhui Wang$^{*}$, Sanrui Yang$^{*}$,
Hao Luan$^{*}$,}\\
{\rm Zhuyang Yu$^{*}$, Jin Wei$^{*}$, Ping Chen$^{*}$,
Xiaoyan Sun$^{\dagger}$, and Jun Dai$^{\dagger}$}\\
{\rm $^{*}$Fudan University}\\
{\rm $^{\dagger}$Worcester Polytechnic Institute}
}

\maketitle

\begin{abstract}

Modern Large Language Model (LLM) agents increasingly construct prompts by aggregating external data segments, such as retrieved passages, user reviews, and documents from independent sources. While this multi-source paradigm enables flexible information fusion, it also exposes LLMs to segment-level poisoning attacks, where an adversary controlling only a small subset of sources can inject malicious segments to steer model outputs toward attacker-desired goals. Existing defenses mainly rely on surface textual patterns, external embeddings, or auxiliary detectors, making them vulnerable to fluent and semantically plausible poisoned segments. Moreover, most defenses focus on binary detection and provide limited support for identifying the responsible segments.
We shift the defensive signal from external text to the LLM's internal representation space. Our key observation is that successful poisoning attacks, including corrupted-evidence and adversarial-instruction attacks, induce structured activation shifts associated with malicious generation. Under calibration, we empirically observe a consistent activation-space pattern, which we term the \emph{poison direction}. Based on this insight, we propose \sys, an internal-state-based framework for detecting and localizing poisoned segments in multi-source LLM inputs. \sys projects high-dimensional MLP activations onto the learned poison direction and uses a lightweight linear SVM to detect contaminated prompts from a small calibration set. Once contamination is detected, \sys invokes \textsc{BinRoL}, which combines recursive replacement ablation, Mahalanobis-distance-based branch pruning, and MAD-based robust leaf detection to efficiently identify poisoned segments. This design does not modify the backend LLM and reduces localization overhead from exhaustive $O(n)$ probing to $O(k \cdot \log n)$ forward passes.
Across three datasets, two attacks, and four open-weight LLMs, \sys achieves low detection errors ($0.01$ FPR and $0.05$ FNR) and strong localization performance ($0.94$ Recall and $0.90$ F1-score). It also remains effective under defense-aware adaptive attacks and can protect black-box APIs through surrogate-based poisoned-segment removal.
\end{abstract}

\section{Introduction}
\label{sec:intro}

Large language models (LLMs) are increasingly used as the reasoning core of autonomous agents that aggregate information from multiple external sources to complete user tasks. In a typical multi-source setting, an agent constructs its prompt by concatenating a task instruction with $n$ data segments, each supplied by an independent external source. This paradigm is common in modern LLM applications. Question-answering agents rely on retrieval-augmented generation (RAG)~\cite{lewis2020retrieval, thakur2021beir} to collect passages from external knowledge bases as supporting evidence, where each passage may originate from a different document whose correctness cannot be directly verified by the agent. Content summarization agents face a similar setting: e-commerce review agents~\cite{amazon2023} aggregate reviews from independent users to generate purchase recommendations, while news briefing agents~\cite{reid2024} synthesize articles from different outlets into a unified summary. Here, the LLM processes external segments whose reliability and intent are not fully controlled by the service provider.

This dependence on external segments creates a fundamental security risk. An adversary who controls even a small fraction of the input sources can inject poisoned segments that substantially steer the model's generation. As a result, the agent may present fabricated or attacker-controlled content as if it were supported by trustworthy evidence. For example, a poisoned passage may cause a question-answering agent to incorrectly state that Mount Fuji is the highest mountain in the world, or manipulate an e-commerce summarizer into producing unwarranted warnings about severe product side effects, potentially causing reputational and financial harm. Recent studies have demonstrated the feasibility of such segment-level poisoning from two complementary perspectives: \emph{knowledge poisoning} and \emph{prompt injection}. Knowledge poisoning attacks~\cite{poisonedrag,prattack,zhang2026adversarial} inject factually incorrect yet semantically plausible segments into external knowledge sources, causing the model to rely on corrupted evidence during generation. For instance, PoisonedRAG~\cite{poisonedrag} shows that optimized poisoned passages can be retrieved for target queries and induce attacker-specified responses. In contrast, prompt injection attacks~\cite{obliinjection,greshake2023not,liu2024formalizing} embed adversarial instructions within external data segments to override the agent's intended task. ObliInjection~\cite{obliinjection} shows that malicious segments can remain effective even when their positions among benign segments are uncertain. Although these paradigms differ in whether they manipulate factual evidence or task instructions, both exploit the same weakness: after prompt integration, a few malicious external segments can corrupt the final LLM output.

\textcolor{blue}{Existing defenses provide only partial protection against segment-level poisoning. Prevention-based methods~\cite{struq,secalign} harden LLMs through fine-tuning or safety alignment, but require modifying the target model, may reduce benign-task utility~\cite{jia2025critical}, and remain vulnerable to adaptive attacks; for example, ObliInjection retains attack success rates of 53\%--78\% against hardened models~\cite{obliinjection}. More importantly, a practical defense for multi-source applications must solve two coupled problems: determining whether an input is contaminated and identifying which external segments are responsible. Detection alone cannot support targeted sanitization, while localization is unreliable without an accurate contamination signal. Existing methods struggle with both objectives because they primarily rely on textual appearance or model-external representations. PPL~\cite{ppl,jain2023baseline} and KAD~\cite{liu2024formalizing} exhibit false negative rates of up to 92.6\% against ObliInjection~\cite{obliinjection}; EVD relies on general-purpose embeddings disconnected from the target LLM's computation; and DataSentinel~\cite{liu2025datasentinel}, despite training a dedicated detector, still misses 79.6\% of ObliInjection-contaminated inputs. Localization methods are similarly limited: PromptLocate~\cite{jia2025promptlocate} uses a text-level oracle to identify malicious instructions, making it suitable for explicit prompt injection but less effective against data-only poisoning containing plausible misinformation. Consequently, existing defenses struggle to reliably detect and localize fluent, semantically coherent poisoned segments.}

\begin{figure}[t]
    \centering
    \includegraphics[width=0.99\linewidth]{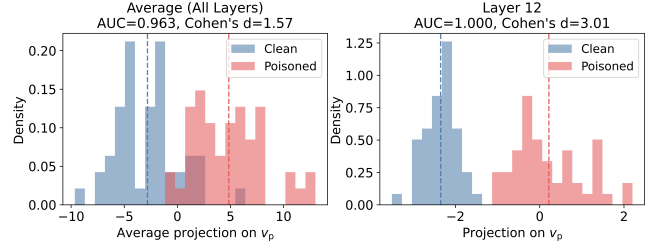}
    \caption{Projection distributions of clean and poisoned activations along the poison direction. Results are shown for Mistral-7B-Instruct-v0.3 on Amazon Reviews under ObliInjection.}
    \label{fig:poison_direction_intro}
\end{figure}

To address this limitation, we shift the defensive signal from external text to the LLM's internal representation space. 
We observe that successful segment-level poisoning attacks, whether based on corrupted evidence or adversarial instructions, induce activation shifts that consistently align along a geometric direction across samples, which we call the \emph{poison direction}. Figure~\ref{fig:poison_direction_intro} illustrates this effect: after projection onto the learned direction, poisoned inputs exhibit a clear shift from clean inputs in both the all-layer average and the most discriminative layer. 
This structured activation-space footprint offers a more direct signal of adversarial influence than surface-level text features. 
However, turning this insight into a practical defense introduces two challenges.
First, modern LLM activations are extremely high-dimensional; for example, the MLP activations of Llama-3.1-8B-Instruct contain 14,336 dimensions per layer. Directly training detectors on such representations is computationally costly and prone to overfitting when only limited calibration data are available. Second, localizing $k$ poisoned segments among $n$ inputs can require many forward passes under segment-wise testing, and dense poisoning can cause malicious signals to mask one another, making attribution difficult.

We propose \sys, an internal-state-based framework for detecting and localizing poisoned segments in multi-source LLM inputs. \sys addresses the two challenges above through two coordinated phases. To address the high-dimensional detection challenge, the \textbf{Detection Phase} learns a poison direction from a small calibration set and projects high-dimensional MLP activations onto this direction, producing a compact layer-wise feature representation. A lightweight linear SVM then classifies whether the assembled prompt is contaminated, while the same calibration process estimates the benign activation distribution for later reuse. To address the efficient localization challenge, once a prompt is flagged as suspicious, the \textbf{Localization Phase} invokes \textsc{BinRoL} (Binary Search with Robust Leaf Detection). \textsc{BinRoL} recursively partitions the input context and applies replacement ablation, substituting candidate segments with neutral text and measuring the resulting reduction in Mahalanobis distance to the benign distribution. This process efficiently narrows down suspicious regions without exhaustive segment-wise testing. At the leaf level, \textsc{BinRoL} applies Median Absolute Deviation (MAD)-based robust outlier detection to identify the final poisoned segments, partially reducing the impact of mutual masking in dense poisoning settings. \sys avoids modifying the backend LLM, relies on internal activation rather than surface text heuristics, and reduces localization overhead from exhaustive $O(n)$ probing to $O(k \cdot \log n)$ forward passes.

We evaluate \sys on HotpotQA, MS-MARCO, and Amazon Reviews across four open-weight LLMs and two multi-source attacks, PoisonedRAG and ObliInjection. \sys consistently outperforms text-level, embedding-based, and localization baselines. 

In summary, our main contributions are as follows:

\begin{itemize}
    \item We propose \sys, an internal-state-based defense that learns a \emph{poison direction} from calibration data in the LLM activation space to detect contaminated multi-source inputs and localize the responsible segments.

    \item We design \textsc{BinRoL}, which combines recursive replacement ablation, Mahalanobis-based branch pruning, and MAD-based robust leaf detection to localize poisoned segments with $O(k \cdot \log n)$ forward passes.

    \item Across three datasets, two attacks, and four open-weight LLMs, \sys outperforms existing baselines with average detection FPR/FNR of $0.01/0.05$ and localization Recall/F1-score of $0.94/0.90$. It also demonstrates robustness against adaptive attacks and characterizes partial transferability across attack settings.
\end{itemize}

\section{Related Work}
\label{sec:related}

\subsection{Large Language Models (LLMs)}

Large Language Models (LLMs) are autoregressive generative models typically based on the Transformer architecture. Given an input sequence $X$, an LLM models the conditional probability of an output sequence $Y=\{y_1,y_2,\dots,y_m\}$ in a token-by-token manner:
\begin{equation}
    P(Y|X) = \prod_{t=1}^{m} P(y_t | X, y_{<t}).
\end{equation}

In multi-source applications, the input $X$ is constructed by concatenating segments from multiple external sources. Let $x_i$ denote the $i$-th segment; then the input can be written as $X=[x_1;x_2;\dots;x_n]$. Through self-attention, the model integrates information across these segments to generate a response aligned with the task objective. Table~\ref{tab:notation} in the Appendix~\ref{sec:app-a} summarizes the key notations used in this paper.




\subsection{Attacks on Multi-Source LLM Applications}

In multi-source LLM applications, adversaries often control only a small fraction of external segments, yet these segments can still influence the final output. Existing attacks mainly follow two paradigms: knowledge poisoning and prompt injection. Knowledge poisoning compromises external knowledge sources by injecting factually incorrect but semantically coherent content, causing the model to rely on corrupted evidence during generation~\cite{zhang2026practical,zhang2026adversarial,shafran2024machine}. PoisonedRAG~\cite{poisonedrag} is a representative RAG poisoning attack that inserts optimized passages into the knowledge base to increase retrieval relevance to target queries and induce attacker-specified responses. Other attacks, such as PR-Attack~\cite{prattack} and BadRAG~\cite{xue2024badrag}, further improve poisoning effectiveness by optimizing triggers, poisoned texts, or retriever-generator interactions. Prompt injection instead embeds adversarial instructions into external segments to override the intended task or instruction hierarchy~\cite{greshake2023not,liu2024formalizing,wang2025webinject}. In multi-source settings, malicious segments may appear at arbitrary positions among benign ones, making position-dependent attacks unreliable. ObliInjection~\cite{obliinjection} addresses this challenge with an order-oblivious objective and the order GCG algorithm, producing malicious segments that remain effective under different segment permutations. Other attacks, such as Neural Exec~\cite{pasquini2024neural} and JudgeDeceiver~\cite{shi2024optimization}, extend instruction hijacking to retrieval-augmented and evaluation-oriented scenarios. In summary, knowledge poisoning corrupts factual evidence, whereas prompt injection manipulates task instructions. We adopt PoisonedRAG and ObliInjection as representative attacks to evaluate detection and localization under evidence-level and instruction-level poisoning.

\subsection{Defenses against Segment-Level Poisoning Attacks}

Existing defenses against segment-level poisoning in LLM applications mainly fall into three categories: prevention, detection, and localization. Prevention-based methods, such as StruQ~\cite{struq} and SecAlign~\cite{secalign}, harden LLMs through instruction tuning or safety alignment to separate trusted instructions from untrusted data. However, they require modifying the target model, may affect benign-task utility~\cite{jia2025critical}, and primarily target instruction-level attacks rather than knowledge poisoning. Detection-based methods identify contaminated inputs or segments without changing the backend LLM. Perplexity-based filtering~\cite{ppl,jain2023baseline} uses language-model likelihood, Known-Answer Detection~\cite{liu2024formalizing,kad} probes instruction-following behavior with predefined checks, embedding-based detectors flag semantic outliers in external representation spaces, and DataSentinel~\cite{liu2025datasentinel} trains a dedicated detector through adversarial optimization. Despite their differences, these methods rely on surface statistics, external embeddings, or auxiliary models rather than the protected LLM's internal computation. Localization-based defenses further aim to identify the malicious segments after contamination is detected. PromptLocate~\cite{jia2025promptlocate} localizes injected prompts through semantic segmentation and oracle-style detection, but is tailored to explicit prompt-injection patterns. General attribution methods, such as feature-removal attribution~\cite{zeiler2014visualizing} and Shapley-value attribution~\cite{lundberg2017unified}, can be adapted for localization, but either perturb the input structure or incur high computational cost. In contrast, \sys leverages internal activations to provide a unified defense for detecting and localizing malicious segments under both knowledge poisoning and prompt injection.

\section{Problem Formulation}
\label{sec:threat}
 
\subsection{Multi-Source Information Fusion in LLM Agents}
\label{sec:threat:data}

In modern LLM agent applications, particularly in \emph{Information Fusion} tasks, the input context is frequently aggregated from multiple independent sources. We refer to the discrete data contributed by each source as a \emph{segment}. Formally, consider $n$ sources, where source $i$ provides a text segment $x_i$. The target context $x^t$ is formed by concatenating these $n$ segments in a specific order determined by the service provider's backend logic:
\begin{equation}
    x^t = x_{i_1} \| x_{i_2} \| \cdots \| x_{i_n},
\end{equation}
where $\{i_1, i_2, \ldots, i_n\}$ is a permutation of $\{1, 2, \ldots, n\}$. The LLM $f$ receives a prompt $p = s^t \| x^t$, where $s^t$ is a system instruction specifying the intended task, and generates a response $r^t = f(s^t \| x^t)$.

This formulation captures knowledge-aggregation applications, including Retrieval-Augmented Generation (RAG)~\cite{lewis2020retrieval, poisonedrag} and multi-document summarization (e.g., e-commerce reviews)~\cite{amazon2023}. A key property of these tasks is the \emph{majority-consensus assumption}: the LLM implicitly functions as a consensus builder, synthesizing information across segments, where benign content is expected to dominate and establish a stable contextual baseline. From a security perspective, the service provider aggregates segments from sources that are not trusted. This design exposes the LLM's consensus mechanism to dense \emph{Sybil attacks}, where a single adversary injects multiple malicious segments that masquerade as independent sources and exert disproportionate influence over the model's output. Even without knowledge of the segment permutation, an adaptive adversary can exploit this vulnerability by introducing a small but strategically crafted subset of adversarial segments to bias the final output.


\subsection{Threat Model}
\label{sec:threat:attacker}

\textbf{Attacker's Goal.}
We consider an adversary who aims to manipulate the output of a multi-source LLM application by injecting malicious content into a subset of external segments. Let $x^t = x_{i_1}\|x_{i_2}\|\cdots\|x_{i_n}$ denote the clean context assembled from $n$ external segments, and let $s^t$ denote the task instruction. The benign response is $r^t = f(s^t \| x^t)$. The attacker contaminates a subset $A \subseteq \{1,\ldots,n\}$ of segments, where $|A|=k$ and $k \geq 1$, producing a contaminated context $x^c$ by replacing each clean segment $x_i$ for $i \in A$ with an adversarial segment $\tilde{x}_i$. The attacker's goal is to induce the LLM to generate an attacker-specified response $r^e$ instead of the intended benign response $r^t$. We consider the attack successful if
\begin{equation}
    f(s^t \| x^c) \simeq r^e,
\end{equation}
where $\simeq$ denotes semantic equivalence. For example, the generated answer may endorse a targeted false fact, or the generated summary may promote a fabricated sentiment.\\
\textbf{Attacker's Capabilities and Knowledge.}
The adversary can forge, upload, or compromise $k$ out of the $n$ external sources, such as malicious reviews, poisoned web pages, or corrupted documents. We assume $k<n/2$, reflecting practical compromise costs and preserving benign-source dominance~\cite{xiang2024certifiably}. We further evaluate larger values of $k$ in Section~\ref{sec:exp:adaptive:dilution_density} and provide a detailed analysis. 

The adversary may have different levels of model access depending on the attack method. For open-weight models, poisoned segments may be crafted with white-box access; for closed-source services, the adversary may rely on black-box queries or surrogate models. However, the adversary does not control the service provider's runtime prompt assembly process. In particular, the adversary does not know the exact benign segments contributed by legitimate sources, nor the final segment permutation $\{i_1,i_2,\ldots,i_n\}$ used to construct the context. The adversary also cannot tamper with the defender's calibration data or observe runtime internal activations. Therefore, effective attacks must remain robust to unknown benign context, uncertain segment ordering, and private activation-space parameters.

\subsection{Defender Model}
\label{sec:threat:defender}

\textbf{Defense Goal.}
The defender aims to prevent poisoned external segments from manipulating the LLM's final response. To this end, the defense performs two tasks before the final response is generated:
\begin{itemize}
    \item \emph{Detection}: Given an assembled prompt $p=s^t\|x$, determine whether the context contains any adversarial segment. This is a prompt-level binary classification task.
    \item \emph{Localization}: If the prompt is detected as contaminated, identify the specific segment indices responsible for the attack. Formally, the defender outputs a suspicious set $\hat{D} \subseteq \{1,\ldots,n\}$. The localized segments can then be removed or replaced before the final response is generated.
\end{itemize}
\textbf{Defense Capabilities and Assumptions.}
We assume the defender operates under the following conditions:
\begin{enumerate}
    \item \emph{Segment Boundary Access}: The defender has access to the segment boundaries used by the multi-source application. This is natural for RAG, review aggregation, and multi-document summarization systems, where the service provider explicitly concatenates discrete external segments into the prompt.

    \item \emph{Diagnostic Activation Access}: For open-weight LLMs, the defender can extract hidden-layer activations from the protected model. For closed-source API models, the defender can instead deploy an open-weight LLM as an upstream diagnostic surrogate to inspect and sanitize the context before forwarding it to the black-box model. We evaluate this surrogate-based deployment in Section~\ref{sec:discuss:surrogate}.

    \item \emph{Calibration Data and Coverage}: \textcolor{blue}{The defender has access to a small calibration set containing $M$ clean prompts and $M$ poisoned prompts, with $M=100$ in our experiments. Poisoned samples are generated using known attacks or red-teaming and are assumed to cover representative attack families and data domains expected at deployment time, rather than exact runtime attack instances. ACTPROBE's performance depends on this coverage and may degrade against unseen attacks with substantially different activation patterns. As attacks evolve, the defender can update the calibration set with new attack samples and recalibrate the detector.}

\end{enumerate}

\section{Our \sys}
\label{sec:method}

In this section, we present \sys, a two-stage framework engineered to detect and localize poisoned segments within multi-source LLM inputs. Following a high-level architectural overview (Section~\ref{sec:method:overview}), we define the poison direction probe that underpins the entire framework (Section~\ref{sec:method:probe}). We subsequently detail the operational mechanics of the Poison Prompt Detection stage (Section~\ref{sec:method:detect}) and the Poison Segment Localization stage (Section~\ref{sec:method:locate}).
\begin{figure*}[h]
    \centering
    \includegraphics[width=0.95\linewidth]{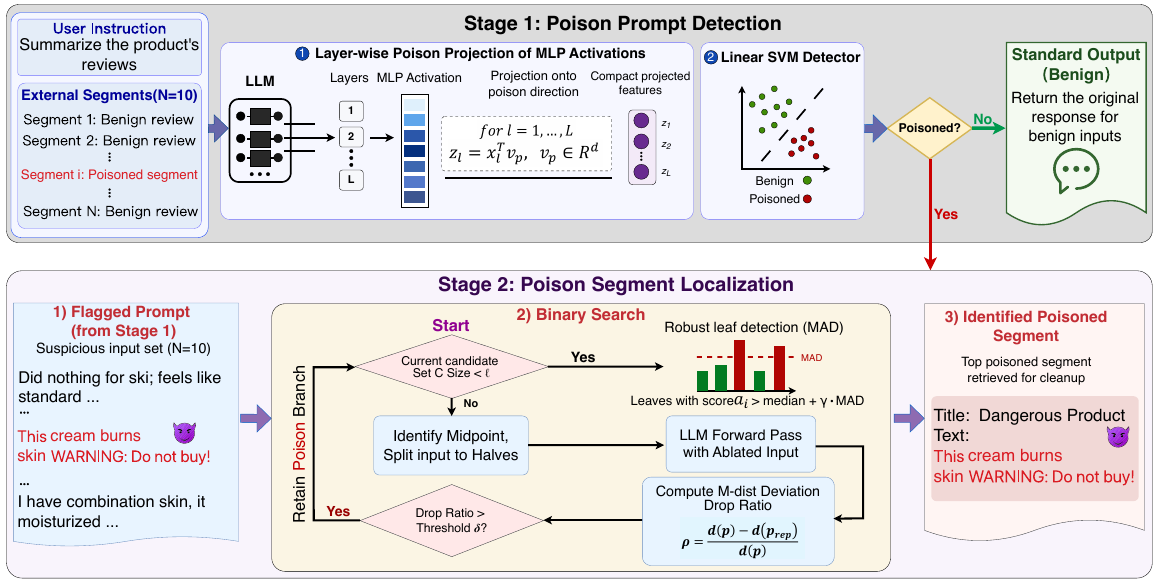}
    \caption{Overview of \sys. The detection phase identifies poisoned prompts via projection-based SVM classification, and the subsequent localization phase isolates malicious segments with \textsc{BinRoL}.}
    \label{fig:overview}
\end{figure*}

\subsection{Overview}
\label{sec:method:overview}

The workflow of \sys consists of offline calibration and online inference, as illustrated in Figure~\ref{fig:overview} and formalized in Algorithm~\ref{alg:pipeline} in the Appendix~\ref{sec:app-a}.\\
\textbf{Offline Calibration.} Leveraging a modest calibration dataset of $M$ clean and $M$ poisoned prompts, \sys establishes its defensive boundaries by learning three core components: (1)~a \emph{poison direction probe} $\mathbf{v}_{\mathrm{p}}$ that projects high-dimensional LLM activations onto a subspace highly sensitive to adversarial manipulation (Section~\ref{sec:method:probe}); (2)~a linear SVM classifier tailored for prompt-level detection (Section~\ref{sec:method:detect}); and (3)~clean distribution parameters $(\boldsymbol{\mu}, \boldsymbol{\Sigma}^{-1})$ derived as a byproduct of the profiling process. Crucially, these parameters are directly reused during the localization stage without incurring any additional computational overhead.\\
\textbf{Online Inference.} Upon receiving a newly assembled prompt $p = s^t \| x_1 \| \cdots \| x_n$, \sys initiates the \emph{Poison Prompt Detection} stage. The prompt's internal activations are extracted, projected via $\mathbf{v}_{\mathrm{p}}$, and evaluated by the SVM. If classified as benign, the inspection terminates, and the LLM permits standard generation. Conversely, if an anomaly is flagged, \sys triggers the \emph{Poison Segment Localization} stage. Here, the \textsc{BinRoL} algorithm (Algorithm~\ref{alg:binrol}) systematically isolates all contaminated segments utilizing a recursive binary search coupled with robust leaf-node anomaly detection, efficiently pinpointing the poisoned sources with an expected cost of only $O(k \cdot \log n)$ forward passes.

\subsection{Poison Direction Probe}
\label{sec:method:probe}

The hidden representations of modern LLMs reside in exceptionally high-dimensional spaces (\eg, the MLP activation dimension is $d=14{,}336$ per layer across $L=32$ transformer blocks in Llama-3.1-8B-Instruct, resulting in a raw feature space of $d \times L \approx 460{,}000$ dimensions). Training anomaly detectors directly on these raw activations is highly problematic for two primary reasons. First, it requires learning millions of parameters, thereby necessitating prohibitively large datasets to avert severe overfitting. This data requirement is fundamentally incompatible with real-world defensive scenarios, where rapidly acquiring extensive labeled samples is practically infeasible. Second, raw activations are entangled with substantial task-relevant semantic noise, which can easily obscure the subtle anomalies induced by malicious inputs. To overcome these statistical and representational bottlenecks, we introduce the \emph{poison direction probe}. This component is designed as a targeted dimensionality reduction mechanism, specifically engineered to isolate the dominant activation shift exclusively induced by poisoning.

Specifically, given a calibration set of $M$ paired clean and poisoned prompts, we extract the last-token hidden activation $\mathbf{h}_{i,l} \in \mathbb{R}^d$ at each layer $l$ for each prompt $i$. We use the last-token activation because it serves as a compact prompt-level representation for decoder-only LLMs immediately before next-token prediction, and has been widely adopted in activation probing and representation engineering~\cite{burns2022discovering,zou2023representation,tan2024revprag}.
The poison direction at layer $l$ is calculated as the normalized mean difference between the adversarial and benign representations:
\begin{equation}
    \mathbf{v}_l = \frac{1}{M} \sum_{i=1}^{M} \left( \mathbf{h}^{\mathrm{p}}_{i,l} - \mathbf{h}^{\mathrm{c}}_{i,l} \right), \quad
    \mathbf{v}_{\mathrm{p}}[l] = \frac{\mathbf{v}_l}{\|\mathbf{v}_l\|},
    \label{eq:vpois}
\end{equation}
where $\mathbf{h}^{\mathrm{p}}_{i,l}$ and $\mathbf{h}^{\mathrm{c}}_{i,l}$ denote the activations of the $i$-th poisoned and clean prompt at layer $l$, respectively. By projecting any given activation $\mathbf{h}_l$ onto this unit vector, we obtain a scalar projection $s_l = \mathbf{h}_l \cdot \mathbf{v}_{\mathrm{p}}[l]$. Consequently, the feature representation for an entire prompt is constructed as:
\begin{equation}
    \mathbf{s} = [s_1, s_2, \ldots, s_L] \in \mathbb{R}^L.
    \label{eq:feature}
\end{equation}

Through this projection, the representation at each layer is distilled from $d$ dimensions down to a single scalar, effectively compressing the full network activation into a compact $L$-dimensional feature vector. Unlike generic dimensionality reduction techniques (\eg, PCA~\cite{abdi2010principal}) that predominantly capture directions of maximum variance, $\mathbf{v}_{\mathrm{p}}$ isolates a mechanistically grounded subspace that explicitly captures the perturbation footprint of adversarial poisoning. This targeted projection preserves the critical adversarial signal while systematically filtering out task-irrelevant semantic noise. Consequently, a higher projection magnitude $s_l$ directly quantifies the severity of the malicious activation shift, endowing the final feature vector $\mathbf{s}$ with both exceptional discriminative power and inherent interpretability. 
This discriminative capability is visually confirmed in Appendix~\ref{sec:app-b} (Figures~\ref{fig:activation_projections_Obliinjection} and \ref{fig:activation_projections_poisonedrag}). 
We empirically verify this advantage in Section~\ref{sec:exp:ablation:proj}, demonstrating that $\mathbf{v}_{\mathrm{p}}$ substantially outperforms random projections and PCA in detection and localization.

\subsection{Poison Prompt Detection}
\label{sec:method:detect}

The Poison Prompt Detection stage determines whether an assembled prompt is contaminated. We formulate this as a binary classification task operating on the compact $L$-dimensional projection features derived in Section~\ref{sec:method:probe}.\\
\textbf{Classifier Design.} The poison direction probe distills the hidden states into a compact feature vector (\eg, $L{=}32$ dimensions for Llama-3.1-8B-Instruct) that is explicitly aligned with the adversarial perturbation. Within this low-dimensional and well-structured representation space, a linear decision boundary is empirically sufficient and computationally efficient. We employ a linear Support Vector Machine (SVM)~\cite{cortes1995support} for its maximum-margin formulation, which provides favorable generalization properties in low-dimensional feature spaces even with limited training data (\eg, $M{=}100$ benign and $M{=}100$ poisoned calibration samples for $L{=}32$ features). We empirically substantiate this design choice by benchmarking the linear SVM against alternative classification algorithms in Section~\ref{sec:exp:ablation}. Concretely, after standardizing the projection features to zero mean and unit variance via a fitted scaler, the classification rule is defined as:
\begin{equation}
    \hat{y} = \mathrm{sign}\!\left(\mathbf{w}^\top \tilde{\mathbf{s}} + b\right),
    \label{eq:svm}
\end{equation}
where $\tilde{\mathbf{s}} \in \mathbb{R}^L$ is the standardized feature vector, $\mathbf{w} \in \mathbb{R}^L$ is the learned weight vector, and $b$ is the bias term. 
Since the features are standardized, the magnitude $|w_l|$ provides a comparable estimate of the contribution of layer $l$ to poisoning detection. We use these layer-wise weights to select the most discriminative layers for the localization stage.\\
\textbf{Baseline Distribution Profiling.} Fitting the SVM concurrently yields a critical statistical byproduct required for the subsequent localization stage: the distributional parameters of benign activations. Using the unstandardized projection features of the $M$ clean calibration prompts, we estimate the empirical mean and covariance:
\begin{equation}
    \boldsymbol{\mu} = \frac{1}{M} \sum_{i=1}^{M} \mathbf{s}^{\mathrm{c}}_i, \quad
    \boldsymbol{\Sigma} = \mathrm{Cov}\!\left(\{\mathbf{s}^{\mathrm{c}}_i\}_{i=1}^{M}\right) + \lambda \mathbf{I},
    \label{eq:cleandist}
\end{equation}
where $\lambda{=}10^{-6}$ serves as a standard Tikhonov regularization term to guarantee the invertibility of $\boldsymbol{\Sigma}$. The parameters $(\boldsymbol{\mu}, \boldsymbol{\Sigma}^{-1})$ define the normative boundary of benign prompts within the projection space. They are stored for direct cross-phase reuse during localization, acting as the reference distribution for Mahalanobis distance computations without incurring any additional calibration overhead. The validity of this parameter reuse is grounded in the structural robustness of the poison direction $\mathbf{v}_{\mathrm{p}}$ against neutral text replacements. \\
\textbf{Online Inference Pipeline.} During active deployment, processing a new prompt requires only a single forward pass through the LLM. The extracted MLP activations are projected onto $\mathbf{v}_{\mathrm{p}}$ to generate $\mathbf{s}$, standardized to $\tilde{\mathbf{s}}$, and evaluated by the SVM. If the prompt is classified as benign, the defensive inspection terminates immediately, imposing negligible computational overhead on the standard generation process. Conversely, if an adversarial perturbation is detected, the system automatically invokes the Poison Segment Localization stage to pinpoint the malicious segments.

\subsection{Poison Segment Localization via \textsc{BinRoL}}
\label{sec:method:locate}

Upon a positive flag from the detection stage, the localization phase determines \emph{which} of the $n$ segments are poisoned. We propose \textsc{BinRoL} (\textbf{Bin}ary search with \textbf{Ro}bust \textbf{L}eaf detection), which decomposes localization into two phases: a coarse binary narrowing phase that efficiently reduces the candidate set, followed by a fine-grained leaf detection phase that precisely isolates the poisoned segments. The procedure is formalized in Algorithm~\ref{alg:binrol} in the Appendix~\ref{sec:app-b}.

\subsubsection{Replacement Ablation}
\label{sec:method:locate:ablation}

A straightforward localization strategy is to directly inspect each segment's activation projection and flag segments with unusually large scores. However, this strategy is unreliable in autoregressive LLMs, where the hidden state at each position is conditioned on all preceding tokens. As a result, the poisoning signal introduced by a contaminated segment can \emph{propagate forward} through the sequence, causing subsequent benign segments to inherit traces of the malicious perturbation. This signal leakage makes it difficult to distinguish the segment that originates the poisoning effect from those that merely transmit its influence.

To address this issue, we adopt \emph{replacement ablation}. 
Specifically, we replace a target segment or segment group with neutral text (\eg, ``No relevant information is available for this query.''; see Appendix~\ref{app:neutral_texts} for the complete list of templates) and perform a new forward pass. 
This operation preserves the segment slot and the surrounding prompt structure while suppressing the semantic and adversarial content of the replaced segment. By comparing the LLM's internal states before and after replacement, we obtain a cleaner estimate of the replaced segment's contribution to the poisoning signal, rather than relying on its raw activation projection, which may be confounded by forward propagation effects. 
We empirically validate the superiority of this replacement-ablation design over direct deletion-based ablation in Section~\ref{sec:exp:ablation:ablmethod}.

\subsubsection{Drop Ratio}
\label{sec:method:locate:mahal}

We measure the effect of replacement ablation by quantifying how much the prompt representation moves toward the clean distribution after candidate segments are replaced. Let $\mathbf{s}_{\mathrm{full}}$ and $\mathbf{s}_{\mathrm{rep}}$ denote the projection features before and after replacement, respectively. We compute the Mahalanobis distance~\cite{mahalanobis2018generalized} to the clean distribution as
\begin{equation}
    d(\mathbf{s}) = \sqrt{(\mathbf{s} - \boldsymbol{\mu})^\top \boldsymbol{\Sigma}^{-1} (\mathbf{s} - \boldsymbol{\mu})},
    \label{eq:mahal}
\end{equation}
where $(\boldsymbol{\mu}, \boldsymbol{\Sigma}^{-1})$ are the benign distribution parameters estimated during the detection stage (Eq.~\ref{eq:cleandist}). The \emph{drop ratio} is then defined as the relative reduction in anomaly score:
\begin{equation}
    \rho = \frac{d(\mathbf{s}_{\mathrm{full}}) - d(\mathbf{s}_{\mathrm{rep}})}{d(\mathbf{s}_{\mathrm{full}})}.
    \label{eq:dropratio}
\end{equation}

A larger $\rho$ indicates that replacing the candidate segments substantially reduces the poisoning-induced deviation, suggesting that these segments contribute strongly to the malicious signal. Since $\rho$ is normalized by the original anomaly score, it mitigates variations caused by different prompt contents and lengths, reducing the need for per-sample threshold calibration.

\subsubsection{The \textsc{BinRoL} Algorithm}
\label{sec:method:locate:binrol}

\textsc{BinRoL} (Algorithm~\ref{alg:binrol}) consists of two complementary procedures: a coarse-grained binary narrowing stage and a fine-grained leaf detection stage.\\
\textbf{Phase 1: BinaryNarrow.}
This procedure recursively reduces the candidate set from $n$ segments to at most $\ell$ segments, where $\ell$ is the leaf size. At each recursion level, the current candidate set $\mathcal{C}$ is split into two halves, $\mathcal{L}$ and $\mathcal{R}$. Each half is independently replaced with neutral text, and the corresponding drop ratios $\rho_{\mathcal{L}}$ and $\rho_{\mathcal{R}}$ are computed. The recursion then follows three branching rules:
\begin{itemize}
    \item Both $\rho_{\mathcal{L}} > \delta$ and $\rho_{\mathcal{R}} > \delta$, both branches are explored, indicating that poisoned segments may exist in both halves.
    \item Only one half exceeds $\delta$, only that branch is explored.
    \item If neither half exceeds $\delta$, the branch with the larger $\rho$ is selected as a fallback to avoid prematurely discarding potential poisoned segments.
\end{itemize}

The branching threshold $\delta$ is tuned on an independent validation set and set to a permissive value (e.g., $\delta=0.02$) to prioritize candidate recall during coarse narrowing, leaving fine-grained discrimination to the robust leaf detection stage.\\
\textbf{Phase 2: RobustLeaf.}
Once the candidate set satisfies $|\mathcal{C}| \leq \ell$, \textsc{BinRoL} switches from group-level narrowing to per-segment attribution. For each segment $i \in \mathcal{P}(\mathcal{C})$, the parent candidate set of the current leaf $\mathcal{C}$, we perform individual replacement ablation and compute its \emph{ablation score}:
\begin{equation}
    a_i = \frac{1}{|\mathcal{L}_{\mathrm{top}}|} 
    \sum_{l \in \mathcal{L}_{\mathrm{top}}} 
    \left(\mathbf{h}^{\mathrm{full}}_l - \mathbf{h}^{\mathrm{rep}}_{i,l}\right) \cdot \mathbf{v}_{\mathrm{p}}[l],
    \label{eq:ablation_score}
\end{equation}
where $\mathbf{h}^{\mathrm{full}}_l$ denotes the MLP activation at layer $l$ for the original prompt, and $\mathbf{h}^{\mathrm{rep}}_{i,l}$ denotes the activation after replacing segment $i$ with neutral text. 
The set $\mathcal{L}_{\mathrm{top}}$ contains the most discriminative layers, selected according to the magnitudes of the SVM weights learned during detection, with its size specified as a validation-tuned hyperparameter in Section~\ref{sec:exp:setup}. Although prompt-level detection uses all layers to capture global evidence, segment-level attribution requires a cleaner and more localized signal. Focusing on $\mathcal{L}_{\mathrm{top}}$ suppresses weakly informative layers that may introduce semantic noise, while also reducing the cost of repeated ablation scoring. A larger $a_i$ indicates that replacing segment $i$ removes a stronger component of the poisoning-induced activation shift, suggesting that this segment contributes more to the signal.

To identify anomalous candidates within a leaf, we apply a robust outlier detection rule based on the Median Absolute Deviation (MAD). 
Rather than computing MAD only within the current leaf $\mathcal{C}$, we
estimate the robust statistics from the parent candidate set
$\mathcal{P}(\mathcal{C})$, i.e., the candidate set before the last binary
split:
\begin{equation}
    \tilde{m} = \mathrm{median}(\{a_i\}_{i \in \mathcal{P}(\mathcal{C})}), \quad 
    \hat{\sigma} = \eta \cdot 
    \mathrm{median}(\{|a_i - \tilde{m}|\}_{i \in \mathcal{P}(\mathcal{C})}),
    \label{eq:mad}
\end{equation}
where $\eta=1.4826$ makes MAD a consistent estimator of the standard
deviation under a Gaussian distribution~\cite{hampel1986robust}. 
A candidate segment $i \in \mathcal{C}$ is flagged as poisoned if
\begin{equation}
    a_i > \tilde{m} + \gamma \cdot \hat{\sigma},
\end{equation}
where $\gamma$ controls the sensitivity of the attribution rule. 
Parent-level MAD provides a more stable threshold than leaf-level MAD
when multiple poisoned segments fall into the same small leaf, while
remaining more localized than global thresholding over all segments. 
If no candidate exceeds the threshold, \textsc{BinRoL} selects the segment
with the largest ablation score as a fallback, preserving localization
recall when attribution signals are weak or ambiguous.\\
\textbf{Complexity.}
In \textsc{BinaryNarrow}, each active node requires two forward passes for
replacement ablation on its left and right halves, and the recursion depth
is at most $\lceil \log_2(n/\ell) \rceil$. 
In \textsc{RobustLeaf}, \sys computes ablation scores over the parent
candidate set $\mathcal{P}(\mathcal{C})$ and applies the MAD threshold to
the current leaf $\mathcal{C}$. Since $\mathcal{P}(\mathcal{C})$ is the
immediate parent of a leaf, its size is bounded by $O(\ell)$ under balanced
splitting. Therefore, when $k$ poisoned segments activate $k$ search paths,
the total localization cost is
\begin{equation}
    O\!\left(k\bigl(\log(n/\ell)+\ell\bigr)\right),
\end{equation}
which reduces to $O(k\log n)$ for a constant leaf size $\ell$.

    


\section{Evaluation}
\label{sec:exp}
 
\subsection{Experimental Setup}
\label{sec:exp:setup}

\textbf{Datasets.}
We evaluate \sys on three datasets spanning diverse multi-source scenarios, with statistics summarized in Table~\ref{tab:datasets} in the Appendix~\ref{sec:app-a}. \emph{HotpotQA}~\cite{yang2018hotpotqa} is a Wikipedia-based multi-hop QA benchmark; we construct each query with 6 Wikipedia segments. \emph{MS-MARCO}~\cite{nguyen2016ms} is a large-scale machine reading comprehension and passage retrieval dataset; we use 10 retrieved passages per query to instantiate RAG-style QA. \emph{Amazon Reviews}~\cite{amazon2023} represents review summarization, where 50 user reviews per product are aggregated as external segments. These datasets cover both knowledge-grounded tasks and general multi-source aggregation. The prompt templates are shown in Figures~\ref{fig:prompt_qa} and~\ref{fig:prompt_review}.
\par
\textbf{Attacks.}
\textcolor{blue}{We evaluate \sys against two representative segment-level poisoning attacks. PoisonedRAG~\cite{poisonedrag} is a knowledge-poisoning attack that injects plausible but factually incorrect passages into the knowledge base, causing retrieved evidence to steer the LLM toward attacker-specified answers. ObliInjection~\cite{obliinjection} is a prompt-injection attack that uses an order-oblivious objective to craft adversarial segments that remain effective under different segment permutations.}
\par
\textbf{Target LLMs.}
We evaluate \sys on four open-weight LLMs with accessible internal MLP activations: \emph{Llama-3.1-8B-Instruct}~\cite{llama3}, \emph{Mistral-7B-Instruct-v0.3}~\cite{mistral}, \emph{GPT-OSS-20B}, and \emph{Qwen3-4B-Instruct-2507}~\cite{qwen}. For closed-source API models, including \emph{GPT-5.4-mini} and \emph{Gemini-3.1-flash-Lite}, we use an open-weight LLM as an upstream diagnostic surrogate, as discussed in Section~\ref{sec:discuss:surrogate}.
\par
\textbf{Baseline Defenses.}
\textcolor{blue}{We compare \sys with representative detection and localization baselines using the same input segments and, where applicable, calibration data. For detection, \emph{PPL}~\cite{ppl,jain2023baseline} flags segments whose language-model perplexity exceeds a predefined threshold; \emph{KAD}~\cite{liu2024formalizing,kad} prepends a known-answer detection prompt to each segment and identifies instruction interference when the expected answer is absent; \emph{EVD}~\cite{jia2025promptlocate} encodes segments with a pre-trained sentence encoder and detects outliers in the resulting embedding space; and \emph{DataSentinel}~\cite{liu2025datasentinel} uses a dedicated detection LLM trained through adversarial game-theoretic optimization, for which we evaluate the officially released model. For localization, \emph{SFA}~\cite{petsiuk2018rise} projects each segment's boundary-token activation onto the learned poison direction $\mathbf{v}_{\mathrm{p}}$ and selects the highest-scoring segment; \emph{FRA}~\cite{zeiler2014visualizing} removes each segment individually and identifies the one causing the largest change in the model's internal features or output behavior; \emph{SVA}~\cite{lundberg2017unified} uses Shapley values to estimate each segment's marginal contribution to the poisoning signal; and \emph{PromptLocate}~\cite{jia2025promptlocate} combines oracle-based contamination detection, contextual inconsistency analysis, and binary group search to locate the injected region.}
\par
\textbf{Calibration and Hyperparameter Tuning.}
\textcolor{blue}{For each dataset--attack setting, we randomly sample 100 base instances without replacement and construct paired benign and poisoned prompts, yielding 100 prompts of each class. Calibration, validation, and evaluation splits are created before prompt construction and are disjoint in query/document identifiers for HotpotQA and MS-MARCO and in product/review identifiers for Amazon Reviews.}
This calibration set is used to learn the poison direction $\mathbf{v}_{\mathrm{p}}$, train the linear SVM detector, and estimate the benign distribution parameters $(\boldsymbol{\mu}, \boldsymbol{\Sigma}^{-1})$. The remaining data are reserved for evaluation. 
To avoid data leakage, we tune all hyperparameters on an independent validation split: $\delta=0.02$, $\gamma=0.8$, $|\mathcal{L}_{\mathrm{top}}|=6$, and dataset-specific leaf sizes $\ell=2,3,10$ for HotpotQA, MS-MARCO, and Amazon Reviews, respectively.
By default, we use a poisoning ratio of $10\%$, i.e., $k=\lceil 0.1 n\rceil$ poisoned segments per input, where $n$ is the total number of segments.
All reported results are averaged over five runs, with experiments and language model queries conducted on NVIDIA A6000 GPUs.
\begin{table*}[t]
\renewcommand{\arraystretch}{1.1}
\centering
\scriptsize
\setlength{\tabcolsep}{3pt} 
\begin{tabular}{l|l|cc|cc||cc|cc||cc|cc||cc|cc}
\toprule
\multirow{3}{*}{\textbf{Dataset}} & \multirow{3}{*}{\textbf{Method}} 
& \multicolumn{4}{c||}{\textbf{Llama-3.1-8B-Instruct}} 
& \multicolumn{4}{c||}{\textbf{Mistral-7B-Instruct-v0.3}} 
& \multicolumn{4}{c||}{\textbf{GPT-OSS-20B}} 
& \multicolumn{4}{c}{\textbf{Qwen3-4B-Instruct-2507}} \\
\cmidrule(lr){3-6} \cmidrule(lr){7-10} \cmidrule(lr){11-14} \cmidrule(lr){15-18}
& & \multicolumn{2}{c|}{\textbf{PoisonedRAG}} & \multicolumn{2}{c||}{\textbf{ObliInjection}} 
& \multicolumn{2}{c|}{\textbf{PoisonedRAG}} & \multicolumn{2}{c||}{\textbf{ObliInjection}} 
& \multicolumn{2}{c|}{\textbf{PoisonedRAG}} & \multicolumn{2}{c||}{\textbf{ObliInjection}} 
& \multicolumn{2}{c|}{\textbf{PoisonedRAG}} & \multicolumn{2}{c}{\textbf{ObliInjection}} \\
\cmidrule(lr){3-4} \cmidrule(lr){5-6} 
\cmidrule(lr){7-8} \cmidrule(lr){9-10} 
\cmidrule(lr){11-12} \cmidrule(lr){13-14}
\cmidrule(lr){15-16} \cmidrule(lr){17-18}
& & FPR & FNR & FPR & FNR 
& FPR & FNR & FPR & FNR 
& FPR & FNR & FPR & FNR 
& FPR & FNR & FPR & FNR \\
\midrule

\multirow{5}{*}{HotpotQA} 
& PPL          & 0.36 & 0.27 & 0.46 & 0.47 & 0.44 & 0.35 & 0.52 & 0.38 & 0.42 & 0.61 & 0.24 & 0.22 & 0.47 & 0.46 & 0.48 & 0.37 \\
& KAD          & 0.58 & 0.38 & 0.44 & 0.43 & 0.84 & 0.16 & 0.79 & 0.08 & 0.71 & 0.33 & 0.68 & 0.44 & 0.72 & 0.28 & 0.48 & 0.37 \\
& EVD          & 0.78 & 0.63 & 0.83 & 0.60 & 0.89 & 0.74 & 0.78 & 0.81 & 0.94 & 0.00 & 0.00 & 0.89 & 0.63 & 0.47 & 0.75 & 0.13 \\
& DataSentinel & 0.03 & 0.44 & 0.28 & 0.02 & 0.01 & 0.43 & 0.28 & \textbf{0.01} & 0.06& 0.39 & 0.28 & \textbf{0.01} & 0.05 & 0.40 & 0.28 & \textbf{0.01} \\
& \textbf{\sys (Ours)} 
& \textbf{0.00} & \textbf{0.02} & \textbf{0.00} & \textbf{0.02}
& \textbf{0.00} & \textbf{0.02} & \textbf{0.00} & 0.02
& \textbf{0.00} & \textbf{0.06} & \textbf{0.00} & 0.03
& \textbf{0.00} & \textbf{0.01} & \textbf{0.00} & 0.03 \\
\midrule

\multirow{5}{*}{MS-MARCO} 
& PPL          & 0.27 & 0.32 & 0.42 & 0.26 & 0.65 & 0.45 & 0.25 & 0.41 & 0.45 & 0.42 & 0.10 & 0.15 & 0.53 & 0.46 & 0.36 & 0.39 \\
& KAD          & 0.48 & 0.48 & 0.55 & 0.56 & 0.89 & 0.08 & 0.89 & 0.09 & 0.74 & 0.23 & 0.70 & 0.54 & 0.58 & 0.39 & 0.76 & 0.20 \\
& EVD          & 0.58 & 0.67 & 0.55 & 0.72 & 0.55 & 0.82 & 0.66 & 0.82 & 0.73 & 0.08 & 0.04 & 0.73 & 0.48 & 0.47 & 0.44 & 0.25 \\
& DataSentinel & 0.05 & 0.77 & 0.06 & 0.55 & 0.06 & 0.78 & 0.05 & 0.53 & 0.09 & 0.73 & 0.05 & 0.58 & 0.08 & 0.72 & 0.05 & 0.55 \\
& \textbf{\sys (Ours)} 
& \textbf{0.00} & \textbf{0.02} & \textbf{0.00} & \textbf{0.01}
& \textbf{0.01} & \textbf{0.02} & \textbf{0.00} & \textbf{0.03}
& \textbf{0.02} & \textbf{0.12} & \textbf{0.00} & \textbf{0.05}
& \textbf{0.00} & \textbf{0.02} & \textbf{0.00} & \textbf{0.01} \\
\midrule

\multirow{5}{*}{Amazon Reviews} 
& PPL          & 0.35 & 0.56 & 0.00 & 0.00 & 0.25 & 0.56 & 0.52 & 0.27 & 0.55 & 0.48 & 0.24 & 0.24 & 0.62 & 0.38 & 0.28 & 0.66 \\
& KAD          & 0.98 & 0.00 & 0.98 & 0.00 & 0.99 & 0.00 & 0.99 & 0.00 & 0.21 & 0.78 & 0.17 & 0.93 & 0.79 & 0.10 & 0.80 & 0.19 \\
& EVD          & 0.84 & 0.83 & 0.65 & 0.74 & 0.74 & 0.89 & 0.56 & 0.72 & 0.40 & 0.74 & 0.94 & 0.11 & 0.45 & 0.68 & 0.00 & 0.79 \\
& DataSentinel & \textbf{0.00} & 1.00 & 0.01 & 0.96 & \textbf{0.01 }& 0.98 & 0.00 & 0.94 & 0.03 & 0.97 & \textbf{0.00} & 0.95 & \textbf{0.03} & 0.97 & \textbf{0.01} & 0.96 \\
& \textbf{\sys (Ours)} 
& 0.04 & \textbf{0.12} & \textbf{0.00} & \textbf{0.02}
& 0.03 & \textbf{0.06} & \textbf{0.00} & \textbf{0.02}
& \textbf{0.02} & \textbf{0.14} & 0.01 & \textbf{0.09}
& 0.05 & \textbf{0.16} & 0.05 & \textbf{0.14} \\
\bottomrule
\end{tabular}
\caption{Comprehensive detection performance of different methods against PoisonedRAG and ObliInjection across three datasets and four LLMs. Best results are highlighted in \textbf{bold}.}
\label{tab:detect_main}
\end{table*}
\par
\textbf{Evaluation Metrics.}
For detection, following DataSentinel~\cite{liu2025datasentinel}, we report False Positive Rate (FPR) and False Negative Rate (FNR), where lower values indicate fewer benign prompts incorrectly flagged and fewer poisoned prompts missed. \textcolor{blue}{We additionally report ROC-AUC to evaluate threshold-independent detection performance, where a higher value indicates better discrimination between clean and poisoned prompts.} For localization, we report Recall and F1-score to measure whether poisoned segments are correctly recovered while limiting false localization.
 
\subsection{Main Results}
\label{sec:exp:main}

\subsubsection{Detection Performance}
\label{sec:exp:detect:main}

We evaluate whether \sys can distinguish contaminated prompts from benign ones across three datasets, two attacks, and four LLM architectures. Table~\ref{tab:detect_main} reports FPR and FNR, while Figure~\ref{fig:roc_auc_results} presents ROC-AUC to assess threshold-independent detection performance.
\sys achieves the best overall balance between false alarms and missed detections, with average FPR and FNR of $0.01$ and $0.05$, respectively, outperforming PPL ($0.38$, $0.38$), KAD ($0.70$, $0.29$), EVD ($0.59$, $0.60$), and DataSentinel ($0.08$, $0.61$). Its ROC-AUC ranges from $0.95$ to $1.00$ across all dataset–attack–model combinations, showing that internal activations provide a stable discrimination signal independent of a specific threshold.
\sys remains stable across datasets. On HotpotQA, its FPR is consistently $0.00$ and its FNR does not exceed $0.06$. It also maintains low errors on MS-MARCO, while keeping FPR and FNR within $0.05$ and $0.16$, respectively, on the more challenging Amazon Reviews dataset. In contrast, DataSentinel suffers substantial missed detections under domain shifts, with FNRs of $0.53$--$0.78$ on MS-MARCO and approximately $0.94$ on Amazon Reviews. KAD can reduce FNR in some settings but incurs FPRs of approximately $0.98$. Overall, \sys avoids both high false alarms and missed detections, providing more reliable detection across settings.

\subsubsection{Localization Performance}
\label{sec:exp:locate:main}

We evaluate whether detected attacks can be traced to the responsible input segments. Table~\ref{tab:locate_main} reports Recall and F1-score across three datasets, two attacks, and four LLM architectures, where F1-score additionally penalizes over-localization. Across all settings, \sys achieves an average Recall of $0.94$ and F1-score of $0.90$, substantially outperforming the strongest baseline, FRA, whose average F1-score is $0.43$.
The advantage is particularly strong on the QA datasets: \sys achieves an average F1-score of $0.95$ on HotpotQA and average Recall and F1-score of $0.97$ and $0.93$ on MS-MARCO, respectively. Although Amazon Reviews is more challenging due to its shorter, noisier, and semantically overlapping segments, \sys obtains the best F1-score in every setting. For example, under ObliInjection on Mistral-7B-Instruct-v0.3, PromptLocate achieves slightly higher Recall than \sys ($0.94$ vs.\ $0.93$) but a substantially lower F1-score ($0.51$ vs.\ $0.91$), indicating over-localization. Overall, \sys more reliably recovers poisoned segments while limiting false attribution.

\begin{table*}[!t]
\renewcommand{\arraystretch}{1.2}
\centering
\scriptsize
\setlength{\tabcolsep}{1.2pt}
\begin{tabular}{l|l|cc|cc||cc|cc||cc|cc||cc|cc}
\toprule
\multirow{3}{*}{\textbf{Dataset}} & \multirow{3}{*}{\textbf{Method}} 
& \multicolumn{4}{c||}{\textbf{Llama-3.1-8B-Instruct}} 
& \multicolumn{4}{c||}{\textbf{Mistral-7B-Instruct-v0.3}} 
& \multicolumn{4}{c||}{\textbf{GPT-OSS-20B}} 
& \multicolumn{4}{c}{\textbf{Qwen3-4B-Instruct-2507}} \\
\cmidrule(lr){3-6} \cmidrule(lr){7-10} \cmidrule(lr){11-14} \cmidrule(lr){15-18}

& & \multicolumn{2}{c|}{\textbf{PoisonedRAG}} & \multicolumn{2}{c||}{\textbf{ObliInjection}}
& \multicolumn{2}{c|}{\textbf{PoisonedRAG}} & \multicolumn{2}{c||}{\textbf{ObliInjection}}
& \multicolumn{2}{c|}{\textbf{PoisonedRAG}} & \multicolumn{2}{c||}{\textbf{ObliInjection}}
& \multicolumn{2}{c|}{\textbf{PoisonedRAG}} & \multicolumn{2}{c}{\textbf{ObliInjection}} \\

\cmidrule(lr){3-4} \cmidrule(lr){5-6}
\cmidrule(lr){7-8} \cmidrule(lr){9-10}
\cmidrule(lr){11-12} \cmidrule(lr){13-14}
\cmidrule(lr){15-16} \cmidrule(lr){17-18}

& & Recall & F1-score & Recall & F1-score
& Recall & F1-score & Recall & F1-score
& Recall & F1-score & Recall & F1-score
& Recall & F1-score & Recall & F1-score \\
\midrule

\multirow{5}{*}{HotpotQA} 
& SFA          & 0.22 & 0.19 & 0.08 & 0.05 & 0.10 & 0.09 & 0.02 & 0.01 & 0.42 & 0.33 & 0.29 & 0.19 & 0.26 & 0.23 & 0.01 & 0.01 \\
& FRA          & 0.49 & 0.36 & 0.76 & 0.51 & 0.46 & 0.36 & 0.67 & 0.45 & 0.77 & 0.52 & 0.66 & 0.44 & 0.79 & 0.52 & 0.23 & 0.16 \\
& SVA          & 0.27 & 0.22 & 0.36 & 0.24 & 0.07 & 0.06 & 0.09 & 0.06 & 0.47 & 0.36 & 0.32 & 0.22 & 0.55 & 0.42 & 0.03 & 0.02 \\
& PromptLocate & 0.71 & 0.45 & 1.00 & 0.35 & 0.69 & 0.45 & 1.00 & 0.36 & 0.71 & 0.45 & 0.99 & 0.36 & 0.71 & 0.44 & 0.99 & 0.36 \\
& \textbf{\sys (Ours)} 
& \textbf{0.97} & \textbf{0.97} & \textbf{1.00} & \textbf{0.90}
& \textbf{1.00} & \textbf{0.97} & \textbf{1.00} & \textbf{0.98}
& \textbf{0.98} & \textbf{0.97} & \textbf{1.00} & \textbf{1.00}
& \textbf{0.98} & \textbf{0.93} & \textbf{0.90} & \textbf{0.87} \\
\midrule

\multirow{5}{*}{MS-MARCO} 
& SFA          & 0.28 & 0.25 & 0.00 & 0.00 & 0.13 & 0.12 & 0.06 & 0.04 & 0.43 & 0.35 & 0.27 & 0.18 & 0.23 & 0.21 & 0.03 & 0.02 \\
& FRA          & 0.64 & 0.48 & 0.92 & 0.61 & 0.64 & 0.48 & 0.92 & 0.61 & 0.79 & 0.56 & 0.79 & 0.53 & 0.87 & 0.60 & 0.83 & 0.55 \\
& SVA          & 0.33 & 0.29 & 0.16 & 0.11 & 0.18 & 0.16 & 0.25 & 0.16 & 0.47 & 0.38 & 0.65 & 0.43 & 0.52 & 0.41 & 0.33 & 0.22 \\
& PromptLocate & 0.42 & 0.36 & 0.97 & 0.40 & 0.43 & 0.36 & 0.98 & 0.38 & 0.45 & 0.37 & 0.97 & 0.39 & 0.45 & 0.38 & 0.99 & 0.41 \\
& \textbf{\sys (Ours)} 
& \textbf{0.98} & \textbf{0.93} & \textbf{1.00} & \textbf{0.94}
& \textbf{0.98} & \textbf{0.96} & \textbf{1.00} & \textbf{0.99}
& \textbf{0.87} & \textbf{0.85} & \textbf{0.99} & \textbf{0.98}
& \textbf{0.98} & \textbf{0.92} & \textbf{0.99} & \textbf{0.87} \\
\midrule

\multirow{5}{*}{Amazon Reviews} 
& SFA          & 0.87 & 0.55 & 0.28 & 0.19 & 0.74 & 0.47 & 0.17 & 0.11 & 0.47 & 0.29 & 0.17 & 0.11 & 0.65 & 0.40 & 0.25 & 0.17 \\
& FRA          & 0.89 & 0.56 & 0.36 & 0.24 & 0.88 & 0.56 & 0.28 & 0.19 & 0.46 & 0.29 & 0.24 & 0.16 & 0.63 & 0.37 & 0.30 & 0.20 \\
& SVA          & 0.91 & 0.57 & 0.46 & 0.31 & 0.86 & 0.55 & 0.20 & 0.14 & 0.45 & 0.28 & 0.26 & 0.17 & 0.66 & 0.41 & 0.44 & 0.29 \\
& PromptLocate & 0.17 & 0.09 & 0.78 & 0.38 & 0.25 & 0.14 & \textbf{0.94} & 0.51 & 0.29 & 0.14 & \textbf{0.91 }& 0.50 & 0.30 & 0.14 & \textbf{0.80} & 0.36 \\
& \textbf{\sys (Ours)} 
& \textbf{0.96} & \textbf{0.94} & \textbf{0.98} & \textbf{0.97}
& \textbf{0.98} & \textbf{0.95} & 0.93 & \textbf{0.91}
& \textbf{0.78} & \textbf{0.74} & 0.72 & \textbf{0.68}
& \textbf{0.80} & \textbf{0.71} & 0.75 & \textbf{0.65} \\
\bottomrule
\end{tabular}
\caption{Poisoned segment localization performance across three datasets and four LLMs. Recall and F1-score are reported as decimal values. Best results are highlighted in \textbf{bold}.}
\label{tab:locate_main}
\end{table*}

\subsection{Robustness against Adaptive Attacks}
\label{sec:exp:adaptive}

To assess the resilience of \sys against adaptive adversaries, we evaluate several evasion strategies. Unless otherwise specified, experiments are conducted on MS-MARCO using Llama-3.1-8B-Instruct and PoisonedRAG. 

\subsubsection{Paraphrase Evasion Attack}
\label{sec:exp:adaptive:para}

Adversarial paraphrasing aims to evade text-level defenses by rewriting poisoned segments while preserving their malicious intent. We evaluate this attack on Llama-3.1-8B-Instruct using MS-MARCO under three paraphrase levels: \textbf{L1} replaces key words with semantic equivalents, \textbf{L2} restructures sentences without changing their meaning, and \textbf{L3} performs semantic reconstruction using an oracle model (\eg, GPT-5.4-mini; see Figure~\ref{fig:prompt_rewrite}) while retaining the adversarial payload. 

Table~\ref{tab:paraphrase} shows that paraphrasing substantially weakens surface-level and external-representation defenses. For detection, PPL's FNR increases from $0.32$ to $0.63$ from the original attack to L3, while EVD remains highly vulnerable, rising from $0.67$ to $0.85$. DataSentinel also exhibits high FNR under paraphrasing, increasing from $0.77$ to $0.82$. In contrast, \sys maintains the lowest or tied-lowest FNR across all settings, ranging from $0.02$ to $0.18$. For localization, PromptLocate drops from $0.36$ F1-score on the original attacks to $0.15$ under L1 and L2, indicating that paraphrased poisoned segments are difficult to attribute using surface inconsistency. FRA and SVA are more stable but remain below $0.60$ F1-score. By comparison, \sys maintains high localization quality, with F1-score between $0.93$ and $0.97$. These results indicate that effective paraphrased attacks still leave activation-space traces that can be captured by the poison-direction projection and localized by \textsc{BinRoL}.

\begin{table}[!t]
\renewcommand{\arraystretch}{1.0}
\centering
\footnotesize
\setlength{\tabcolsep}{5pt}
\begin{tabular}{l|l|cccc}
\toprule
\textbf{Task Category} & \textbf{Method} & \textbf{Orig} & \textbf{L1} & \textbf{L2} & \textbf{L3} \\
\midrule

\multirow{5}{*}{\shortstack[l]{\textbf{Detection} \\ (Metric: FNR $\downarrow$)}} 
& PPL          & 0.32 & 0.50 & 0.55 & 0.63 \\
& KAD          & 0.48 & 0.44 & 0.46 & 0.43 \\
& EVD          & 0.67 & 0.80 & 0.80 & 0.85 \\
& DataSentinel & 0.77  & 0.81 & 0.81 & 0.82 \\
\cmidrule(lr){2-6}
& \textbf{\sys (Ours)}  & \textbf{0.02} & \textbf{0.13} & \textbf{0.09} & \textbf{0.18} \\
\midrule
\multirow{5}{*}{\shortstack[l]{\textbf{Localization} \\ (Metric: F1-score $\uparrow$)}} 
& SFA          & 0.25 & 0.08 & 0.07 & 0.07 \\
& FRA          & 0.48 & 0.57 & 0.56 & 0.58 \\
& SVA          & 0.29 & 0.44 & 0.45 & 0.47 \\
& PromptLocate & 0.36 & 0.15 & 0.15 & 0.19 \\
\cmidrule(lr){2-6}
& \textbf{\sys (Ours)}  & \textbf{0.94} & \textbf{0.97} & \textbf{0.96} & \textbf{0.93} \\
\bottomrule
\end{tabular}
\caption{Robustness against paraphrase evasion using Llama-3.1-8B-Instruct on the MS-MARCO dataset.}
\label{tab:paraphrase}
\end{table}

\subsubsection{Signal Dilution and Poisoning Density}
\label{sec:exp:adaptive:dilution_density}

We evaluate two adaptive strategies that alter the distribution of malicious content. Signal dilution embeds a short malicious payload into a larger benign passage, with the adversarial content ratio decreasing from $100\%$ to $20\%$. Density-based adaptation increases the number of poisoned segments from $k=1$ to $k=5$ with $n=10$, where $k=5$ represents a stress test with half of the context contaminated. All experiments use Llama-3.1-8B-Instruct on MS-MARCO.

\begin{figure}[t]
    \centering
    \includegraphics[width=\columnwidth]{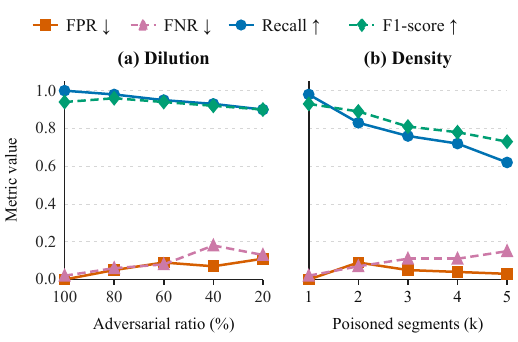}
    \caption{Robustness of \sys under signal dilution and increasing poisoning density. Lower FPR/FNR and higher Recall/F1-score are better.}
    \label{fig:dilution_density}
\end{figure}

As shown in Figure~\ref{fig:dilution_density}(a), signal dilution primarily affects detection. At the strongest dilution level, FPR increases from $0.00$ to $0.11$ and FNR from $0.02$ to $0.13$, with FNR peaking at $0.18$ for the $40\%$ ratio. Localization remains stable, retaining Recall and F1-score of $0.90$ at the $20\%$ ratio. In contrast, increasing poisoning density has a greater impact on localization, as shown in Figure~\ref{fig:dilution_density}(b). From $k=1$ to $k=5$, FNR increases from $0.02$ to $0.15$, while Recall and F1-score decrease from $0.98$ and $0.93$ to $0.62$ and $0.73$, respectively. This degradation occurs because dense poisoning weakens the minority-outlier assumption of MAD-based leaf detection, allowing multiple poisoned segments within the same region to shift the MAD reference and mask weaker signals. Nevertheless, \sys retains an F1-score of $0.73$ even when half of the context is contaminated. 

\textcolor{blue}{Beyond adaptive paraphrasing, signal dilution, and increased poisoning density, we evaluate a stronger defense-aware attacker that jointly targets the SVM detection score and the \textsc{BinRoL} localization signal. With all defense parameters frozen, \sys retains average detection FPR/FNR of $0.03/0.12$ and localization Recall/F1-score of $0.90/0.83$. Detailed attack construction and results are provided in Appendix~\ref{app:adaptive_attack}.}

\subsection{Transferability of \sys}
\label{sec:exp:transfer}

We evaluate the transferability of \sys across unseen attacks, data distributions, and LLM architectures. Cross-attack and cross-dataset experiments use Llama-3.1-8B-Instruct to isolate the transfer variable (Sections~\ref{sec:exp:transfer:attack} and~\ref{sec:exp:transfer:dataset}), while cross-architecture adaptation is evaluated separately in Section~\ref{sec:exp:transfer:model}.

\subsubsection{Cross-Attack Transferability}
\label{sec:exp:transfer:attack}

To characterize the transferability of the learned poison direction $\mathbf{v}_{\mathrm{p}}$ across attack families, we evaluate the empirical zero-shot transfer of \sys between PoisonedRAG (PR, knowledge poisoning) and ObliInjection (OI, prompt injection). Experiments use Llama-3.1-8B-Instruct on MS-MARCO. As shown in Table~\ref{tab:cross_attack}, transfer is asymmetric. When calibrated on PR and evaluated on OI, \sys achieves a localization F1-score of $0.79$, while detection FNR increases to $0.35$. In the reverse direction, FNR increases to $0.64$ and localization F1-score decreases to $0.66$. This asymmetry suggests that PR calibration captures broader activation shifts induced by corrupted evidence, whereas OI calibration emphasizes more attack-specific instruction triggers. Thus, the learned poison direction exhibits partial cross-attack transfer rather than universal attack independence, and performance may degrade against unseen attacks with substantially different activation patterns. Appendix~\ref{app:cross_goal} further evaluates transfer between distinct attack objectives, including answer manipulation and tool-use hijacking.

\begin{table}[!t]
\renewcommand{\arraystretch}{1.0}
\centering
\footnotesize
\setlength{\tabcolsep}{2pt}
\begin{tabular}{l|l|cc|cc}
\toprule
\multirow{2}{*}{\shortstack[l]{\textbf{Calibration} \\ \textbf{(Train)}}} & 
\multirow{2}{*}{\shortstack[l]{\textbf{Evaluation} \\ \textbf{(Test)}}} & 
\multicolumn{2}{c|}{\textbf{Detection Phase}} & 
\multicolumn{2}{c}{\textbf{Localization Phase}} \\
\cmidrule(lr){3-4} \cmidrule(lr){5-6}
 & & FPR ($\downarrow$) & FNR ($\downarrow$) & Recall ($\uparrow$) & F1-score ($\uparrow$) \\
\midrule
PR & PR & 0.00 & 0.02 & 0.98 & 0.93 \\
PR & OI & 0.21 & 0.35 & 0.79 & 0.79 \\
\midrule
OI & OI & 0.00 & 0.01 & 1.00 & 0.94 \\
OI & PR & 0.00 & 0.64 & 0.66 & 0.66 \\
\bottomrule
\end{tabular}
\caption{Cross-attack transferability of \sys on Llama-3.1-8B-Instruct using MS-MARCO. PR denotes PoisonedRAG and OI denotes ObliInjection.}
\label{tab:cross_attack}
\end{table}



\subsubsection{Cross-Dataset Transferability}
\label{sec:exp:transfer:dataset}

To evaluate zero-shot transfer under shifts in contextual semantics and segment length, we calibrate \sys on one dataset and evaluate it on others using Llama-3.1-8B-Instruct under PoisonedRAG. As shown in Table~\ref{tab:cross_dataset}, \sys transfers well between the two QA datasets: calibration on HotpotQA yields an FNR of $0.01$ and localization F1-score of $0.96$ on MS-MARCO. Calibration on either QA dataset also transfers effectively to Amazon Reviews for detection, maintaining an FNR of $0.00$, although localization performance declines under the domain shift. In contrast, calibration on Amazon Reviews transfers less effectively to QA, with FNR increasing to $0.18$ on HotpotQA and $0.23$ on MS-MARCO. This asymmetry suggests that contextually rich QA data provide broader activation coverage, whereas the narrower review domain produces a more specialized poison direction. Thus, diverse calibration data can improve transfer across heterogeneous applications.

\begin{table}[!t]
\renewcommand{\arraystretch}{1.15}
\centering
\footnotesize
\setlength{\tabcolsep}{2pt}
\begin{tabular}{l|cc|cc|cc}
\toprule
\multirow{2}{*}{\textbf{Calibration Data}} & 
\multicolumn{2}{c|}{\textbf{Test: HotpotQA}} & 
\multicolumn{2}{c|}{\textbf{Test: MS-MARCO}} & 
\multicolumn{2}{c}{\textbf{Test: Amazon}} \\
\cmidrule(lr){2-3} \cmidrule(lr){4-5} \cmidrule(lr){6-7}
 & FNR & F1-score & FNR & F1-score & FNR & F1-score \\
\midrule
HotpotQA       & 0.02 & 0.97 & 0.01 & 0.96 & 0.00 & 0.69 \\
MS-MARCO       & 0.02 & 0.95 & 0.02 & 0.93 & 0.00 & 0.74 \\
Amazon Reviews & 0.18 & 0.70 & 0.23 & 0.84 & 0.12 & 0.94 \\
\bottomrule
\end{tabular}
\caption{Cross-dataset transferability of \sys against PoisonedRAG on Llama-3.1-8B-Instruct.}
\label{tab:cross_dataset}
\end{table}



\subsubsection{Cross-Architecture Adaptation (Few-Shot)}
\label{sec:exp:transfer:model}

Because hidden dimensions differ across LLM architectures, the poison direction $\mathbf{v}_{\mathrm{p}}$ cannot be transferred directly and must be re-estimated for each model. We therefore re-calibrate \sys on each target architecture using only $M=50$ samples and evaluate it on MS-MARCO under PoisonedRAG. As shown in Table~\ref{tab:cross_architecture}, few-shot calibration achieves an FNR of $0.07$ and F1-score of $0.94$ on Mistral-7B-Instruct-v0.3. On GPT-OSS-20B, \sys retains an F1-score of $0.78$, although FNR increases to $0.14$. These results show that PoisonedRAG-induced activation shifts remain sufficiently structured for efficient adaptation across open-weight LLMs with limited re-calibration overhead.

\begin{table}[t]
\renewcommand{\arraystretch}{1.15}
\centering
\footnotesize
\setlength{\tabcolsep}{2.5pt}
\begin{tabular}{l|cc|cc}
\toprule
\multirow{2}{*}{\textbf{Target LLM Architecture}} & 
\multicolumn{2}{c|}{\textbf{Detection Phase}} & 
\multicolumn{2}{c}{\textbf{Localization Phase}} \\
\cmidrule(lr){2-3} \cmidrule(lr){4-5}
& FPR ($\downarrow$) & FNR ($\downarrow$) & Recall ($\uparrow$) & F1-score ($\uparrow$) \\
\midrule
Llama-3.1-8B-Instruct    & 0.00 & 0.08 & 0.98 & 0.88 \\
Mistral-7B-Instruct-v0.3 & 0.00 & 0.07 & 0.97 & 0.94 \\
Qwen3-4B-Instruct-2507   & 0.01 & 0.06 & 0.99 & 0.75 \\
GPT-OSS-20B              & 0.02 & 0.14 & 0.84 & 0.78 \\
\bottomrule
\end{tabular}
\caption{Few-shot cross-architecture adaptation of \sys. The poison direction $\mathbf{v}_{\mathrm{p}}$ is re-calibrated on each target LLM using MS-MARCO samples with PoisonedRAG.}
\label{tab:cross_architecture}
\end{table}


\subsection{Ablation Studies}
\label{sec:exp:ablation}

We evaluate the contribution of each \sys component using Llama-3.1-8B-Instruct on MS-MARCO under ObliInjection, unless otherwise specified. In the following tables, HQA, MAR, and AMZ denote HotpotQA, MS-MARCO, and Amazon Reviews, respectively.



\subsubsection{Classifier and Decision Logic Comparison}
\label{sec:exp:ablation:classifier}

We compare Linear SVM with training-free rules (Best Single Layer, Mean Fusion, and Max Fusion) and supervised classifiers, including Logistic Regression (LR)~\cite{cox1958regression}, Random Forest (RF)~\cite{breiman2001random}, and RBF-SVM~\cite{cortes1995support}. As shown in Table~\ref{tab:classifier}, Linear SVM achieves a tied-best average FNR of $0.02$ and the lowest FNR on MS-MARCO ($0.01$). Training-free rules are less stable, with average FNRs of $0.03$--$0.05$; Max Fusion reaches $0.08$ on MS-MARCO. RF also obtains an average FNR of $0.02$, but Linear SVM offers comparable performance with a simpler decision boundary. RBF-SVM performs worst, with an average FNR of $0.11$, indicating possible overfitting under limited calibration data. These results show that projection onto the poison direction produces a sufficiently structured signal for lightweight linear classification.

\begin{table}[t]
\renewcommand{\arraystretch}{1.0}
\centering
\footnotesize
\setlength{\tabcolsep}{5pt}
\begin{tabular}{l|l|ccc|c}
\toprule
\textbf{Category} & \textbf{Method / Classifier} & \textbf{HQA} & \textbf{MAR} & \textbf{AMZ} & \textbf{Avg} \\
\midrule
\multirow{3}{*}{\shortstack[l]{Heuristic \\ Rules}} 
& Best Single Layer  & 0.02 & 0.04 & 0.04 & 0.03 \\
& Mean Fusion        & 0.04 & 0.06 & 0.04 & 0.05 \\
& Max Fusion         & 0.05 & 0.08 & 0.03 & 0.05 \\
\midrule
\multirow{4}{*}{\shortstack[l]{Supervised \\ Classifiers}} 
& Logistic Regression (LR) & 0.02 & 0.04 & 0.02 & 0.03 \\
& Random Forest (RF)       & 0.02 & 0.03 & 0.02 & 0.02 \\
& Non-linear SVM (RBF)     & 0.12 & 0.12 & 0.08 & 0.11 \\
\cmidrule(lr){2-6}
& \textbf{Linear SVM (Ours)} & 0.02 & \textbf{0.01} & 0.02 & \textbf{0.02} \\
\bottomrule
\end{tabular}
\caption{Detection-phase classifier and decision-logic comparison on Llama-3.1-8B-Instruct. The metric is FNR ($\downarrow$).}
\label{tab:classifier}
\end{table}



\subsubsection{Efficacy of the Poison Direction $\mathbf{v}_{\mathrm{p}}$}
\label{sec:exp:ablation:proj}

We assess whether the learned poison direction captures attack-relevant signals by comparing its projection with Random Direction, PCA top-$k$ components, and unreduced MLP activations on Llama-3.1-8B-Instruct using MS-MARCO. As shown in Table~\ref{tab:projection}, $\mathbf{v}_{\mathrm{p}}$ achieves the best performance, with an FNR of $0.01$, Recall of $1.00$, and F1-score of $0.94$. PCA yields an FNR of $0.45$ and F1-score of $0.14$, suggesting that variance-dominant components primarily capture benign semantic variation. Random Direction also provides weak localization, with an F1-score of $0.36$. Raw activations retain useful information and achieve an F1-score of $0.68$, but incur substantially higher computational overhead. These results indicate that the learned direction effectively reduces task-irrelevant variation while preserving poisoning-related activation shifts.


\begin{table}[t]
\renewcommand{\arraystretch}{1.0}
\centering
\footnotesize
\setlength{\tabcolsep}{3pt}
\begin{tabular}{l|cc|cc}
\toprule
\multirow{2}{*}{\textbf{Dimensionality Reduction}} & 
\multicolumn{2}{c|}{\textbf{Detection Phase}} & 
\multicolumn{2}{c}{\textbf{Localization Phase}} \\
\cmidrule(lr){2-3} \cmidrule(lr){4-5}
& FPR ($\downarrow$) & FNR ($\downarrow$) & Recall ($\uparrow$) & F1-score ($\uparrow$) \\
\midrule
Random Direction         & 0.04 & 0.19 & 0.44 & 0.36 \\
PCA (top-$k$ components) & 0.00 & 0.45 & 0.20 & 0.14 \\
No Reduction (Raw acts)  & 0.00 & 0.02 & 0.72 & 0.68 \\
\midrule
\textbf{$\mathbf{v}_{\mathrm{p}}$ Projection (Ours)} 
                         & \textbf{0.00} & \textbf{0.01} & \textbf{1.00} & \textbf{0.94} \\
\bottomrule
\end{tabular}
\caption{Impact of dimensionality reduction methods on detection and localization on MS-MARCO.}
\label{tab:projection}
\end{table}



\subsubsection{Necessity of Replacement Ablation}
\label{sec:exp:ablation:ablmethod}

During the coarse-grained search, \textsc{BinRoL} masks candidate segments to isolate their contribution to the poisoning signal. We compare replacement ablation, which substitutes candidates with neutral text, against deletion ablation using Llama-3.1-8B-Instruct. As shown in Table~\ref{tab:ablmethod}, replacement consistently performs better across datasets, improving average localization F1-score from $0.84$ to $0.94$ and from $0.83$ to $0.94$ on HotpotQA. Deletion alters sequence structure and token positions, introducing unrelated activation changes, whereas replacement preserves positional context and enables more reliable attribution.

\begin{table}[t]
\renewcommand{\arraystretch}{1.0}
\centering
\footnotesize
\setlength{\tabcolsep}{6pt}
\begin{tabular}{l|ccc|c}
\toprule
\textbf{Ablation Strategy} & \textbf{HQA} & \textbf{MAR} & \textbf{AMZ} & \textbf{Average} \\
\midrule
Deletion (Physical Masking) & 0.83 & 0.82 & 0.88 & 0.84 \\
\textbf{Replacement (Ours)} & \textbf{0.94} & \textbf{0.90} & \textbf{0.97} & \textbf{0.94} \\
\bottomrule
\end{tabular}
\caption{Replacement versus deletion ablation for localization using Llama-3.1-8B-Instruct. The metric is F1-score ($\uparrow$).}
\label{tab:ablmethod}
\end{table}



\subsubsection{Robustness of MAD-based Leaf Detection}
\label{sec:exp:ablation:leaf}

We compare MAD-based leaf attribution in \textsc{BinRoL} with a conventional $z$-score baseline using Llama-3.1-8B-Instruct on MS-MARCO, under sparse ($k=1$) and dense ($k=3$) poisoning with $n=10$. As shown in Table~\ref{tab:leaf}, all variants achieve a Recall of $1.00$ for $k=1$. Under dense poisoning, however, the $z$-score baseline with $z=0.50$ drops to a Recall of $0.40$ and an F1-score of $0.54$. MAD with $\gamma=0.80$ achieves the highest Recall of $0.62$ and a tied-best F1-score of $0.68$ for $k=3$, while retaining an F1-score of $0.94$ for $k=1$. This robustness stems from median-based statistics. When multiple poisoned candidates occur in the same leaf, their high ablation scores can shift the mean and inflate the standard deviation, raising the $z$-score threshold and masking poisoned segments. By relying on the median, MAD is less sensitive to extreme scores when poisoned candidates do not dominate the leaf, making it more reliable for multi-segment attribution.

\begin{table}[t]
\renewcommand{\arraystretch}{1.0}
\centering
\footnotesize
\setlength{\tabcolsep}{2pt}
\begin{tabular}{l|cc|cc}
\toprule
\multirow{2}{*}{\textbf{Leaf Detection Strategy}} & 
\multicolumn{2}{c|}{\textbf{Sparse ($k{=}1$)}} & 
\multicolumn{2}{c}{\textbf{Dense ($k{=}3$)}} \\
\cmidrule(lr){2-3} \cmidrule(lr){4-5}
 & Recall ($\uparrow$) & F1-score ($\uparrow$) 
 & Recall ($\uparrow$) & F1-score ($\uparrow$) \\
\midrule
$z$-score ($z{=}0.5$)   & 1.00 & 1.00 & 0.40 & 0.54 \\
$z$-score ($z{=}1.0$)   & 1.00 & 0.99 & 0.32 & 0.46 \\
MAD ($\gamma{=}0.3$)    & 1.00 & 0.67 & 0.53 & 0.64 \\
MAD ($\gamma{=}0.5$)    & 1.00 & 0.75 & 0.60 & 0.68 \\
\midrule
\textbf{MAD ($\gamma{=}0.80$, Ours)} 
                         & \textbf{1.00} & \textbf{0.94} & \textbf{0.62} & \textbf{0.68} \\
\bottomrule
\end{tabular}
\caption{Localization performance under different outlier detection strategies on MS-MARCO.}
\label{tab:leaf}
\end{table}

\section{Discussion and Limitations}
\label{sec:discuss}

\subsection{Surrogate-Based Black-Box Defense}
\label{sec:discuss:surrogate}

\sys requires internal MLP activations, which are unavailable for closed-source API models. To support this setting, we deploy an open-weight LLM as an \emph{upstream diagnostic surrogate}. The surrogate detects and localizes poisoned segments, removes them from the prompt, and forwards the sanitized prompt to the target API for final generation, without accessing or modifying the target model's internal states.
End-to-end evaluations against ObliInjection show that this surrogate defense consistently suppresses attacks while preserving task utility. Across all settings, it reduces average ASR from $0.29$ to $0.02$ and increases ACC from $0.71$ to $0.87$, whereas budget-matched random removal provides little benefit. Detailed metric definitions, experimental settings, and per-setting results are provided in Appendix~\ref{sec:app:blackbox-surrogate}.

\subsection{System Efficiency and Practicality}
\label{sec:discuss:efficiency}

\textbf{Data-efficient and maintainable calibration.}
\sys uses a small labeled calibration set to learn the poison direction $\mathbf{v}_{\mathrm{p}}$, train the detector, and estimate the benign distribution parameters $(\boldsymbol{\mu}, \boldsymbol{\Sigma}^{-1})$. We use $M=100$ samples per class, corresponding to 100 benign and 100 poisoned prompts. Benign samples can be collected from operational data, while poisoned samples can be synthesized using known attacks or red-teaming. A sensitivity analysis with $M\in\{10,25,50,100,200\}$ shows that detection improves substantially up to $M=100$ and largely saturates thereafter, while localization remains stable across calibration sizes. Detailed results are provided in Appendix~\ref{app:calibration_size}. Calibration is performed offline, and representative samples from anticipated attacks and data domains can be pooled to train a single detector for each protected model or application context. The calibration set can be periodically updated as new attacks emerge, while runtime detection overhead remains independent of the number of attack types represented during calibration.
\par
\textbf{Computational scalability of localization.}
Table~\ref{tab:overhead_full} in the Appendix~\ref{sec:app-b} compares \textsc{BinRoL} with four localization baselines under $n=50$. SFA requires one forward pass but achieves a low F1-score of $0.25$. FRA performs segment-wise deletion, requiring 51 forward passes and achieving an F1-score of $0.48$. SVA has exponential complexity, while PromptLocate depends on oracle LLM calls and thus has non-comparable overhead. In contrast, \textsc{BinRoL} requires 21 forward passes, achieving a $2.43\times$ speedup over FRA and a higher F1-score of $0.93$. These results show that \textsc{BinRoL} improves both localization efficiency
and accuracy, making \sys practical for multi-source LLM agents.

\subsection{Limitations and Future Work}
\label{sec:discuss:limitations}

\textcolor{blue}{The current \sys targets segment-level poisoning whose influence remains attributable through segment replacement. Although effective under both sparse and dense poisoning, localization accuracy decreases as the number of poisoned segments grows, motivating improved attribution under higher poisoning densities.}
Ultra-long contexts may diffuse poisoning signals across layers and token positions, reducing the effectiveness of last-token probing. Future work could explore multi-layer and token-wise probes to capture earlier transient signals and improve long-context robustness.




\section{Conclusion}
\label{sec:conclusion}

We presented \sys, an internal-state-based defense for detecting and localizing poisoned segments in multi-source LLM inputs. \sys leverages a learned poison direction to detect contamination and uses \textsc{BinRoL} to efficiently localize poisoned segments through replacement ablation and robust leaf detection. Experiments across datasets, attacks, and LLM architectures show that \sys outperforms existing detection and localization baselines, remains robust against adaptive evasion, and can protect black-box APIs via an upstream surrogate model. 





\cleardoublepage

\bibliographystyle{plainurl}
\bibliography{ref1}


\section*{Ethical Considerations}
\label{sec:ethical-considerations}

We structure our ethical analysis around the affected stakeholders, the potential benefits and risks of the research, the safeguards adopted during evaluation, and the justification for publication.
\par
\textbf{Stakeholders and Research Process.}
Relevant stakeholders include users of LLM-based applications, model and service providers, contributors to external data sources, and the research community. Segment-level poisoning may expose users to misinformation, manipulated responses, or unintended tool use, while imposing additional security and operational costs on providers. Our experiments use public benchmark datasets, established attack methods, and synthetically constructed attack variants. We recruited no human participants, collected no new personal information, and made no attempt to identify or re-identify dataset contributors. Open-weight models were evaluated locally. Black-box APIs were queried only through standard inference interfaces using benchmark inputs, without modifying persistent service state. Tool-use hijacking was evaluated through sandboxed mock execution, and no real external actions were performed.
\par
\textbf{Impact of the Research.}
The primary benefit of \sys is improved protection against segment-level poisoning. In addition to detecting contaminated prompts, it localizes malicious segments, enabling targeted sanitization while preserving benign context. Its surrogate-based deployment further extends this protection to closed-source LLM APIs. However, the research also presents dual-use risks: descriptions of defense-aware adaptation and tool-use hijacking may help adversaries refine evasion attempts. False positives or incorrect localization may remove useful information or unfairly implicate benign sources. Moreover, internal activations may encode sensitive properties of deployment inputs if retained without appropriate safeguards.
\par
\textbf{Mitigations.}
We restrict all attacks to controlled benchmark environments and do not poison production agents, knowledge bases, or user content. The tool-hijacking evaluation contains no credentials, live-service automation, or externally consequential actions. Experimental results are reported only in aggregate. For practical deployment, we recommend access controls for internal activations, minimal activation retention, periodic calibration updates, and human review before taking consequential actions against users or data sources. Localization results should be treated as evidence of suspicious influence rather than proof that a source contributor acted maliciously.
\par
\textbf{Ethical Principles and Publication Justification.}
We considered beneficence, respect for persons, justice, and respect for law and the public interest. The work aims to reduce manipulation and misinformation without exposing participants or interfering with live systems. Its targeted sanitization design also limits unnecessary removal of benign content. Although adaptive evaluations provide incremental offensive knowledge, they are necessary to characterize the defense honestly and rely on established attack paradigms rather than introducing a new capability against otherwise secure systems. We therefore conclude that the defensive and reproducibility benefits of conducting and publishing this research outweigh the residual risks under the safeguards described above.

\section*{Open Science}

To facilitate independent validation, we provide the implementation of \sys, together with preprocessing utilities, experimental configurations, and evaluation scripts. These artifacts support reproduction of our main detection and localization experiments and are available at \url{https://anonymous.4open.science/r/Actprob-5573/}.

\appendix
\section{Additional Details of Experiment}
\label{sec:app-a}
\subsection{Neutral Text Templates for Replacement Ablation}
\label{app:neutral_texts}

As detailed in Section~\ref{sec:method:locate}, our \textsc{BinRoL} algorithm employs replacement ablation to mitigate signal leakage and precisely isolate the contribution of individual segments to the adversarial activation shift. To ensure our localization mechanism is robust and not biased toward any specific phrasing, we utilized a diverse pool of semantically neutral strings. During the ablation process, the target segment is uniformly replaced by one of the following neutral templates:
\begin{itemize}[
    leftmargin=*,
    topsep=3pt,
    itemsep=1pt,
    parsep=0pt,
    partopsep=0pt
]
    \item ``No relevant information is available for this query.''
    \item ``This document does not contain useful information.''
    \item ``The content of this source is not applicable.''
\end{itemize}
Our empirical evaluations demonstrate that the exact choice of neutral text from this pool does not significantly impact the drop ratio ($\rho$), corroborating the structural robustness of the poison direction probe ($\mathbf{v}_{\mathrm{p}}$).

\begin{table}[H]
\renewcommand{\arraystretch}{1.05}
\centering
\footnotesize
\begin{tabular}{c|l}
\toprule
\textbf{Symbol} & \textbf{Description} \\
\midrule
$f$ & Target LLM \\
$p$, $s^t$ & Full assembled prompt, target instruction \\
$x_i$ & Retrieved segment from the $i$-th source \\
$n$, $k$ & Total segments, number of poisoned segments \\
$L$, $d$ & Number of transformer layers, MLP hidden dimension \\
$\mathbf{h}_l \in \mathbb{R}^d$ & Last-token MLP activation at layer $l$ \\
$\mathbf{v}_{\mathrm{p}}[l] \in \mathbb{R}^d$ & Poison direction vector at layer $l$ (unit norm) \\
$\mathbf{s} \in \mathbb{R}^L$ & Projection feature: $\mathbf{s} = [\mathbf{h}_l \cdot \mathbf{v}_{\mathrm{p}}[l]]_{l=1}^{L}$ \\
$\mathbf{w} \in \mathbb{R}^L$ & Linear SVM weight vector \\
$\boldsymbol{\mu}$, $\boldsymbol{\Sigma}$ & Benign distribution mean and covariance \\
$\rho$, $a_i$ & Drop ratio, fine-grained ablation score \\
$\delta$, $\gamma$, $\ell$ & Drop threshold, MAD coefficient, leaf size \\
\bottomrule
\end{tabular}
\caption{Summary of Notations}
\label{tab:notation}
\end{table}

\begin{table}[H]
\renewcommand{\arraystretch}{1.0}
\centering
\footnotesize
\setlength{\tabcolsep}{3.5pt}
\begin{tabular}{l|c|c|c}
\toprule
\textbf{Dataset} & \textbf{Task} & \shortstack{\textbf{\#Segs}\\\textbf{/Input}} & \textbf{Avg. Seg. Len.} \\
\midrule
HotpotQA~\cite{yang2018hotpotqa} & Multi-hop QA & 6  & $\sim$140 \\
MS-MARCO~\cite{nguyen2016ms}     & RAG QA       & 10 & $\sim$130 \\
Amazon Reviews~\cite{amazon2023} & Review Summ. & 50 & $\sim$40 \\
\bottomrule
\end{tabular}
\caption{Dataset statistics used in our evaluation.}
\label{tab:datasets}
\end{table}

\begin{figure}[H]
    \centering
    \includegraphics[width=0.99\linewidth]{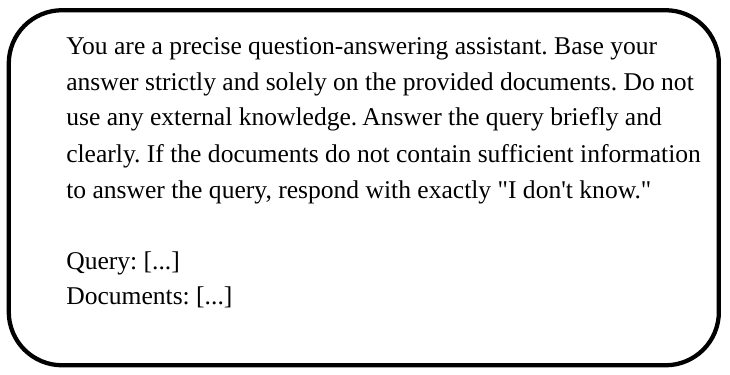}
    \caption{The prompt template for the RAG-based question-answering task.}
    \label{fig:prompt_qa}
\end{figure}

\begin{figure}[H]
    \centering
    \includegraphics[width=0.99\linewidth]{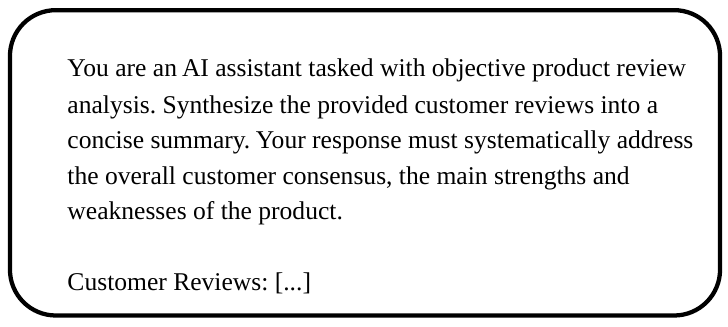}
    \caption{The prompt template for the e-commerce review summarization task.}
    \label{fig:prompt_review}
\end{figure}

\begin{figure}[H]
    \centering
    \includegraphics[width=0.99\linewidth]{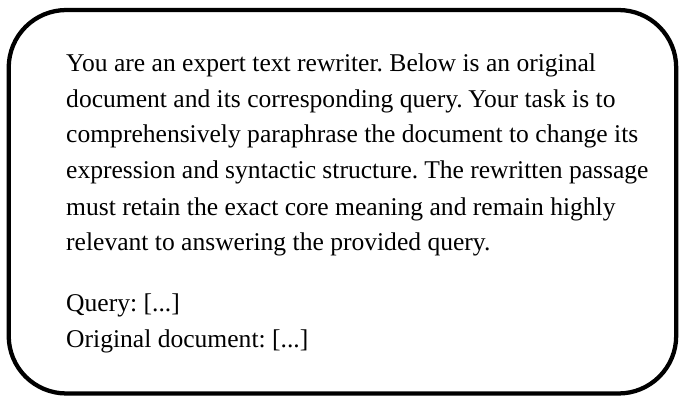}
    \caption{The prompt template used for generating paraphrased adversarial samples.}
    \label{fig:prompt_rewrite}
\end{figure}

\begin{algorithm}[t]
\caption{The \sys Operational Pipeline}
\label{alg:pipeline}
\small
\textbf{Input:} LLM $f$, calibration set with $M$ clean and $M$ poisoned prompts, assembled prompt $p = s^t \| x_1 \| \cdots \| x_n$, number of selected layers $\kappa$, drop threshold $\delta$, MAD scaling constant $\eta$, MAD coefficient $\gamma$, leaf size $\ell$, neutral text $z$ \\
\textbf{Output:} Detection label $\hat{y}$, localized poisoned segments $\mathcal{D}$
 
\begin{algorithmic}[1]
 
\STATE \textit{// Phase I: Offline Calibration} \COMMENT{Executes once prior to deployment}
\FOR{each calibration prompt $p_i$}
    \STATE Forward pass through $f$; extract last-token MLP activation $\mathbf{h}_{i,l}$ at each layer $l$
\ENDFOR

\FOR{$l = 1, \ldots, L$}
    \STATE Compute poison direction: 
    $\mathbf{v}_l \leftarrow 
    \frac{1}{M}\sum_{i=1}^{M}\mathbf{h}^{\mathrm{p}}_{i,l}
    -
    \frac{1}{M}\sum_{i=1}^{M}\mathbf{h}^{\mathrm{c}}_{i,l}$
    \STATE Normalize to unit vector: $\mathbf{v}_{\mathrm{p}}[l] \leftarrow \mathbf{v}_l / \|\mathbf{v}_l\|$
\ENDFOR

\FOR{each calibration prompt $p_i$}
    \STATE Compute projection features: $\mathbf{s}_i \leftarrow [\mathbf{h}_{i,l} \cdot \mathbf{v}_{\mathrm{p}}[l]]_{l=1}^{L}$
\ENDFOR

\STATE Fit scaler on $\{\mathbf{s}_i\}$; train linear SVM to obtain weight vector $\mathbf{w}$
\STATE Estimate benign distribution from unstandardized clean features:
$\boldsymbol{\mu} \leftarrow \mathrm{mean}(\{\mathbf{s}_i^{\mathrm{c}}\})$; 
$\boldsymbol{\Sigma}^{-1} \leftarrow (\mathrm{cov}(\{\mathbf{s}_i^{\mathrm{c}}\}) + \lambda \mathbf{I})^{-1}$
 
\vspace{0.3em}
\STATE \textit{// Phase II: Online Inference} \COMMENT{Executes for each incoming prompt}
\STATE Forward pass on $p$; extract last-token MLP activations $\mathbf{h}_l$
\STATE Compute projection features: $\mathbf{s} \leftarrow [\mathbf{h}_l \cdot \mathbf{v}_{\mathrm{p}}[l]]_{l=1}^{L}$
\STATE $\hat{y} \leftarrow \mathrm{SVM}(\mathrm{scaler}(\mathbf{s}))$

\IF{$\hat{y} == \text{Benign}$}
    \RETURN $\text{Benign}$, $\emptyset$
\ENDIF

\STATE $\mathcal{D} \leftarrow \textsc{BinRoL}(f, p, \{x_1,\ldots,x_n\}, \mathbf{v}_{\mathrm{p}}, \boldsymbol{\mu}, \boldsymbol{\Sigma}^{-1}, \mathbf{w}, \kappa, \delta, \eta, \gamma, \ell, z)$ \COMMENT{Alg.~\ref{alg:binrol}}
\RETURN $\text{Poisoned}$, $\mathcal{D}$
 
\end{algorithmic}
\end{algorithm}

\section{Additional Experimental Results}
\label{sec:app-b}

\subsection{End-to-End Black-Box Surrogate Defense}
\label{sec:app:blackbox-surrogate}

\textcolor{blue}{To evaluate \sys on closed-source LLM APIs, an open-weight surrogate detects and removes poisoned segments before forwarding the sanitized prompt to the target API. Table~\ref{tab:a1-blackbox-surrogate-transfer} reports ASR and ACC. ASR measures whether the output contains or implies the attacker-specified target fact, manifested as induced answers on HotpotQA and MS-MARCO and injected malicious opinions on Amazon Reviews. ACC measures semantic correctness against reference answers for the two QA datasets and overall summary faithfulness, coverage, and factual correctness for Amazon Reviews. ASR and ACC are evaluated independently rather than as complementary outcomes; an Amazon Reviews summary may retain the main benign content while also including the injected opinion, so their sum can exceed $1$. Random Removal uses the same deletion budget as \sys to isolate the benefit of accurate localization. As shown in Table~\ref{tab:a1-blackbox-surrogate-transfer}, No Defense and Random Removal both yield an average ASR of $0.29$, with ACC values of $0.71$ and $0.70$, respectively. In contrast, \sys reduces the average ASR to $0.02$ while increasing ACC to $0.87$, with ASR remaining at or below $0.05$ across all surrogate--target--dataset combinations. These gains demonstrate that accurately localizing and removing poisoned segments, rather than arbitrary deletion, enables effective upstream protection for closed-source LLM APIs.}

\begin{algorithm}[H]
\caption{\textsc{BinRoL}: Binary Search with Robust Leaf Detection}
\label{alg:binrol}
\small
\textbf{Input:} LLM $f$, full assembled prompt $p$, segments $\{x_i\}_{i=1}^n$, poison direction $\mathbf{v}_{\mathrm{p}}$, benign distribution $(\boldsymbol{\mu}, \boldsymbol{\Sigma}^{-1})$, SVM weights $\mathbf{w}$, number of selected layers $\kappa$, drop threshold $\delta$, MAD scaling constant $\eta$, MAD coefficient $\gamma$, leaf size $\ell$, neutral text $z$ \\
\textbf{Output:} Set of detected poison positions $\mathcal{D}$
 
\begin{algorithmic}[1]
 
\STATE \textit{// Initialization}
\STATE $\mathcal{L}_{\mathrm{top}} \leftarrow \text{top-}\kappa \text{ indices of } |\mathbf{w}|$; \quad $\mathcal{D} \leftarrow \emptyset$
\STATE $\mathcal{C}_0 \leftarrow \{1, \ldots, n\}$
\STATE Forward pass on $p$; extract last-token MLP activations
from all $L$ layers and compute
$\mathbf{s}_{\mathrm{full}}
=
[\mathbf{h}^{\mathrm{full}}_l\cdot\mathbf{v}_{\mathrm{p}}[l]]_{l=1}^{L}
\in\mathbb{R}^{L}$
\STATE $d_{\mathrm{full}} \leftarrow d(\mathbf{s}_{\mathrm{full}})$ \COMMENT{Eq.~\ref{eq:mahal}}
\STATE \textsc{BinaryNarrow}($\mathcal{C}_0, \mathcal{C}_0, \mathcal{D}$)
\RETURN $\mathcal{D}$
 
\vspace{0.3em}
\STATE \textit{// Procedure 1: Coarse-grained binary narrowing}
\STATE \textbf{procedure} \textsc{BinaryNarrow}($\mathcal{C}, \mathcal{P}, \mathcal{D}$)
\IF{$|\mathcal{C}| \leq \ell$}
    \STATE \textsc{RobustLeaf}($\mathcal{C}, \mathcal{P}, \mathcal{D}$); \textbf{return}
\ENDIF

\STATE Split $\mathcal{C}$ evenly into halves $\mathcal{C}_{\mathrm{L}}$ and $\mathcal{C}_{\mathrm{R}}$

\FOR{each half $\mathcal{H} \in \{\mathcal{C}_{\mathrm{L}}, \mathcal{C}_{\mathrm{R}}\}$}
    \STATE Replace segments $\{x_i\}_{i \in \mathcal{H}}$ with $z$ in $p$
    \STATE Compute drop ratio $\rho_{\mathcal{H}}$ using all-layer projection features. \COMMENT{Eq.~\ref{eq:dropratio}}
\ENDFOR

\STATE $\mathcal{H}^{*} \leftarrow \{\mathcal{H} \in \{\mathcal{C}_{\mathrm{L}}, \mathcal{C}_{\mathrm{R}}\} \mid \rho_{\mathcal{H}} > \delta\}$

\IF{$\mathcal{H}^{*} \neq \emptyset$}
    \STATE \textbf{for each} $\mathcal{H} \in \mathcal{H}^{*}$ \textbf{do} \textsc{BinaryNarrow}($\mathcal{H}, \mathcal{C}, \mathcal{D}$)
\ELSE
    \STATE $\mathcal{H}_{\max} \leftarrow \arg\max_{\mathcal{H} \in \{\mathcal{C}_{\mathrm{L}}, \mathcal{C}_{\mathrm{R}}\}} \rho_{\mathcal{H}}$
    \STATE \textsc{BinaryNarrow}($\mathcal{H}_{\max}, \mathcal{C}, \mathcal{D}$) \COMMENT{Fallback branch}
\ENDIF
 
\vspace{0.3em}
\STATE \textit{// Procedure 2: Fine-grained robust leaf detection}
\STATE \textbf{procedure} \textsc{RobustLeaf}($\mathcal{C}, \mathcal{P}, \mathcal{D}$)

\FOR{each $i \in \mathcal{P}$}
    \STATE Replace $x_i$ with $z$ in $p$; compute ablation score $a_i$ using layers $\mathcal{L}_{\mathrm{top}}$ \COMMENT{Eq.~\ref{eq:ablation_score}}
\ENDFOR

\STATE $\tilde{m} \leftarrow \mathrm{median}(\{a_i\}_{i\in\mathcal{P}})$; \quad
$\hat{\sigma} \leftarrow \eta \cdot \mathrm{median}(\{|a_i-\tilde{m}|\}_{i\in\mathcal{P}})$

\STATE $\mathcal{D}_{\mathrm{leaf}} \leftarrow \{i \in \mathcal{C} \mid a_i > \tilde{m} + \gamma \cdot \hat{\sigma}\}$ \COMMENT{Apply parent-level MAD to leaf candidates}

\STATE \textbf{if} $\mathcal{D}_{\mathrm{leaf}} = \emptyset$ \textbf{then} $\mathcal{D}_{\mathrm{leaf}} \leftarrow \{\arg\max_{i \in \mathcal{C}} a_i\}$ \COMMENT{Fallback}

\STATE $\mathcal{D} \leftarrow \mathcal{D} \cup \mathcal{D}_{\mathrm{leaf}}$
 
\end{algorithmic}
\end{algorithm}

\begin{figure*}[t]
    \centering
    \includegraphics[width=\textwidth]{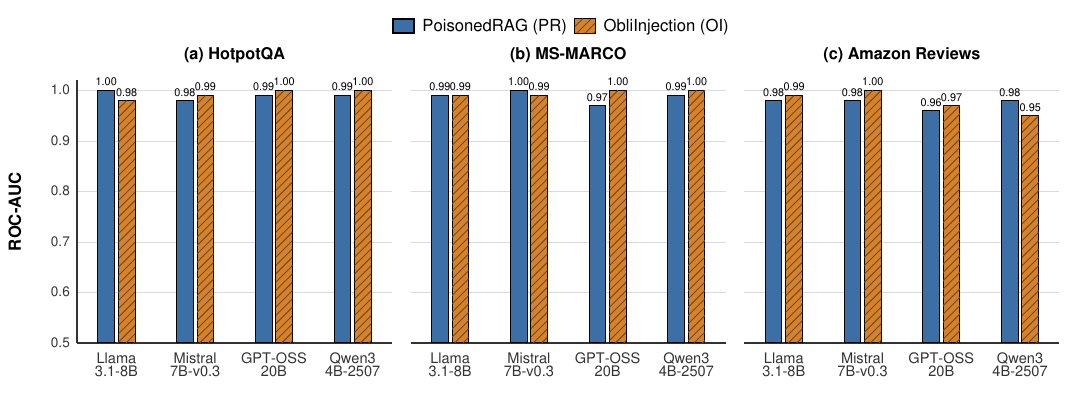}
    \caption{ROC-AUC of \sys across three datasets, two attacks, and four LLMs. PR denotes PoisonedRAG, and OI denotes ObliInjection. Higher values indicate better detection performance.}
    \label{fig:roc_auc_results}
\end{figure*}

\newcommand{\metricpair}{\multicolumn{1}{c}{ASR $\downarrow$} & \multicolumn{1}{c}{ACC $\uparrow$}}
\newcommand{\resultblank}{--}

\begin{table*}[!t]
  \renewcommand{\arraystretch}{1.05}
  \centering
  \scriptsize
  \setlength{\tabcolsep}{1.2pt}
  \resizebox{\textwidth}{!}{%
  \begin{tabular}{ll|cc|cc|cc||cc|cc|cc}
      \toprule
      \multirow{3}{*}{\textbf{Dataset}} &
      \multirow{3}{*}{\textbf{Surrogate Model}} &
      \multicolumn{6}{c||}{\textbf{GPT-5.4-mini}} &
      \multicolumn{6}{c}{\textbf{Gemini-3.1-flash-Lite}} \\
      \cmidrule(lr){3-8}\cmidrule(lr){9-14}

      & &
      \multicolumn{2}{c|}{\textbf{No Defense}} &
      \multicolumn{2}{c|}{\textbf{Random Removal}} &
      \multicolumn{2}{c||}{\textbf{ACTPROBE}} &
      \multicolumn{2}{c|}{\textbf{No Defense}} &
      \multicolumn{2}{c|}{\textbf{Random Removal}} &
      \multicolumn{2}{c}{\textbf{ACTPROBE}} \\
      \cmidrule(lr){3-4}\cmidrule(lr){5-6}\cmidrule(lr){7-8}
      \cmidrule(lr){9-10}\cmidrule(lr){11-12}\cmidrule(lr){13-14}

      & &
      \metricpair & \metricpair & \metricpair &
      \metricpair & \metricpair & \metricpair \\
      \midrule

      \multirow{4}{*}{\textbf{HotpotQA}}
      & Llama-3.1-8B-Instruct
      & \multirow{4}{*}{0.22} & \multirow{4}{*}{0.68}
      & 0.18 & 0.70 & 0.00 & 0.78
      & \multirow{4}{*}{0.19} & \multirow{4}{*}{0.71}
      & 0.21 & 0.74 & 0.01 & 0.80 \\

      & Mistral-7B-Instruct-v0.3
      & & & 0.21 & 0.68 & 0.02 & 0.82
      & & & 0.18 & 0.75 & 0.00 & 0.82 \\

      & GPT-OSS-20B
      & & & 0.18 & 0.71 & 0.01 & 0.83
      & & & 0.23 & 0.72 & 0.00 & 0.86 \\

      & Qwen3-4B-Instruct-2507
      & & & 0.20 & 0.67 & 0.00 & 0.85
      & & & 0.20 & 0.69 & 0.02 & 0.80 \\
      \midrule

      \multirow{4}{*}{\textbf{MS-MARCO}}
      & Llama-3.1-8B-Instruct
      & \multirow{4}{*}{0.28} & \multirow{4}{*}{0.64}
      & 0.27 & 0.66 & 0.01 & 0.88
      & \multirow{4}{*}{0.24} & \multirow{4}{*}{0.70}
      & 0.25 & 0.71 & 0.00 & 0.84 \\

      & Mistral-7B-Instruct-v0.3
      & & & 0.24 & 0.67 & 0.04 & 0.82
      & & & 0.20 & 0.72 & 0.02 & 0.81 \\

      & GPT-OSS-20B
      & & & 0.20 & 0.68 & 0.03 & 0.79
      & & & 0.28 & 0.67 & 0.04 & 0.80 \\

      & Qwen3-4B-Instruct-2507
      & & & 0.28 & 0.62 & 0.03 & 0.84
      & & & 0.22 & 0.68 & 0.02 & 0.81 \\
      \midrule

      \multirow{4}{*}{\makecell{\textbf{Amazon Reviews}}}
      & Llama-3.1-8B-Instruct
      & \multirow{4}{*}{0.37} & \multirow{4}{*}{0.80}
      & 0.39 & 0.83 & 0.00 & 1.00
      & \multirow{4}{*}{0.46} & \multirow{4}{*}{0.70}
      & 0.41 & 0.72 & 0.00 & 1.00 \\

      & Mistral-7B-Instruct-v0.3
      & & & 0.42 & 0.84 & 0.02 & 0.98
      & & & 0.42 & 0.69 & 0.03 & 0.96 \\

      & GPT-OSS-20B
      & & & 0.41 & 0.72 & 0.01 & 0.93
      & & & 0.43 & 0.62 & 0.02 & 0.95 \\

      & Qwen3-4B-Instruct-2507
      & & & 0.43 & 0.78 & 0.03 & 0.93
      & & & 0.47 & 0.63 & 0.05 & 0.94 \\
      \bottomrule
  \end{tabular}%
  }
  \caption{End-to-end black-box surrogate defense performance against ObliInjection across datasets, surrogate models, and target APIs. Attack Success Rate (ASR $\downarrow$) and task Accuracy (ACC $\uparrow$) are reported.}
  \label{tab:a1-blackbox-surrogate-transfer}
\end{table*}

\subsection{Robustness against Defense-Aware Adaptive Attacks}
\label{app:adaptive_attack}

\textcolor{blue}{We evaluate \sys against a defense-aware attacker with knowledge of both its detection and localization mechanisms. For each poisoned input, the attacker constructs multiple variants with the same malicious objective and selects among successful variants those that jointly reduce the SVM detection score and the replacement-ablation signal used by \textsc{BinRoL}. Experiments cover three datasets, two attacks, and two representative LLMs. The poison direction, detector, thresholds, and all other defense parameters remain frozen throughout evaluation.}
\textcolor{blue}{As shown in Table~\ref{tab:adaptive_attack}, adaptation increases the average FPR/FNR from $0.01/0.03$ to $0.03/0.12$ and decreases localization Recall/F1-score from $0.98/0.95$ to $0.90/0.83$. The largest degradation occurs for PR on HotpotQA, where FNR rises from $0.02/0.02$ to $0.21/0.25$ and Recall falls from $0.97/1.00$ to $0.88/0.80$ on the two LLMs. In contrast, adaptive OI maintains zero FPR, FNRs of $0.03$--$0.10$, and F1-scores of $0.76$--$0.94$. Thus, defense-aware adaptation weakens \sys, particularly under PR, but does not eliminate its detection and localization capabilities.}

\begin{table*}[!t]
\renewcommand{\arraystretch}{1.0}
\centering
\footnotesize
\setlength{\tabcolsep}{4.5pt}
\begin{tabular}{l|l|l|cc|cc||cc|cc}
\toprule
\multirow{3}{*}{\textbf{Dataset}} &
\multirow{3}{*}{\textbf{Attack}} &
\multirow{3}{*}{\shortstack{\textbf{Attack}\\\textbf{Version}}} &
\multicolumn{4}{c||}{\textbf{Llama-3.1-8B-Instruct}} &
\multicolumn{4}{c}{\textbf{Mistral-7B-Instruct-v0.3}} \\
\cmidrule(lr){4-7}
\cmidrule(lr){8-11}

& & &
\multicolumn{2}{c|}{\textbf{Detection}} &
\multicolumn{2}{c||}{\textbf{Localization}} &
\multicolumn{2}{c|}{\textbf{Detection}} &
\multicolumn{2}{c}{\textbf{Localization}} \\
\cmidrule(lr){4-5}
\cmidrule(lr){6-7}
\cmidrule(lr){8-9}
\cmidrule(lr){10-11}

& & &
FPR ($\downarrow$) & FNR ($\downarrow$) &
Recall ($\uparrow$) & F1-score ($\uparrow$) &
FPR ($\downarrow$) & FNR ($\downarrow$) &
Recall ($\uparrow$) & F1-score ($\uparrow$) \\
\midrule

\multirow{4}{*}{HotpotQA}
& \multirow{2}{*}{PR}
& Original
& 0.00 & 0.02 & 0.97 & 0.97
& 0.00 & 0.02 & 1.00 & 0.97 \\
& & Adaptive
& 0.07 & 0.21 & 0.88 & 0.84
& 0.02 & 0.25 & 0.80 & 0.72 \\
\cmidrule(lr){2-11}

& \multirow{2}{*}{OI}
& Original
& 0.00 & 0.02 & 1.00 & 0.90
& 0.00 & 0.02 & 1.00 & 0.98 \\
& & Adaptive
& 0.00 & 0.05 & 0.96 & 0.89
& 0.00 & 0.07 & 0.92 & 0.84 \\
\midrule

\multirow{4}{*}{MS-MARCO}
& \multirow{2}{*}{PR}
& Original
& 0.00 & 0.02 & 0.98 & 0.93
& 0.01 & 0.02 & 0.98 & 0.96 \\
& & Adaptive
& 0.08 & 0.06 & 0.97 & 0.94
& 0.09 & 0.16 & 0.88 & 0.83 \\
\cmidrule(lr){2-11}

& \multirow{2}{*}{OI}
& Original
& 0.00 & 0.01 & 1.00 & 0.94
& 0.00 & 0.03 & 1.00 & 0.99 \\
& & Adaptive
& 0.00 & 0.03 & 0.96 & 0.92
& 0.00 & 0.08 & 0.99 & 0.80 \\
\midrule

\multirow{4}{*}{Amazon Reviews}
& \multirow{2}{*}{PR}
& Original
& 0.04 & 0.12 & 0.96 & 0.94
& 0.03 & 0.06 & 0.98 & 0.95 \\
& & Adaptive
& 0.02 & 0.18 & 0.87 & 0.81
& 0.05 & 0.20 & 0.75 & 0.71 \\
\cmidrule(lr){2-11}

& \multirow{2}{*}{OI}
& Original
& 0.00 & 0.02 & 0.98 & 0.97
& 0.00 & 0.02 & 0.93 & 0.91 \\
& & Adaptive
& 0.00 & 0.07 & 0.97 & 0.94
& 0.00 & 0.10 & 0.81 & 0.76 \\

\bottomrule
\end{tabular}
\caption{Detection and localization performance of \sys against original and adaptive PR and OI attacks. PR denotes PoisonedRAG and OI denotes ObliInjection.}
\label{tab:adaptive_attack}
\end{table*}

\begin{table*}[h]
\renewcommand{\arraystretch}{1.05}
\centering
\footnotesize
\setlength{\tabcolsep}{4pt}
\begin{tabular}{l|l|l|l|cc|cc}
\toprule
\multirow{2}{*}{\textbf{Attack}} &
\multicolumn{3}{c|}{\textbf{Cross-Goal Pair}} &
\multicolumn{2}{c|}{\textbf{Detection Phase}} &
\multicolumn{2}{c}{\textbf{Localization Phase}} \\
\cmidrule(lr){2-4}
\cmidrule(lr){5-6}
\cmidrule(lr){7-8}

&
\shortstack{\textbf{Evaluation}\\\textbf{Goal}} &
\shortstack{\textbf{Calibration}\\\textbf{Goal}} &
\shortstack{\textbf{Transfer}\\\textbf{Type}} &
FPR ($\downarrow$) &
FNR ($\downarrow$) &
Recall ($\uparrow$) &
F1-score ($\uparrow$) \\
\midrule

\multirow{4}{*}{PR} &
\multirow{2}{*}{\shortstack[l]{Answer\\Manipulation}} &
Answer Manipulation &
Same-goal &
0.00 & 0.02 & 0.98 & 0.93 \\

&
&
Tool-Use Hijacking &
Cross-goal &
0.00 & 1.00 & 0.15 & 0.12 \\

\cmidrule(lr){2-8}

&
\multirow{2}{*}{\shortstack[l]{Tool-Use\\Hijacking}} &
Answer Manipulation &
Cross-goal &
0.02 & 0.73 & 0.06 & 0.05 \\

&
&
Tool-Use Hijacking &
Same-goal &
0.00 & 0.05 & 0.91 & 0.96 \\

\midrule

\multirow{4}{*}{OI} &
\multirow{2}{*}{\shortstack[l]{Answer\\Manipulation}} &
Answer Manipulation &
Same-goal &
0.00 & 0.01 & 1.00 & 0.94 \\

&
&
Tool-Use Hijacking &
Cross-goal &
0.00 & 1.00 & 0.01 & 0.01 \\

\cmidrule(lr){2-8}

&
\multirow{2}{*}{\shortstack[l]{Tool-Use\\Hijacking}} &
Answer Manipulation &
Cross-goal &
0.02 & 0.99 & 0.44 & 0.28 \\

&
&
Tool-Use Hijacking &
Same-goal &
0.00 & 0.05 & 0.91 & 0.96 \\

\bottomrule
\end{tabular}

\caption{Cross-goal transferability of \sys on
Llama-3.1-8B-Instruct using MS-MARCO under PoisonedRAG and
ObliInjection. PR and OI denote PoisonedRAG and ObliInjection,
respectively.}
\label{tab:cross_goal_transfer}
\end{table*}

\subsection{Cross-Goal Transferability}
\label{app:cross_goal}

\textcolor{blue}{We examine whether a poison direction learned for one attack goal transfers to another. For both PoisonedRAG (PR) and ObliInjection (OI), we consider two goals: \emph{answer manipulation}, which steers the model toward an attacker-specified answer, and \emph{tool-use hijacking}, which induces the selection of an attacker-specified tool. Experiments are conducted on Llama-3.1-8B-Instruct using MS-MARCO. For each attack mechanism, \sys is calibrated on one goal and evaluated on either the same or the other goal. To instantiate tool-use hijacking, each clean prompt contains the top-10 tools retrieved from a benign tool repository. Following the method introduced in~\cite{shi2025prompt}, we add an optimized malicious tool description to the repository and rerun retrieval. We retain only instances in which the malicious tool enters the top-10 results and successfully induces the attacker-specified tool selection without defense.}

\textcolor{blue}{As shown in Table~\ref{tab:cross_goal_transfer}, \sys performs strongly under same-goal calibration. For answer manipulation, it achieves FNRs of $0.02$ and $0.01$ and localization F1-scores of $0.93$ and $0.94$ under PR and OI, respectively. For tool-use hijacking, it obtains an FNR of $0.05$ and an F1-score of $0.96$ under both attacks. In contrast, cross-goal transfer is limited. Calibrating on tool-use hijacking and evaluating answer manipulation increases the FNR to $1.00$, while the F1-score drops to $0.12$ for PR and $0.01$ for OI. In the reverse direction, the FNR reaches $0.73$ for PR and $0.99$ for OI, with F1-scores of $0.05$ and $0.28$, respectively. The FPR remains at or below $0.02$ across all settings, indicating that the degradation is primarily caused by missed cross-goal attacks rather than increased false alarms. These results show that the learned poison direction captures goal-conditioned activation shifts; therefore, deployments covering heterogeneous attack goals should include representative calibration samples for each goal or maintain separate goal-specific poison directions.}

\subsection{Effect of Calibration Set Size}
\label{app:calibration_size}

\textcolor{blue}{We examine the sensitivity of \sys to the calibration-set size by varying the number of samples per class as $M\in\{10,25,50,100,200\}$, where each setting contains $M$ clean and $M$ poisoned prompts. All other experimental settings are kept fixed. Following the main evaluation, detection and localization are evaluated separately.
As shown in Table~\ref{tab:calibration_size}, increasing $M$ substantially improves detection. For PR, the FNR decreases from $0.21$ at $M=10$ to $0.02$ at $M=100$, while for OI it decreases from $0.25$ to $0.01$. Further increasing $M$ to $200$ provides little additional improvement, indicating that detection performance approaches saturation at $M=100$. Meanwhile, localization remains stable across calibration sizes: PR achieves Recall between $0.95$ and $0.98$ and F1-score between $0.89$ and $0.94$, while OI maintains a Recall of $1.00$ and F1-score between $0.94$ and $0.96$. These results support $M=100$ as a practical balance between calibration cost, detection accuracy, and localization stability.}

\begin{table}[H]
\renewcommand{\arraystretch}{1.5}
\centering
\footnotesize
\setlength{\tabcolsep}{2pt} 
\begin{tabular}{l|c|c|c|c}
\toprule
\shortstack{\textbf{Localization}\\\textbf{Strategy}} 
& \textbf{Complexity} 
& \shortstack{\textbf{Total FP}\\\textbf{/ Calls}} 
& \textbf{Speedup} 
& \shortstack{\textbf{F1-score}} \\
\midrule
SFA     & $O(1)$              & 1 FP             & \textbf{51.00$\times$} & 0.25 \\
FRA     & $O(n)$              & 51 FPs           & 1.00$\times$           & 0.48 \\
SVA     & $O(2^n)$ Exact      & Exponential      & $<$1.00$\times$        & 0.29 \\
PromptLocate & $O(k \log n)$ Oracle & Oracle-dependent & Variable              & 0.36 \\
\midrule
\textbf{\textsc{BinRoL} (Ours)} 
        & \textbf{$O(k \log n)$} & \textbf{21 FPs} & \textbf{2.43$\times$} & \textbf{0.93} \\
\bottomrule
\end{tabular}
\caption{Efficiency and Performance Comparison of Localization Strategies ($n{=}50, k{=}5$). FP denotes Forward Passes. Results are evaluated on Llama-3.1-8B-Instruct and Amazon Reviews.}
\label{tab:overhead_full}
\end{table}

\begin{table}[!t]
\renewcommand{\arraystretch}{1.1}
\centering
\footnotesize
\setlength{\tabcolsep}{3pt}
\begin{tabular}{l|c|cc|cc}
\toprule
\multirow{2}{*}{\textbf{Attack}} &
\multirow{2}{*}{\shortstack{\textbf{Calibration}\\\textbf{Size }$\boldsymbol{M}$}} &
\multicolumn{2}{c|}{\textbf{Detection Phase}} &
\multicolumn{2}{c}{\textbf{Localization Phase}} \\
\cmidrule(lr){3-4} \cmidrule(lr){5-6}
& & FPR ($\downarrow$) & FNR ($\downarrow$)
  & Recall ($\uparrow$) & F1-score ($\uparrow$) \\
\midrule

\multirow{6}{*}{PR}
& 10  & 0.00 & 0.21 & 0.96 & 0.94 \\
& 25  & 0.01 & 0.11 & 0.95 & 0.92 \\
& 50  & 0.00 & 0.06 & 0.95 & 0.89 \\
& 100 & 0.00 & 0.02 & 0.98 & 0.93 \\
& 200 & 0.00 & 0.01 & 0.98 & 0.92 \\

\midrule

\multirow{6}{*}{OI}
& 10  & 0.00 & 0.25 & 1.00 & 0.96 \\
& 25  & 0.00 & 0.13 & 1.00 & 0.95 \\
& 50  & 0.00 & 0.09 & 1.00 & 0.96 \\
& 100 & 0.00 & 0.01 & 1.00 & 0.94 \\
& 200 & 0.00 & 0.01 & 1.00 & 0.95 \\

\bottomrule
\end{tabular}
\caption{Effect of calibration-set size on \sys using
Llama-3.1-8B-Instruct and MS-MARCO.
PR denotes PoisonedRAG and OI denotes ObliInjection.}
\label{tab:calibration_size}
\end{table}

\begin{figure*}[t]
    \centering

    \centering
    \small \textbf{(a) Llama-3.1-8B-Instruct} \par\medskip

    \begin{subfigure}{0.28\textwidth}
        \centering
        \includegraphics[width=\linewidth]{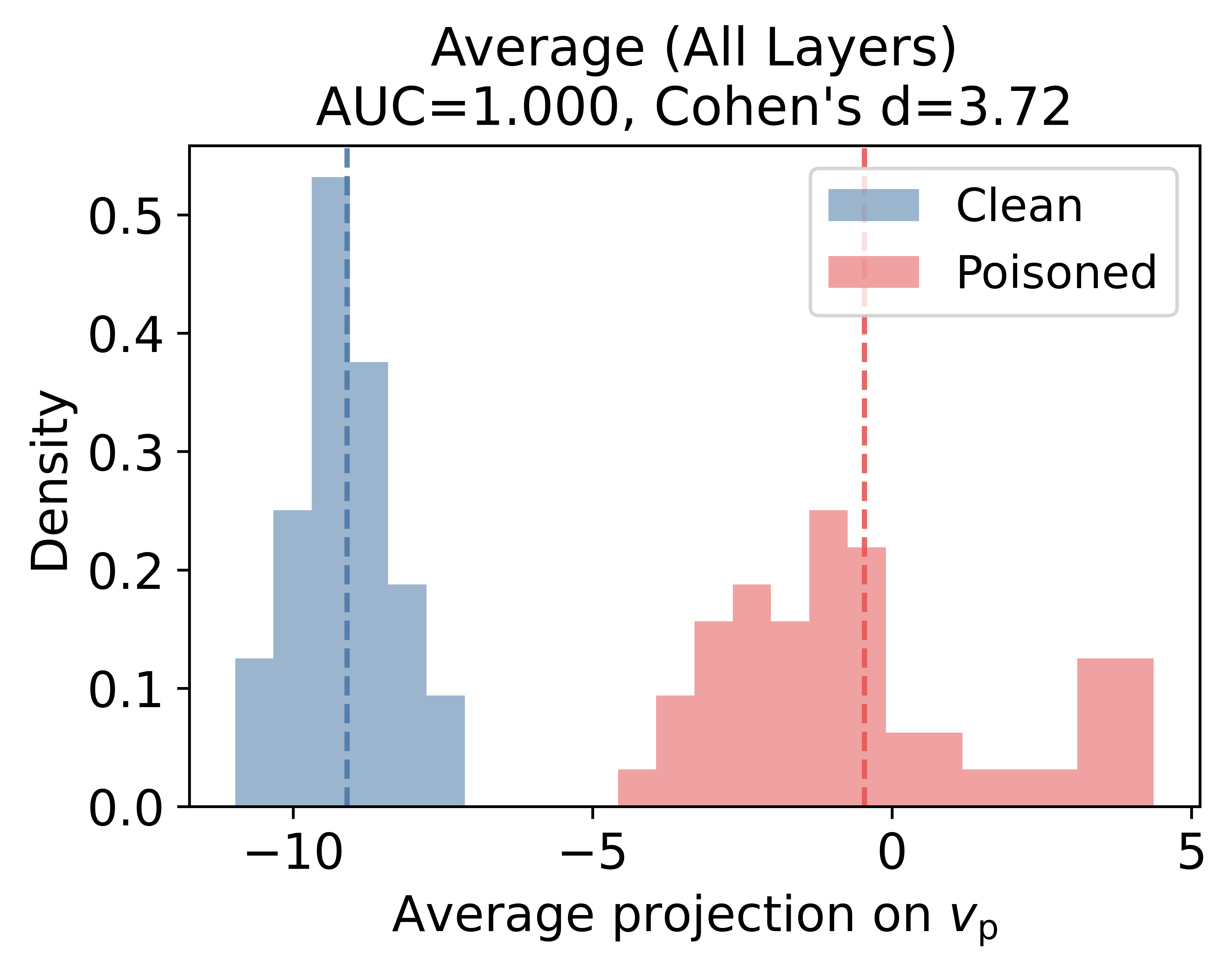}
        \caption{HotpotQA}
        \label{fig:llama_hotpot}
    \end{subfigure}
    \hfill
    \begin{subfigure}{0.28\textwidth}
        \centering
        \includegraphics[width=\linewidth]{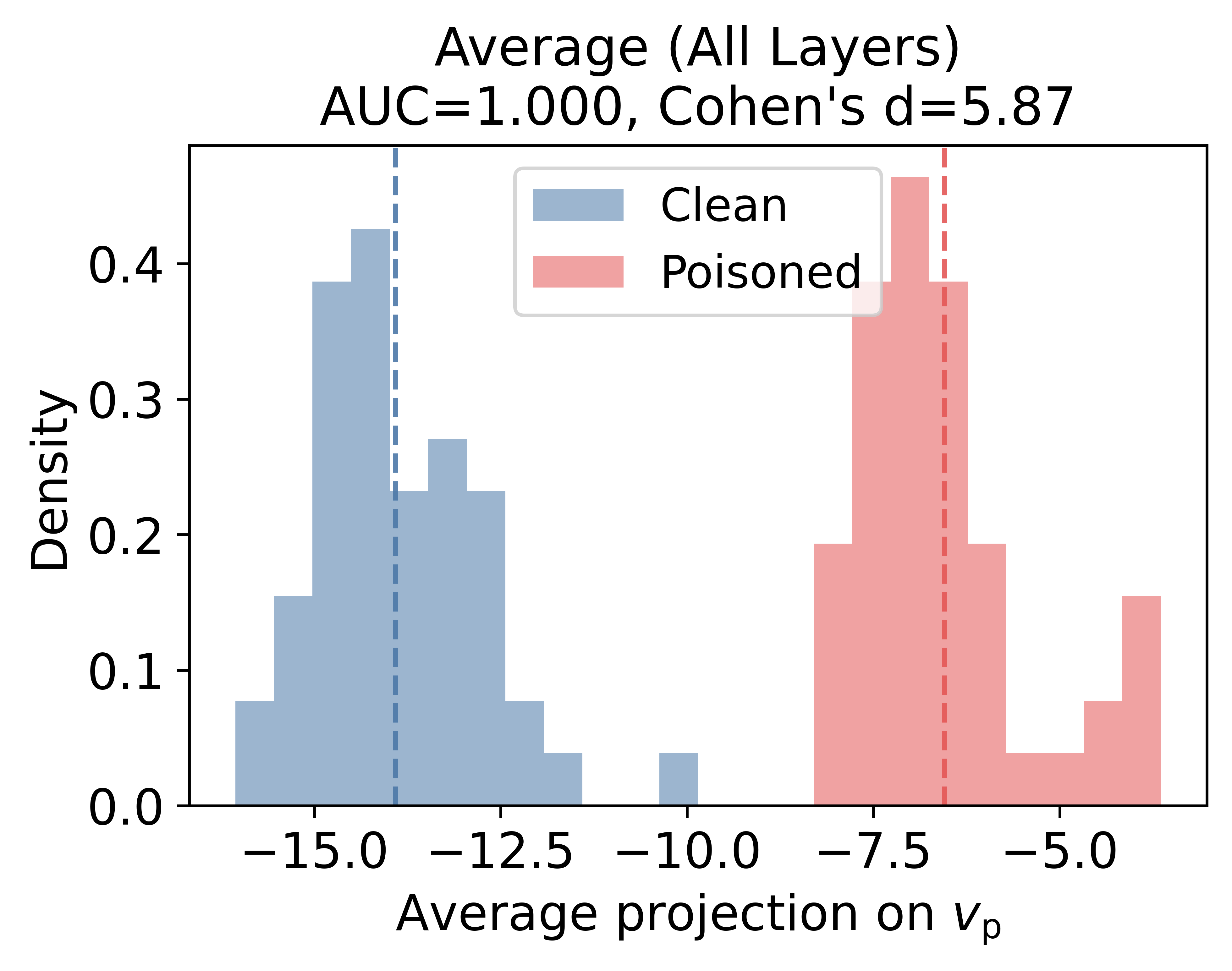}
        \caption{MS-MARCO}
        \label{fig:llama_msmarco}
    \end{subfigure}
    \hfill
    \begin{subfigure}{0.28\textwidth}
        \centering
        \includegraphics[width=\linewidth]{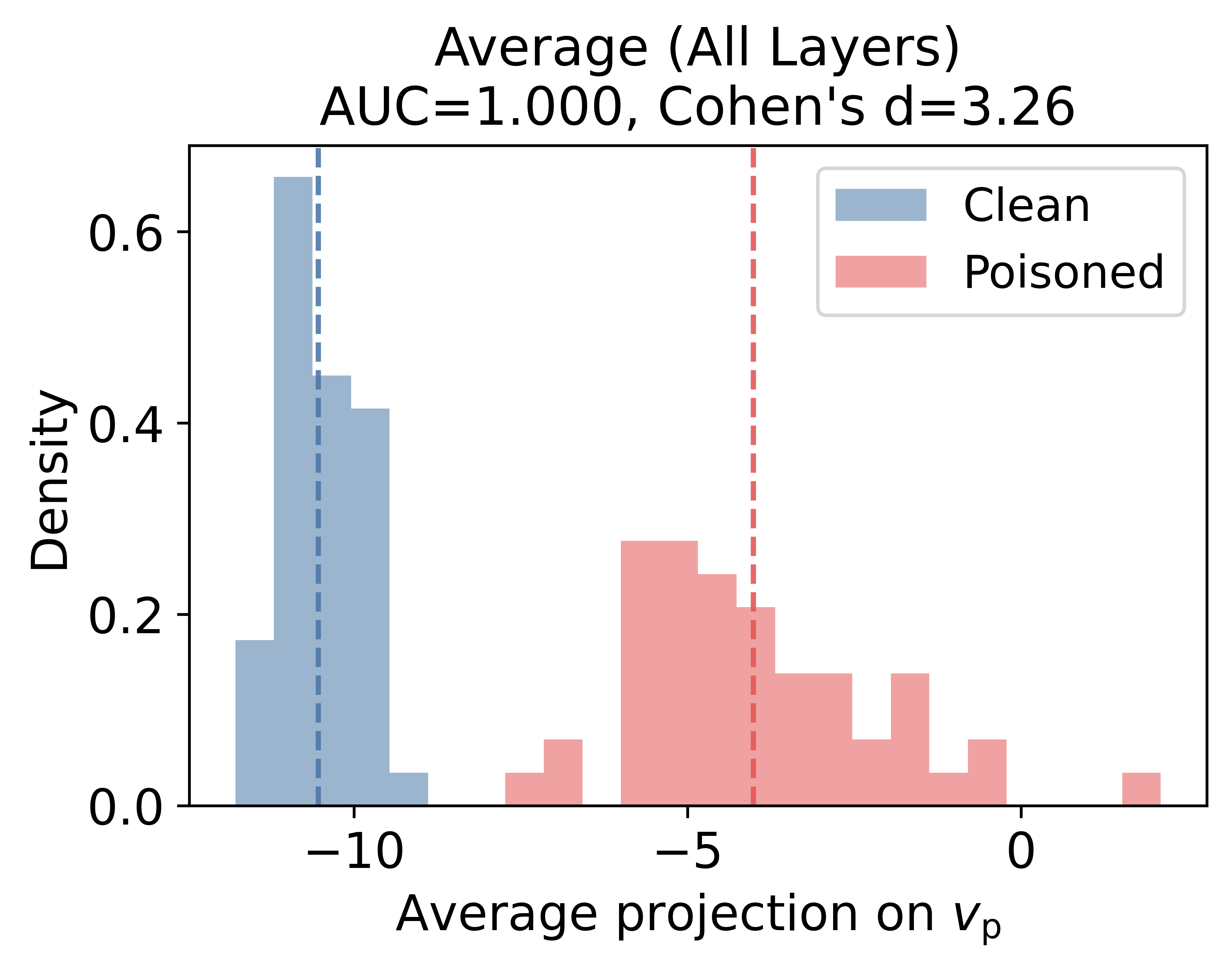}
        \caption{Amazon Reviews}
        \label{fig:llama_reviews}
    \end{subfigure}

    \vspace{1pt} 

    \centering
    \small \textbf{(b) Mistral-7B-Instruct-v0.3} \par\medskip

    \begin{subfigure}{0.28\textwidth}
        \centering
        \includegraphics[width=\linewidth]{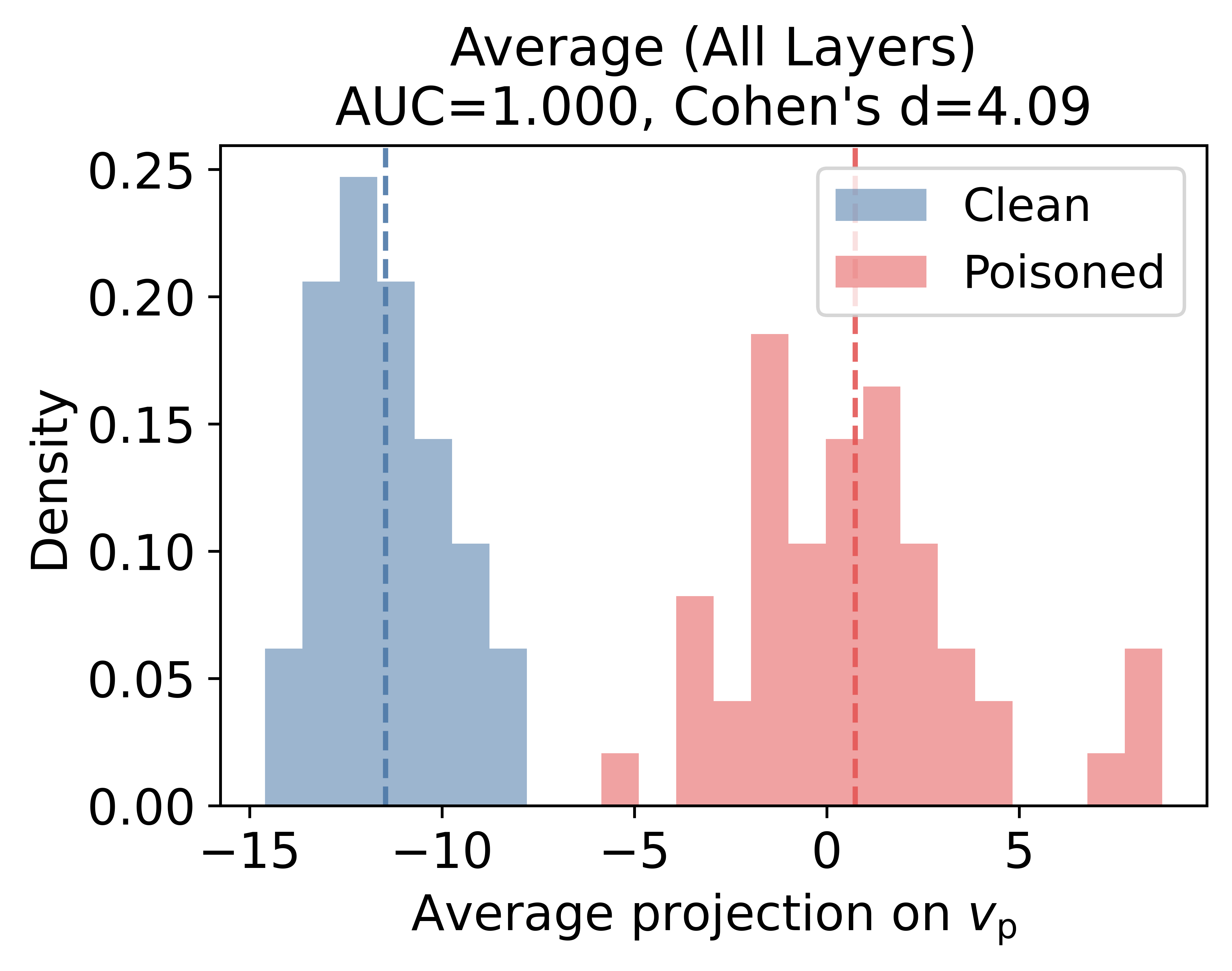}
        \caption{HotpotQA}
        \label{fig:mistral_hotpot}
    \end{subfigure}
    \hfill
    \begin{subfigure}{0.28\textwidth}
        \centering
        \includegraphics[width=\linewidth]{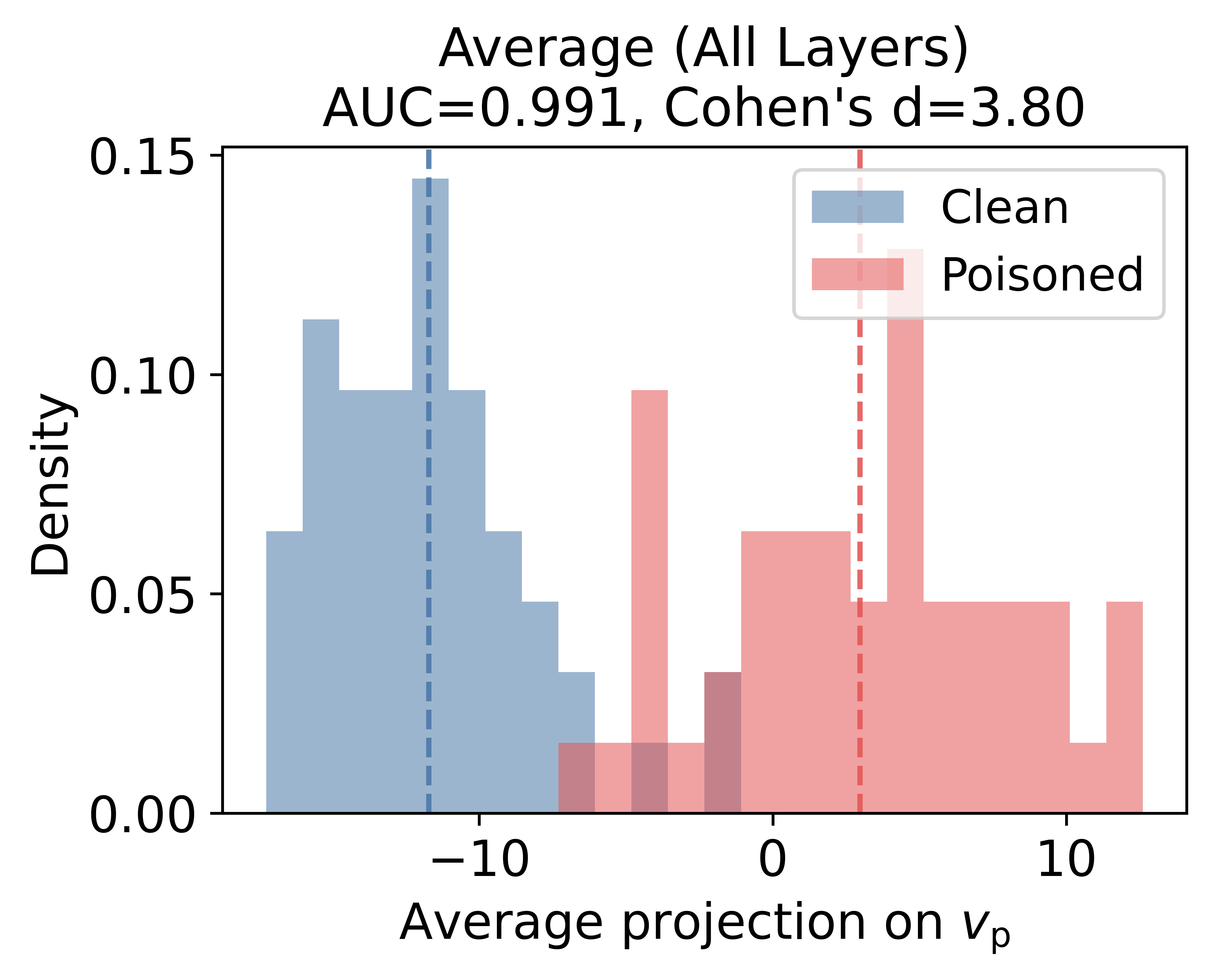}
        \caption{MS-MARCO}
        \label{fig:mistral_msmarco}
    \end{subfigure}
    \hfill
    \begin{subfigure}{0.28\textwidth}
        \centering
        \includegraphics[width=\linewidth]{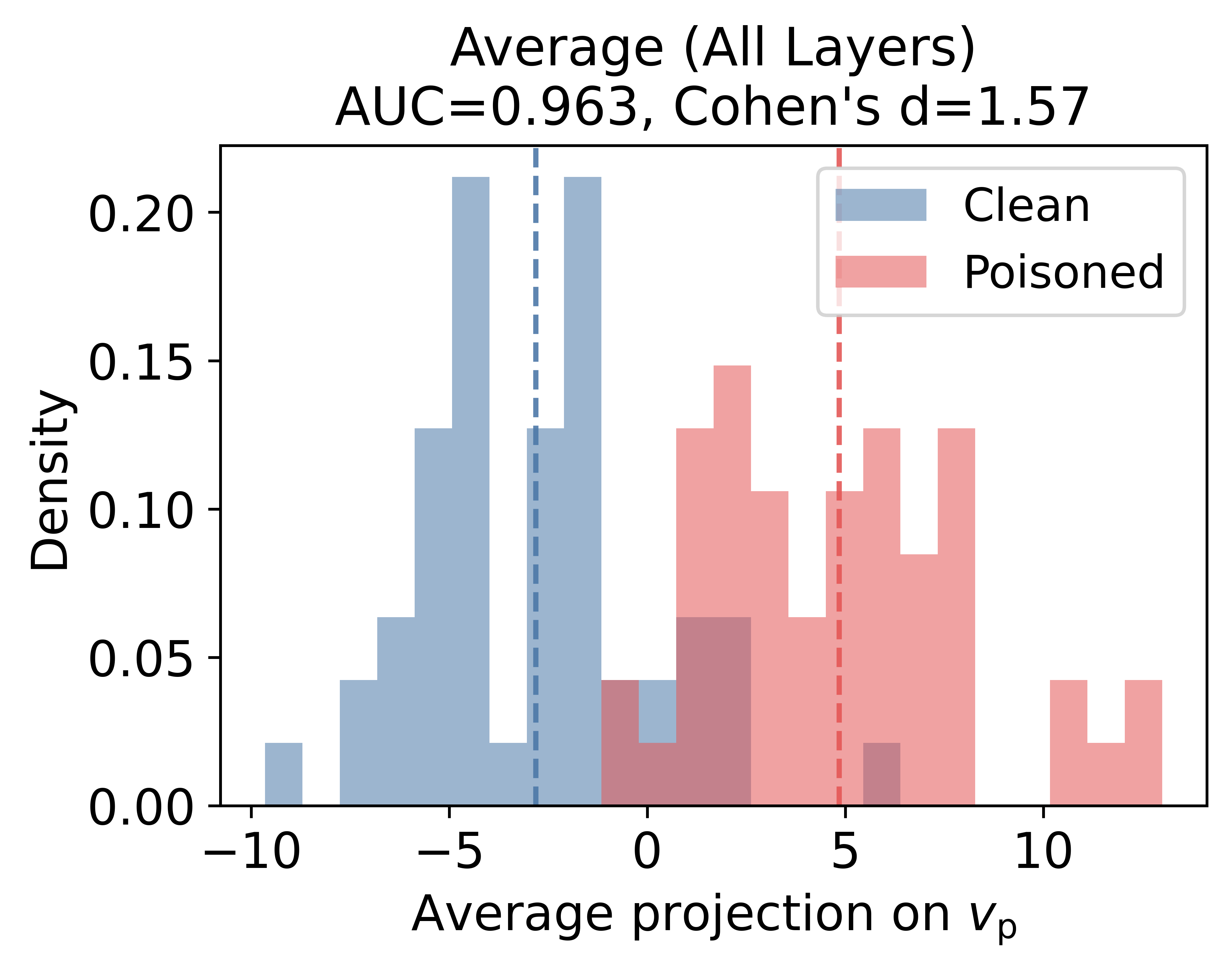}
        \caption{Amazon Reviews}
        \label{fig:mistral_reviews}
    \end{subfigure}

    \vspace{1pt}

    \centering
    \small \textbf{(c) GPT-OSS-20B} \par\medskip

    \begin{subfigure}{0.28\textwidth}
        \centering
        \includegraphics[width=\linewidth]{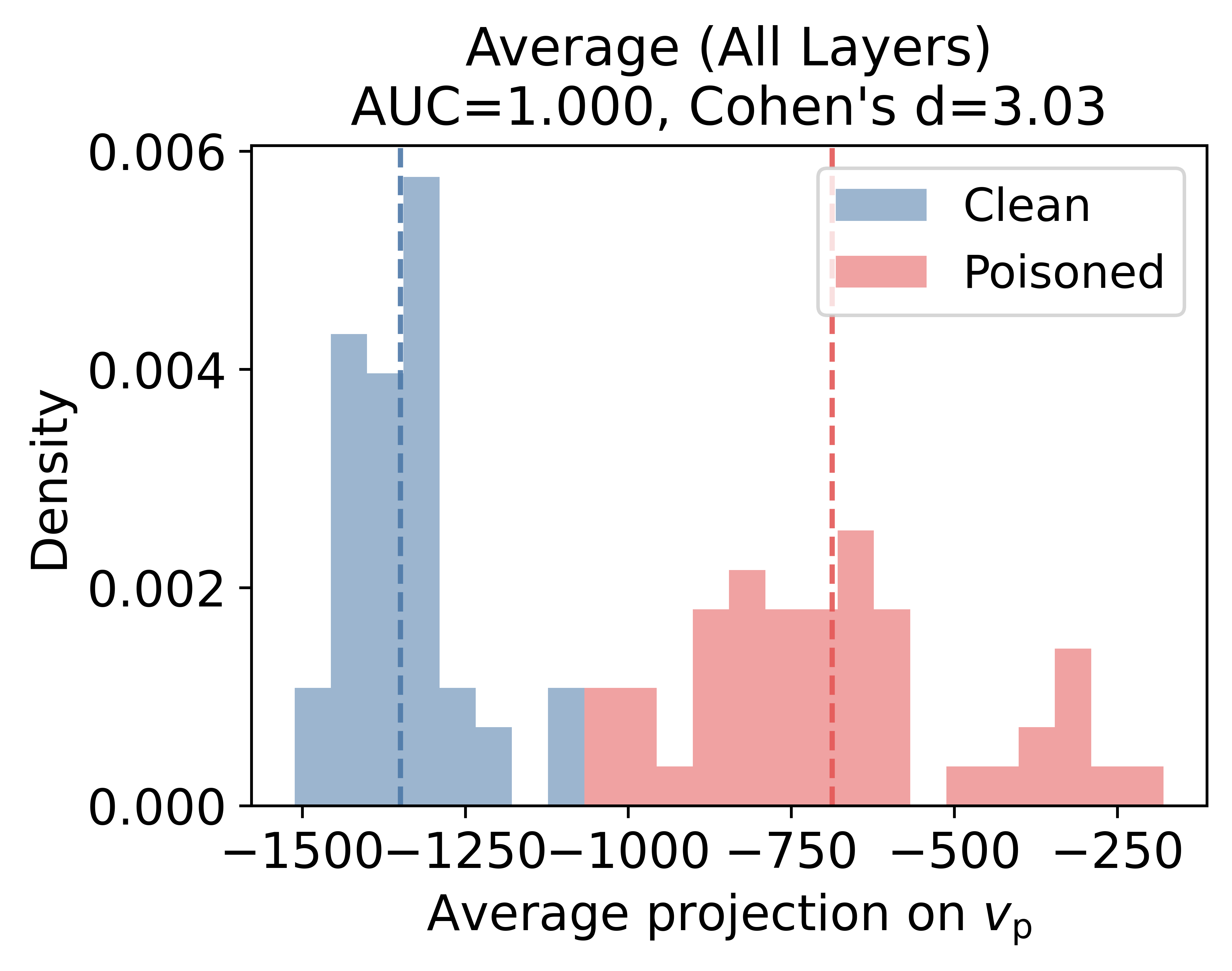}
        \caption{HotpotQA}
        \label{fig:gpt_hotpot}
    \end{subfigure}
    \hfill
    \begin{subfigure}{0.28\textwidth}
        \centering
        \includegraphics[width=\linewidth]{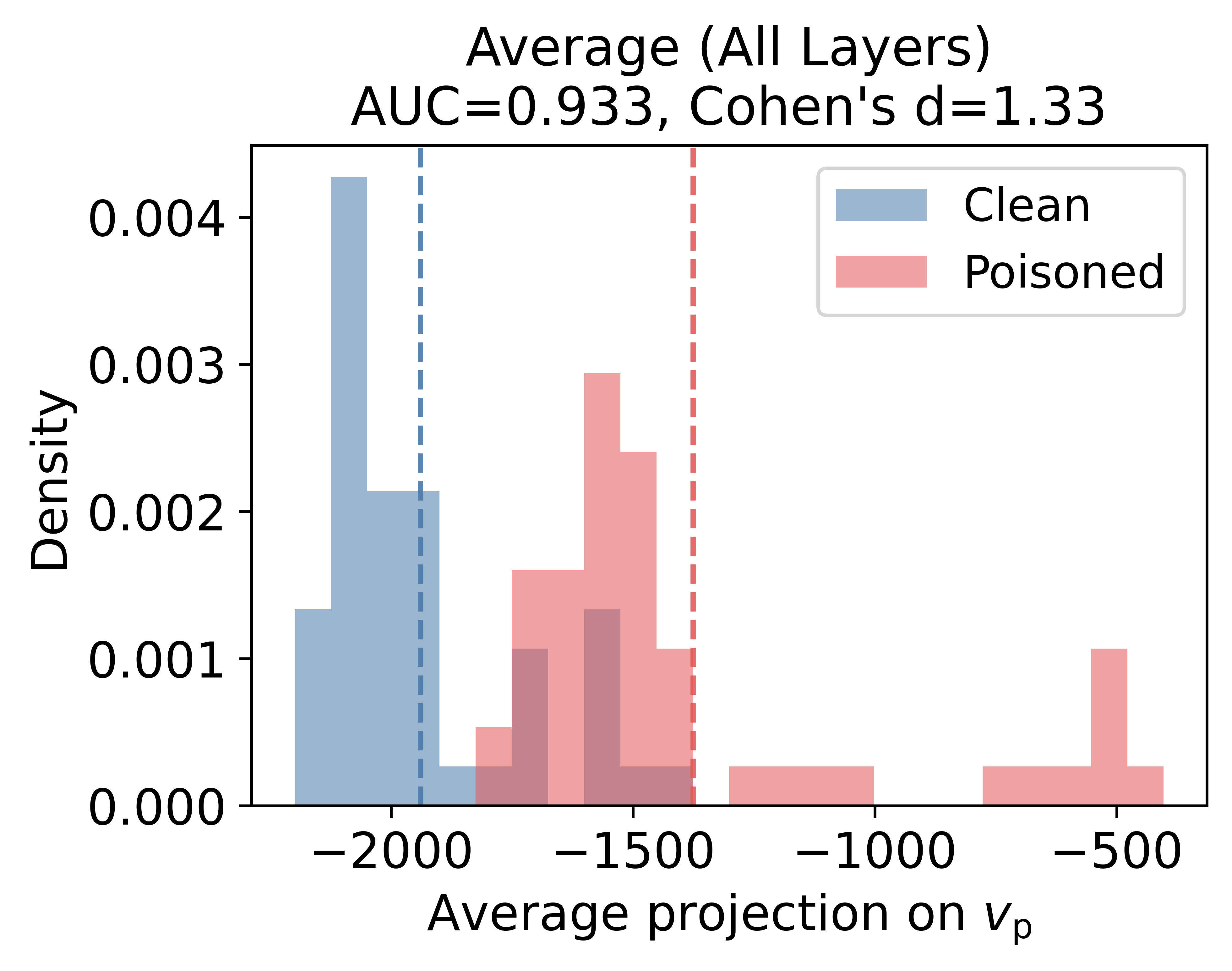}
        \caption{MS-MARCO}
        \label{fig:gpt_msmarco}
    \end{subfigure}
    \hfill
    \begin{subfigure}{0.28\textwidth}
        \centering
        \includegraphics[width=\linewidth]{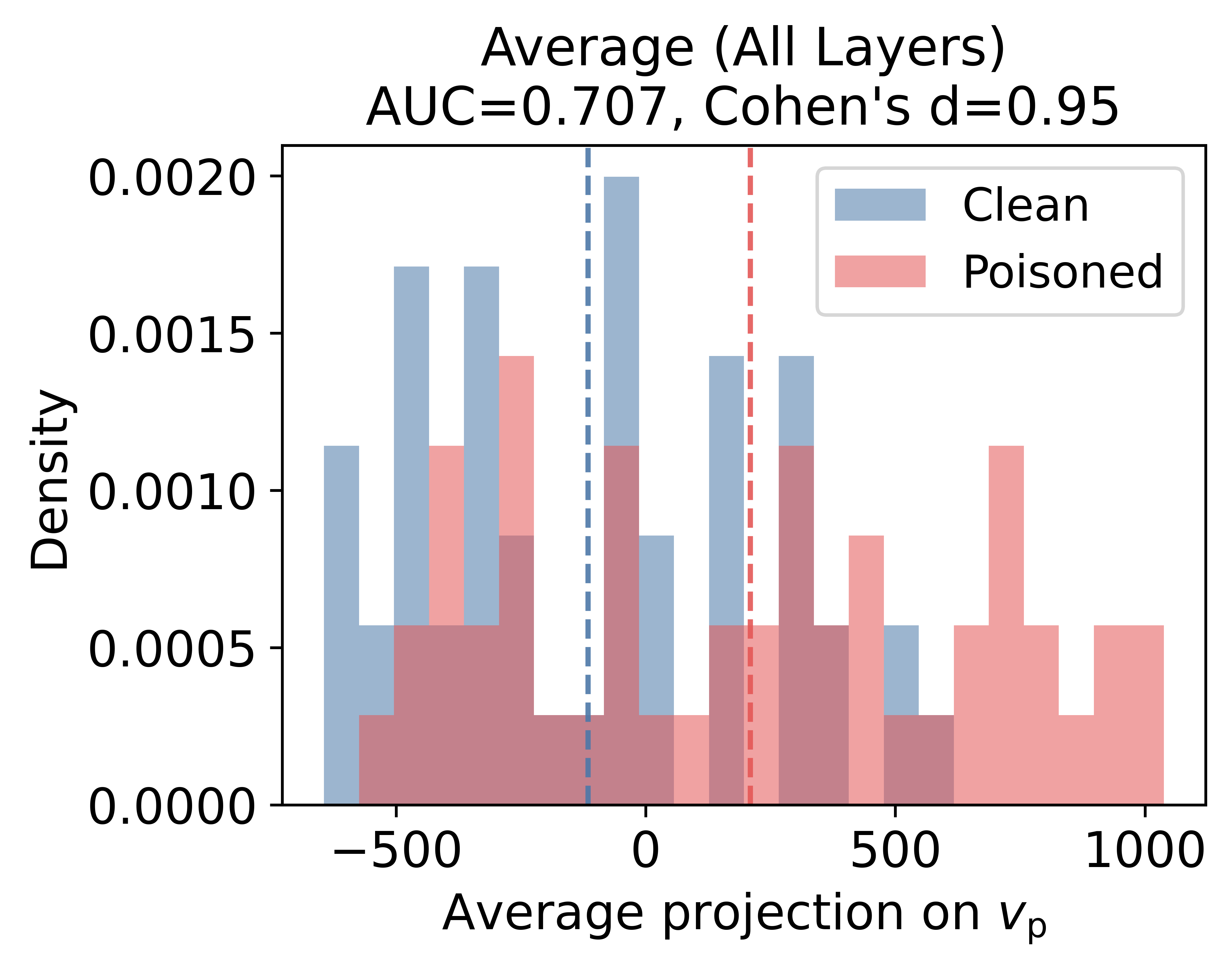}
        \caption{Amazon Reviews}
        \label{fig:gpt_reviews}
    \end{subfigure}

    \vspace{1pt}

    \centering
    \small \textbf{(d) Qwen3-4B-Instruct-2507} \par\medskip

    \begin{subfigure}{0.28\textwidth}
        \centering
        \includegraphics[width=\linewidth]{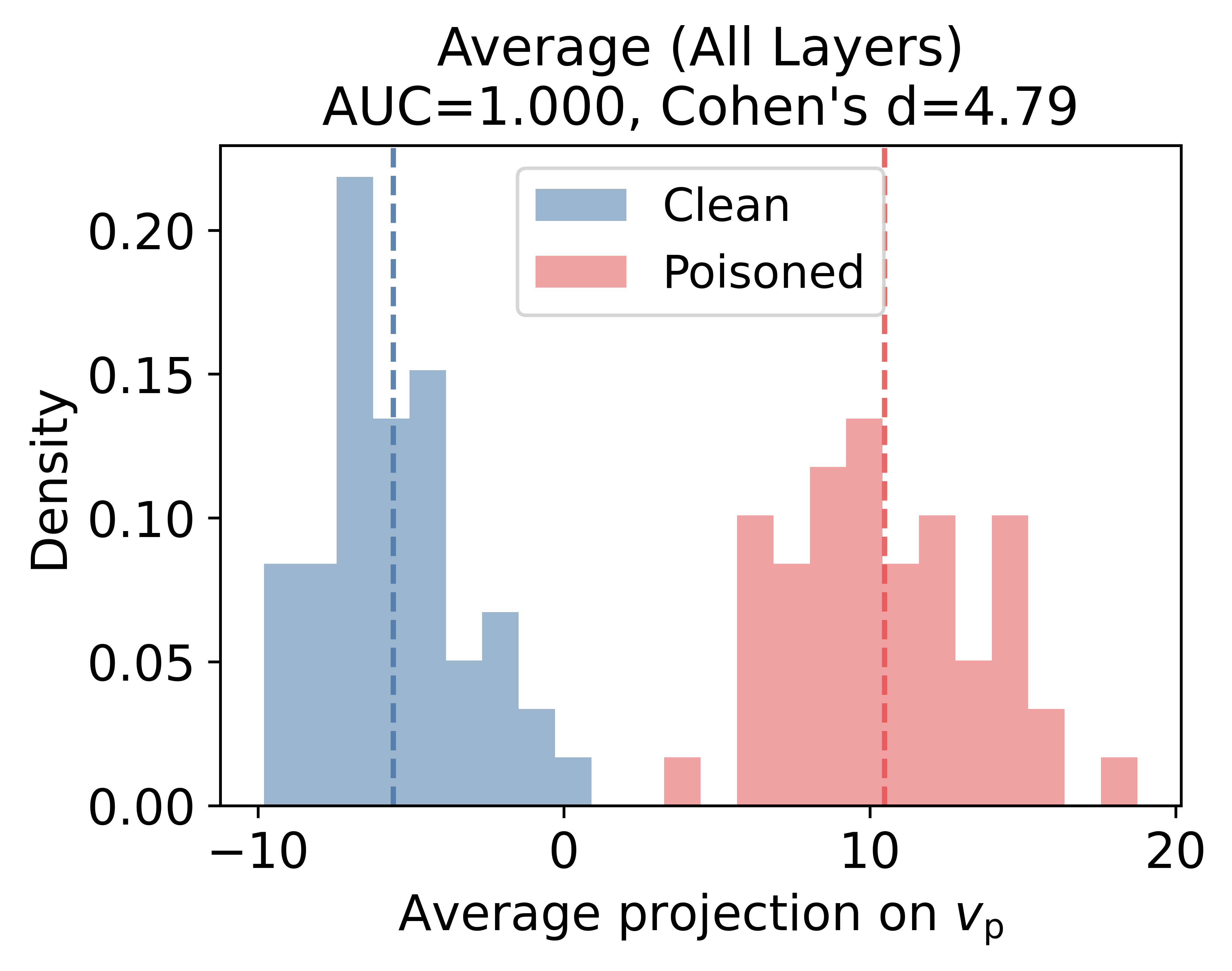}
        \caption{HotpotQA}
        \label{fig:qwen_hotpot}
    \end{subfigure}
    \hfill
    \begin{subfigure}{0.28\textwidth}
        \centering
        \includegraphics[width=\linewidth]{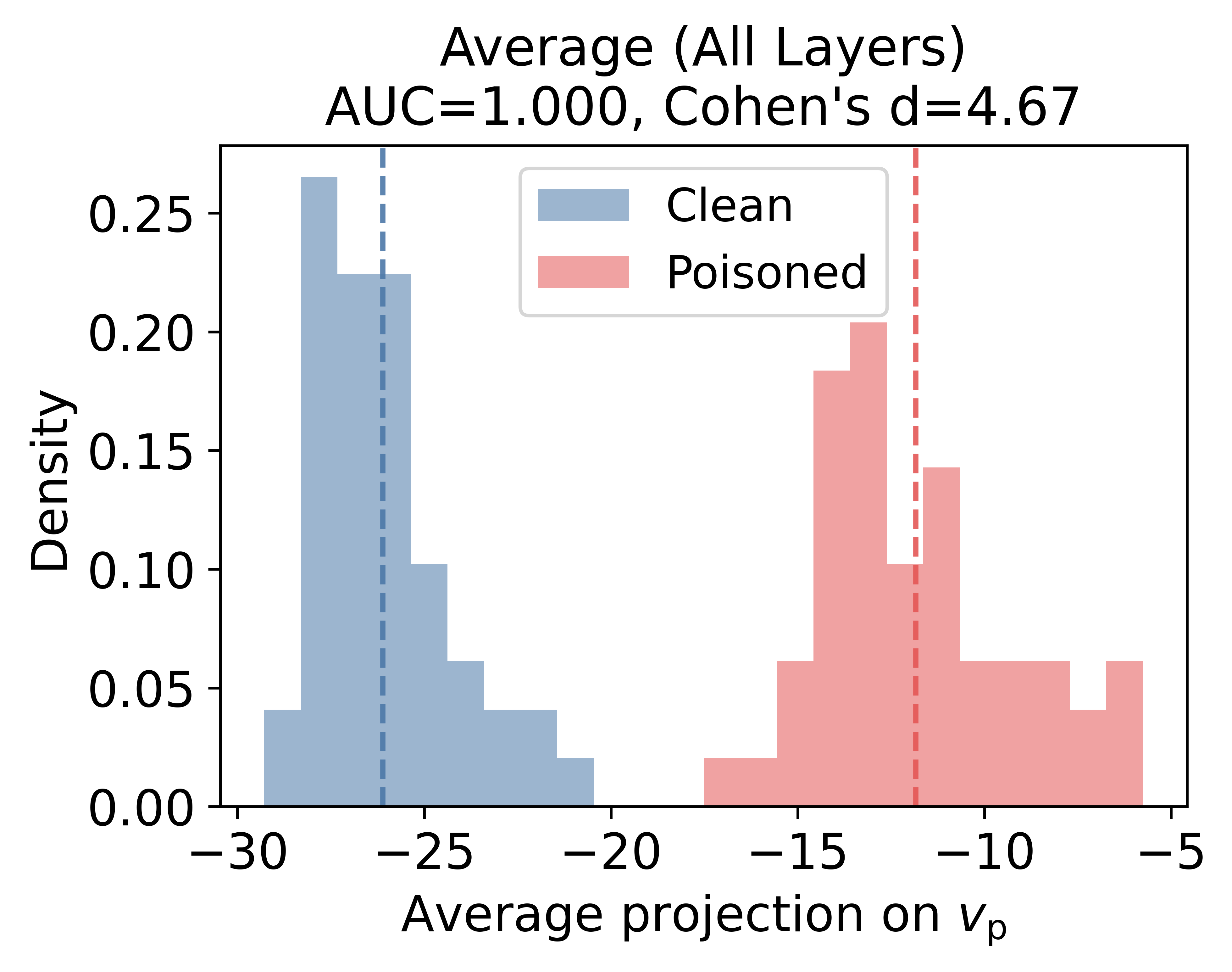}
        \caption{MS-MARCO}
        \label{fig:qwen_msmarco}
    \end{subfigure}
    \hfill
    \begin{subfigure}{0.28\textwidth}
        \centering
        \includegraphics[width=\linewidth]{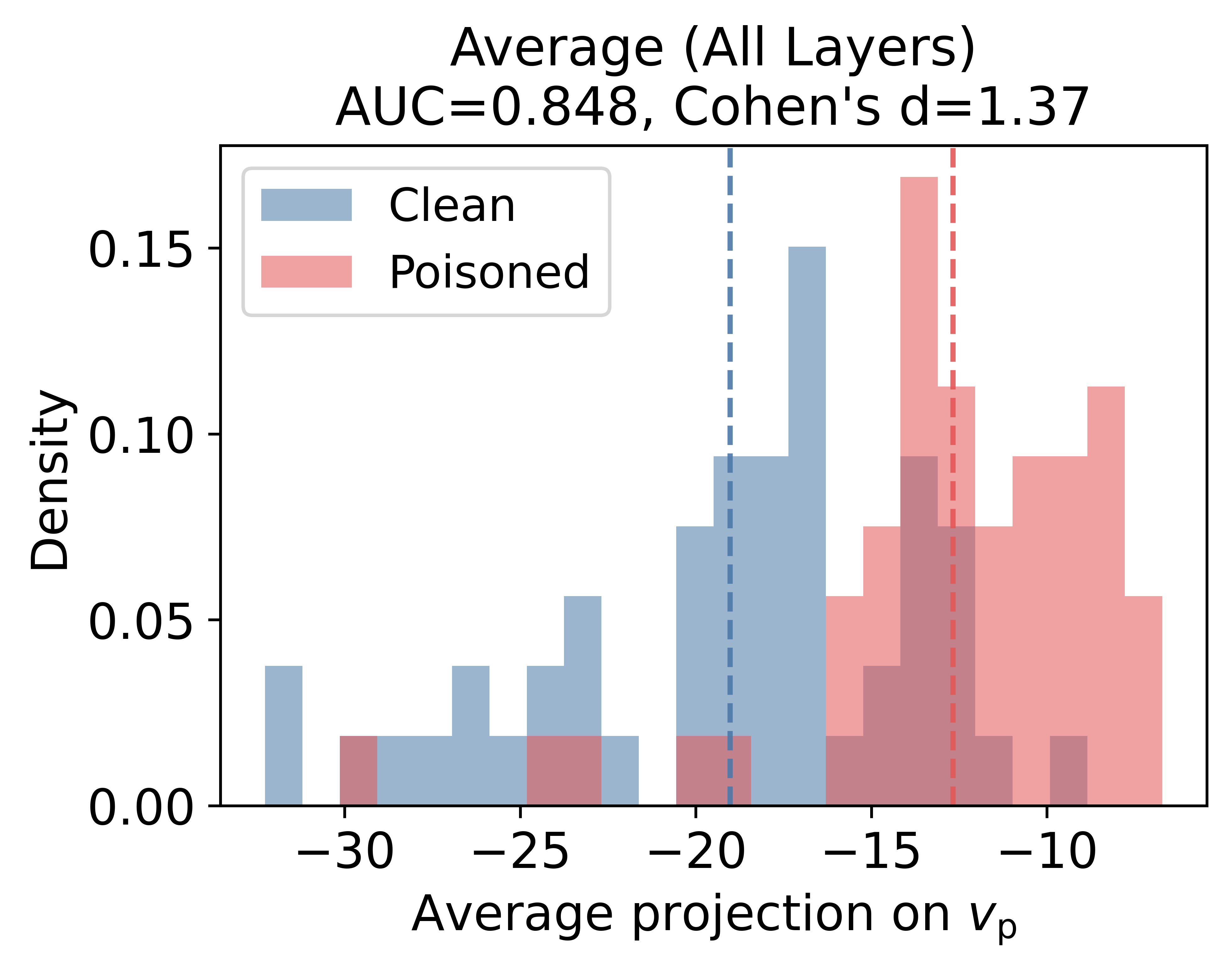}
        \caption{Amazon Reviews}
        \label{fig:qwen_reviews}
    \end{subfigure}

    \caption{Projection distributions of clean and poisoned activations along the poison direction. Results are shown for four different LLM backbones across three datasets under the ObliInjection attack.}
    \label{fig:activation_projections_Obliinjection}
\end{figure*}

\begin{figure*}[t]
    \centering

    \centering
    \small \textbf{(a) Llama-3.1-8B-Instruct} \par\medskip

    \begin{subfigure}{0.28\textwidth}
        \centering
        \includegraphics[width=\linewidth]{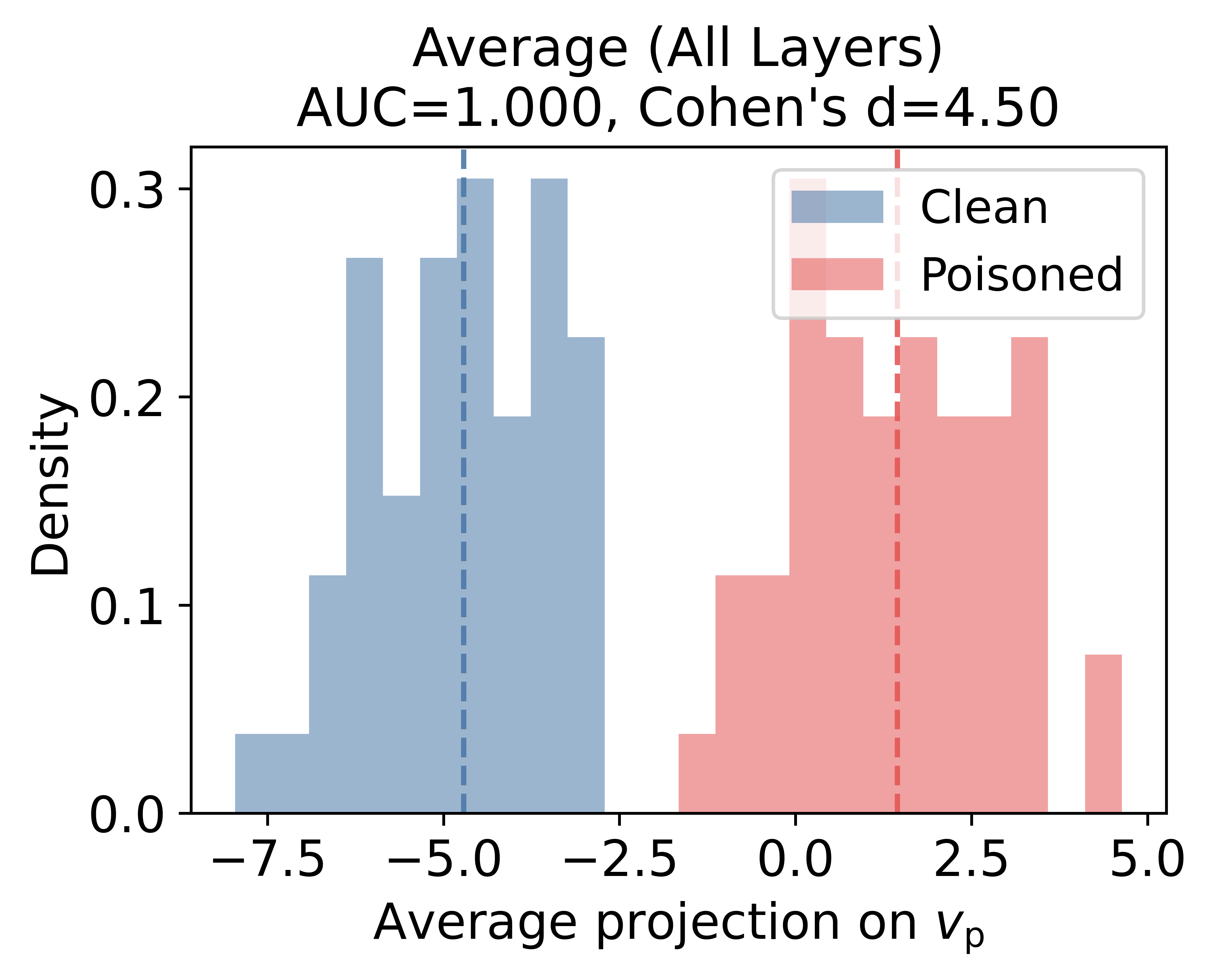}
        \caption{HotpotQA}
        \label{fig:llama_hotpot_pr}
    \end{subfigure}
    \hfill
    \begin{subfigure}{0.28\textwidth}
        \centering
        \includegraphics[width=\linewidth]{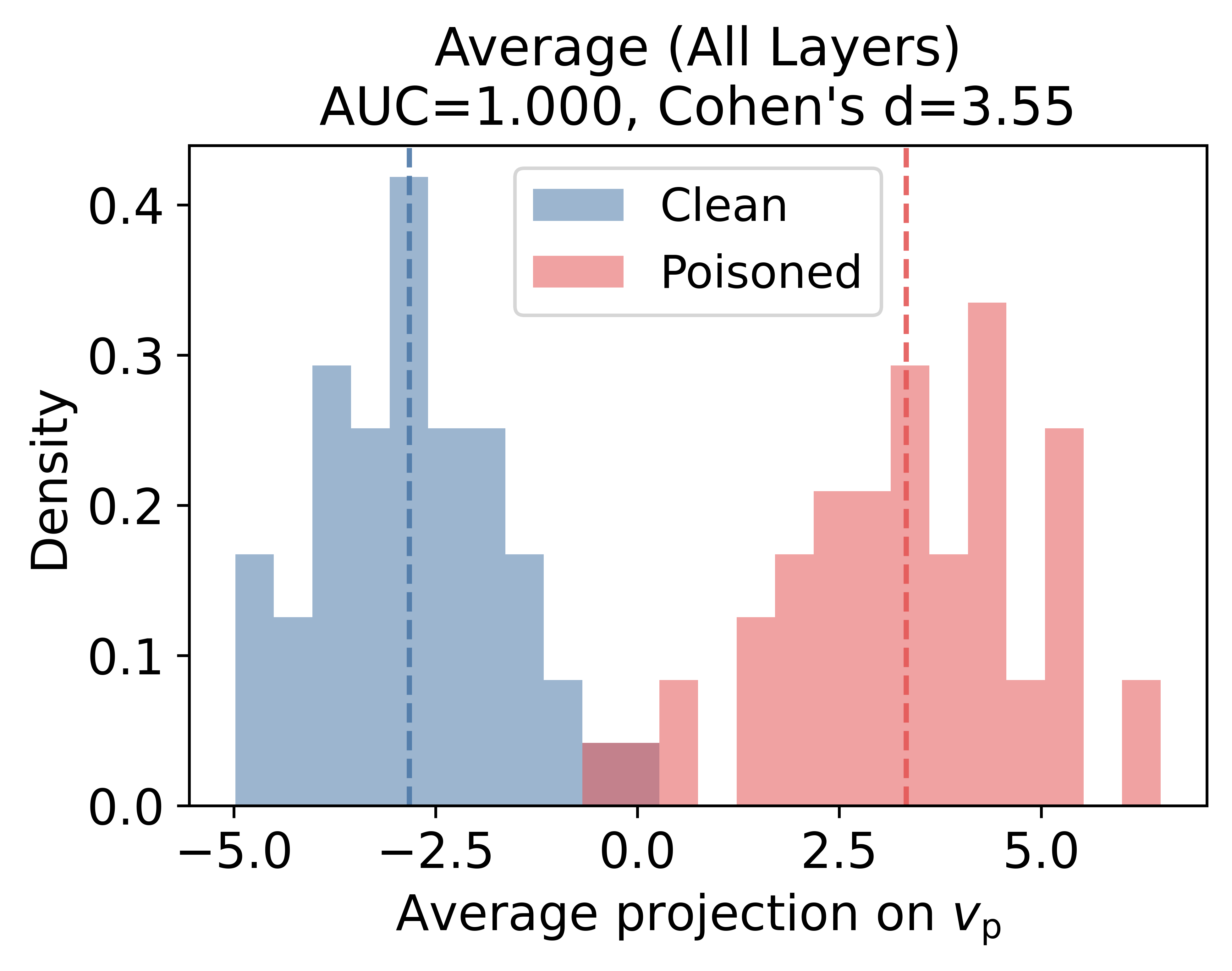}
        \caption{MS-MARCO}
        \label{fig:llama_msmarco_pr}
    \end{subfigure}
    \hfill
    \begin{subfigure}{0.28\textwidth}
        \centering
        \includegraphics[width=\linewidth]{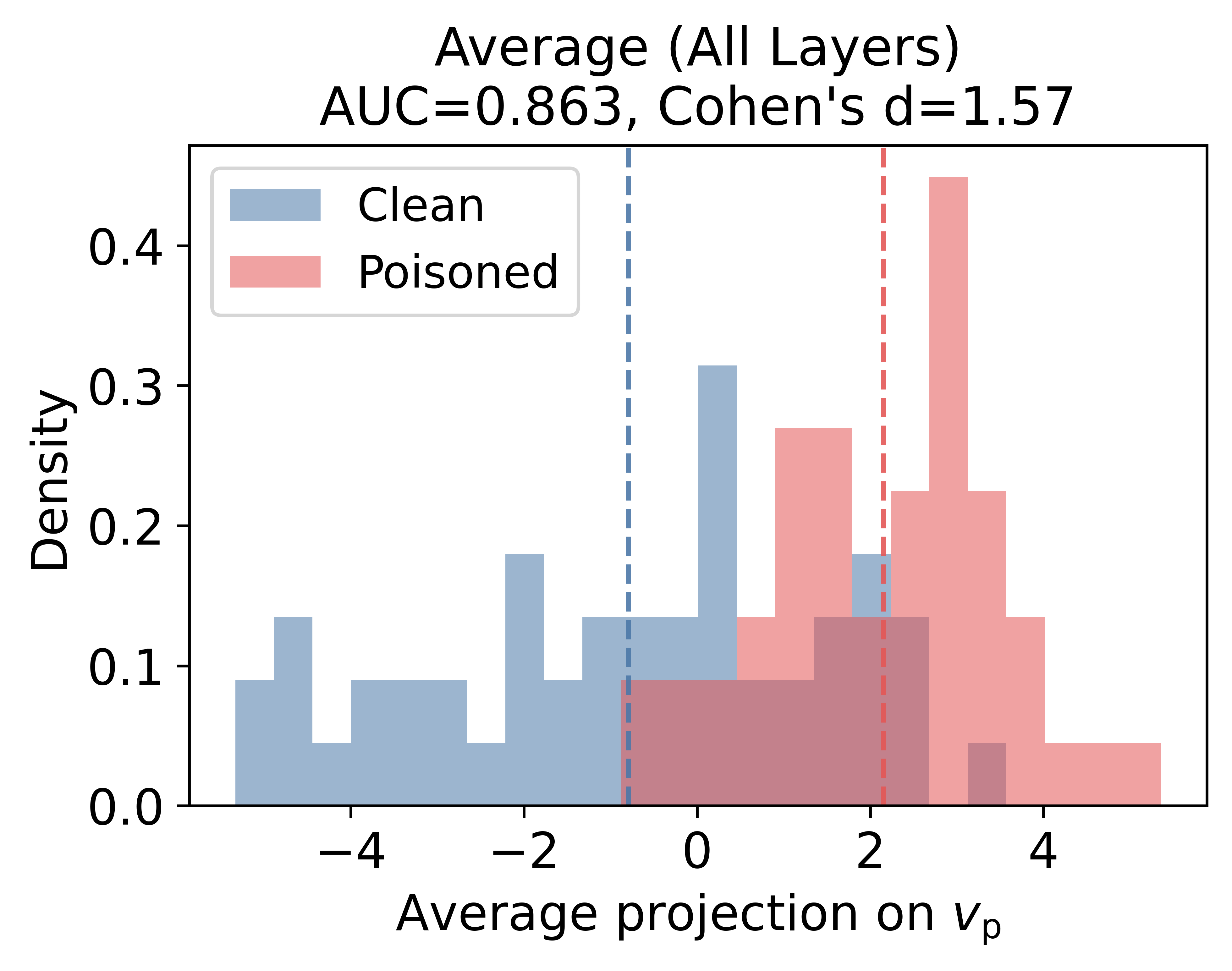}
        \caption{Amazon Reviews}
        \label{fig:llama_reviews_pr}
    \end{subfigure}

    \vspace{1pt} 

    \centering
    \small \textbf{(b) Mistral-7B-Instruct-v0.3} \par\medskip

    \begin{subfigure}{0.28\textwidth}
        \centering
        \includegraphics[width=\linewidth]{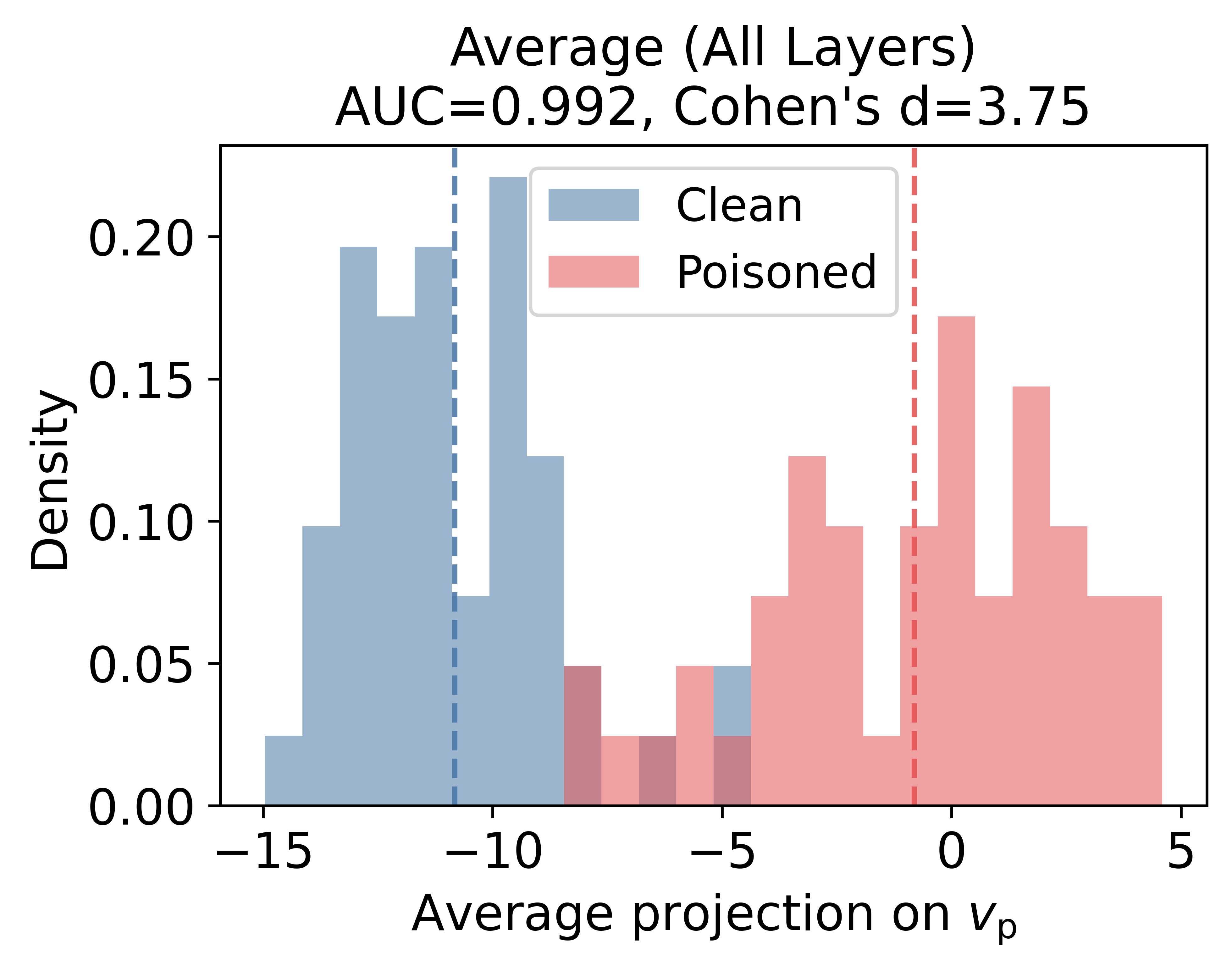}
        \caption{HotpotQA}
        \label{fig:mistral_hotpot_pr}
    \end{subfigure}
    \hfill
    \begin{subfigure}{0.28\textwidth}
        \centering
        \includegraphics[width=\linewidth]{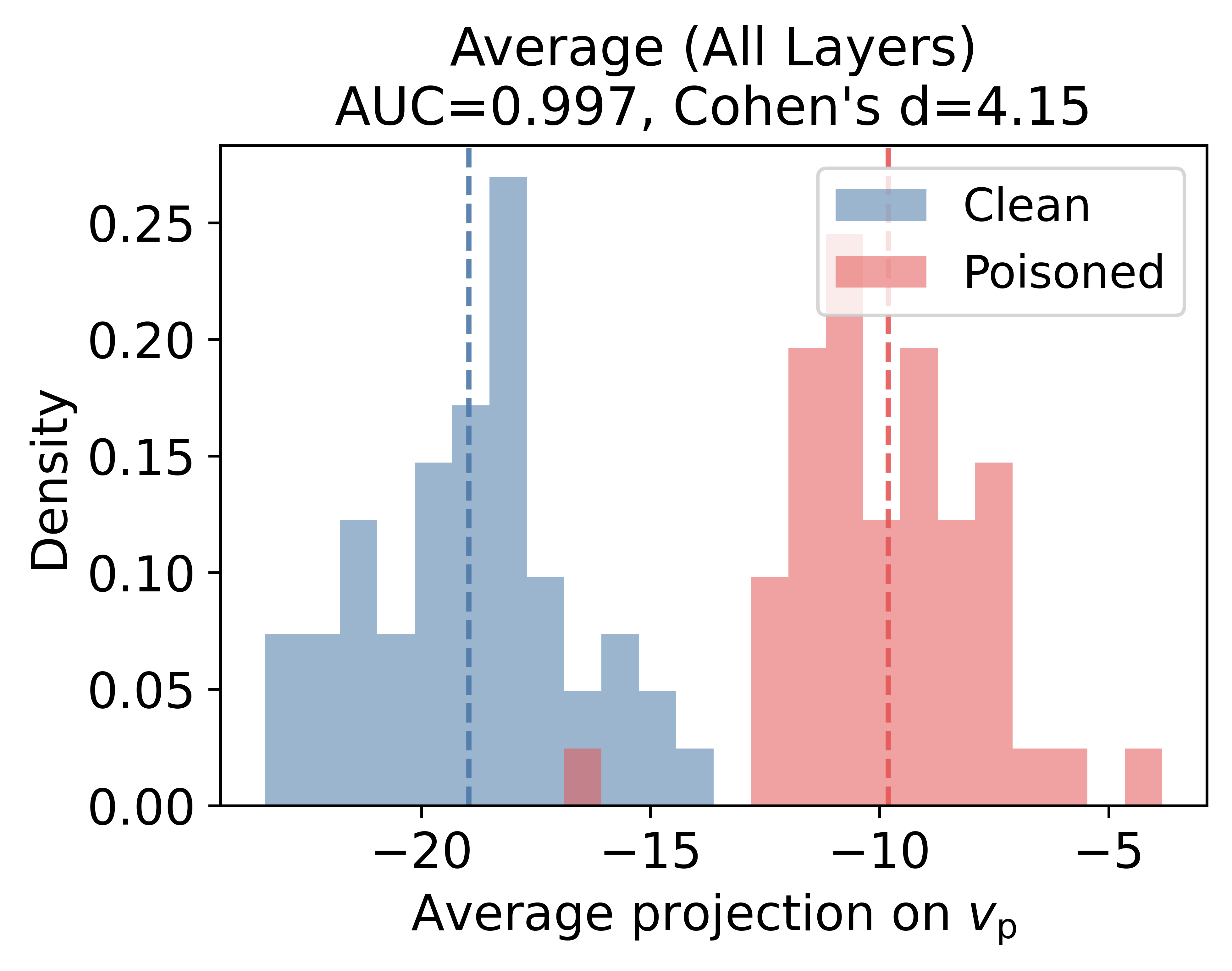}
        \caption{MS-MARCO}
        \label{fig:mistral_msmarco_pr}
    \end{subfigure}
    \hfill
    \begin{subfigure}{0.28\textwidth}
        \centering
        \includegraphics[width=\linewidth]{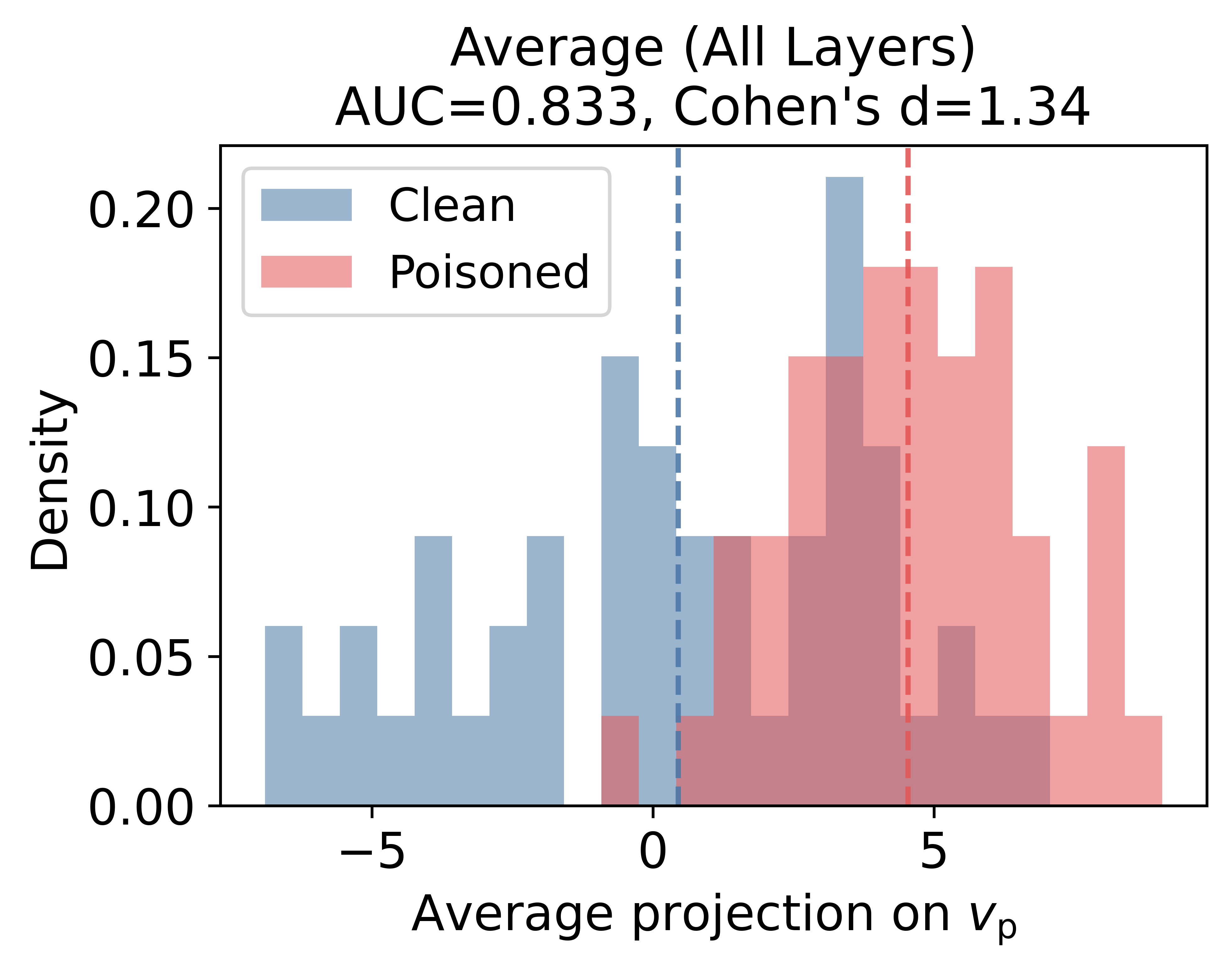}
        \caption{Amazon Reviews}
        \label{fig:mistral_reviews_pr}
    \end{subfigure}

    \vspace{1pt}

    \centering
    \small \textbf{(c) GPT-OSS-20B} \par\medskip

    \begin{subfigure}{0.28\textwidth}
        \centering
        \includegraphics[width=\linewidth]{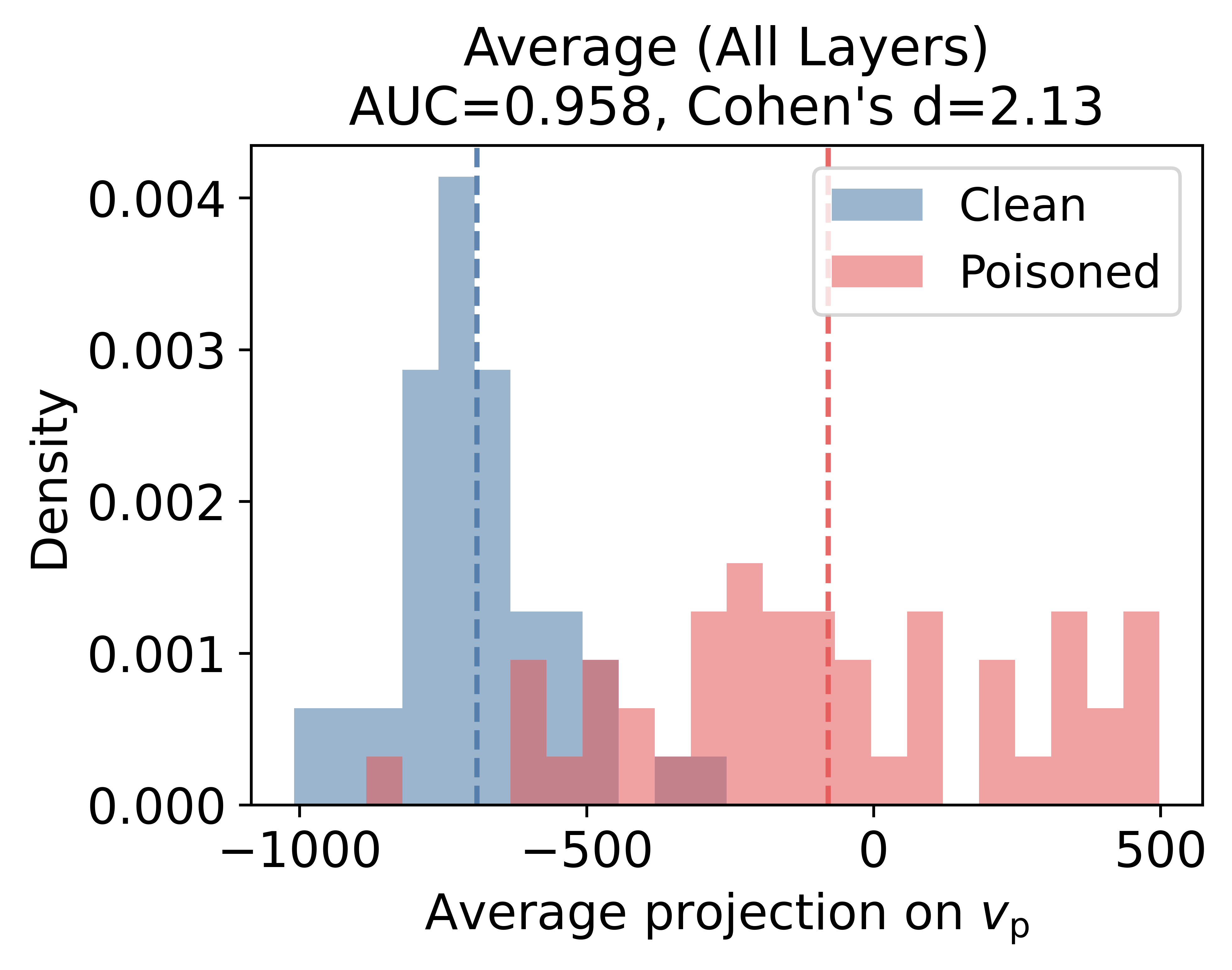}
        \caption{HotpotQA}
        \label{fig:gpt_hotpot_pr}
    \end{subfigure}
    \hfill
    \begin{subfigure}{0.28\textwidth}
        \centering
        \includegraphics[width=\linewidth]{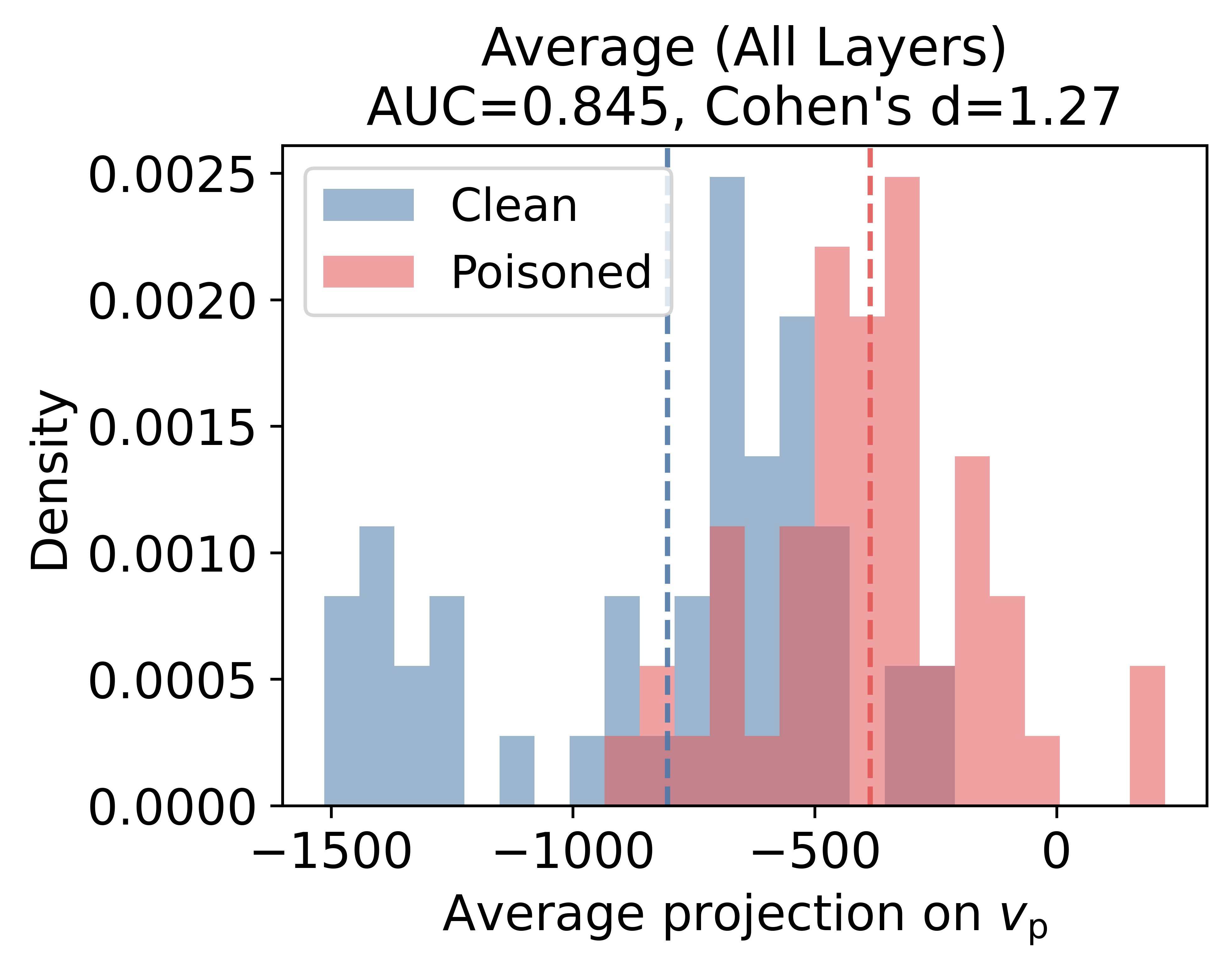}
        \caption{MS-MARCO}
        \label{fig:gpt_msmarco_pr}
    \end{subfigure}
    \hfill
    \begin{subfigure}{0.28\textwidth}
        \centering
        \includegraphics[width=\linewidth]{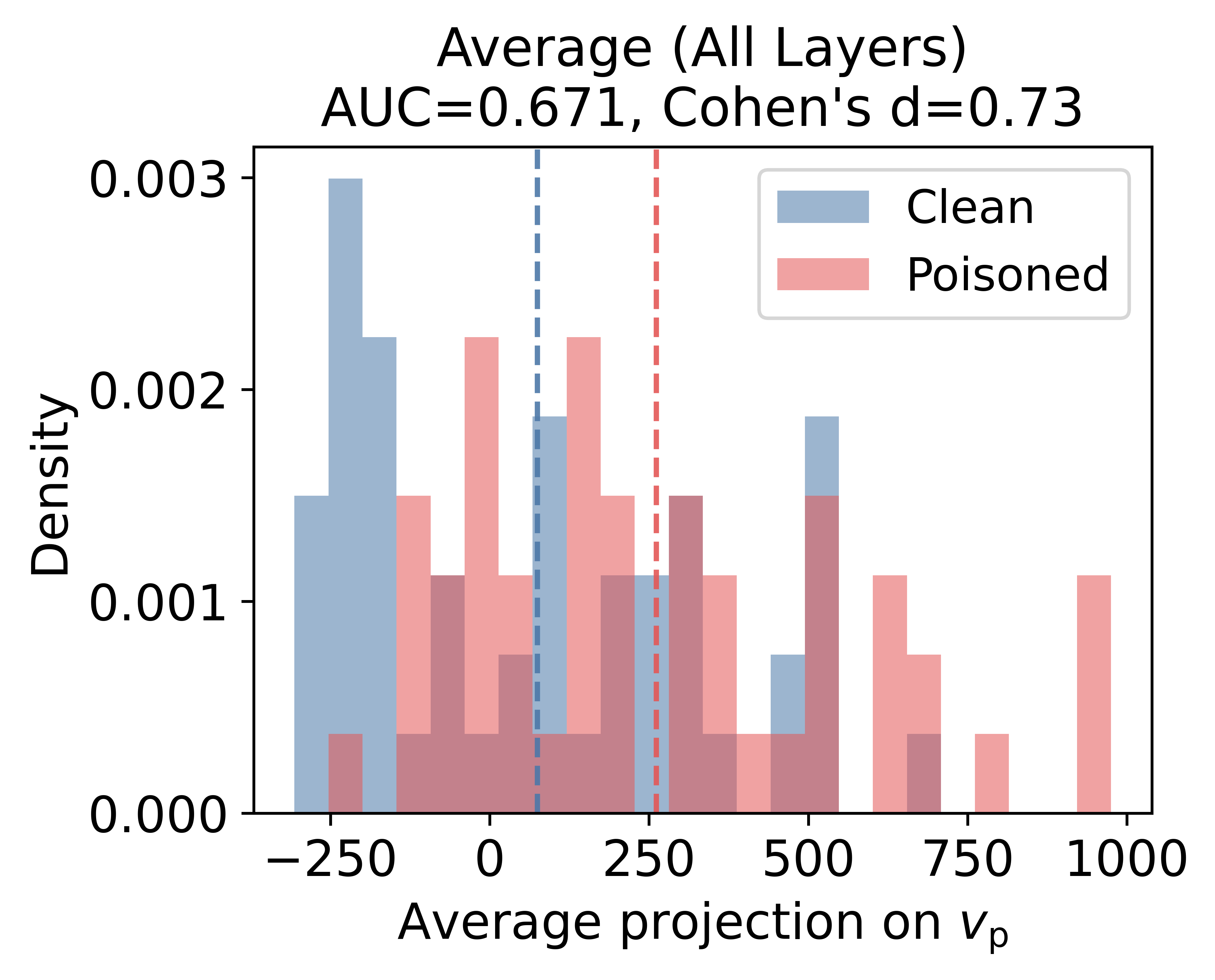}
        \caption{Amazon Reviews}
        \label{fig:gpt_reviews_pr}
    \end{subfigure}

    \vspace{1pt}

    \centering
    \small \textbf{(d) Qwen3-4B-Instruct-2507} \par\medskip

    \begin{subfigure}{0.28\textwidth}
        \centering
        \includegraphics[width=\linewidth]{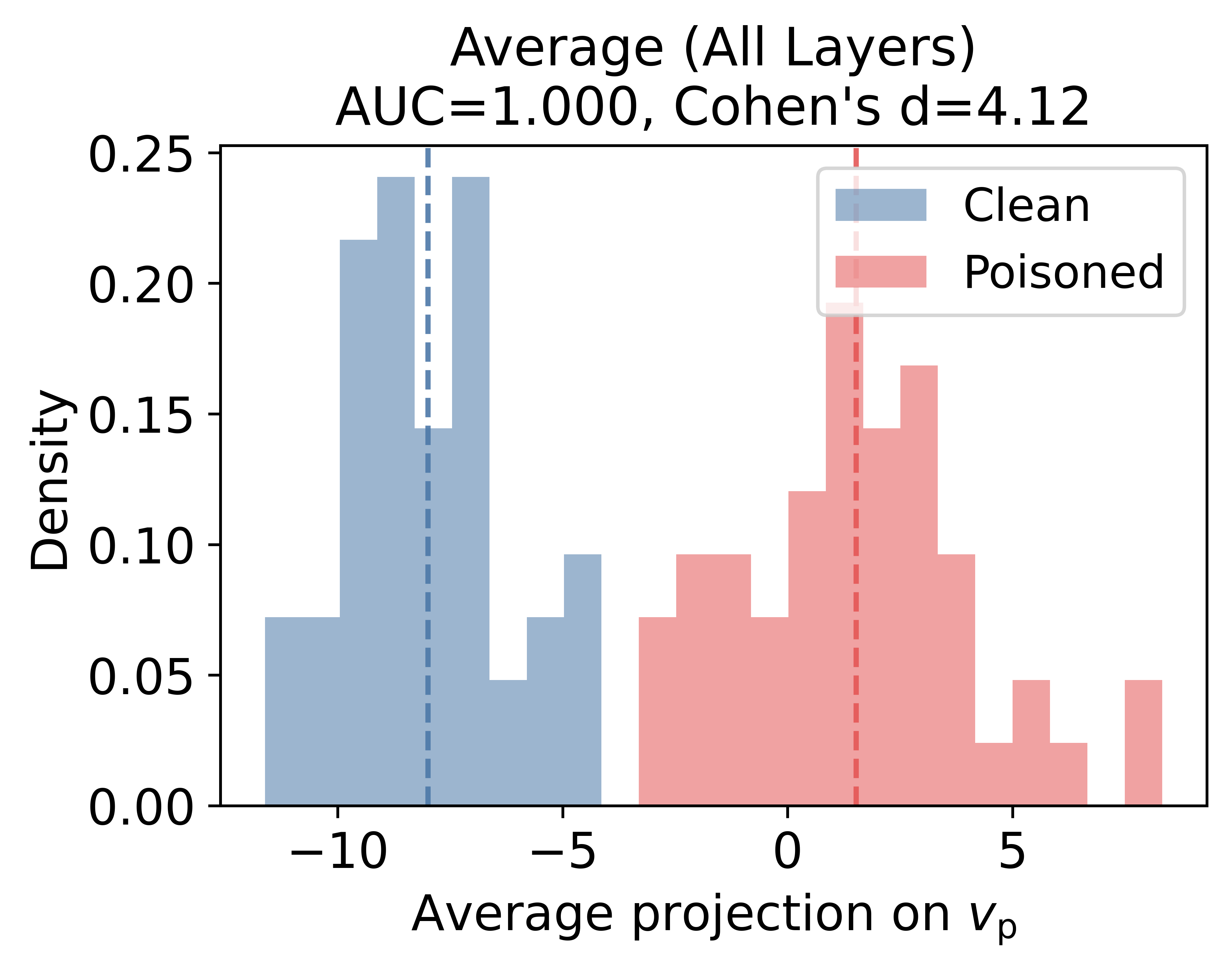}
        \caption{HotpotQA}
        \label{fig:qwen_hotpot_pr}
    \end{subfigure}
    \hfill
    \begin{subfigure}{0.28\textwidth}
        \centering
        \includegraphics[width=\linewidth]{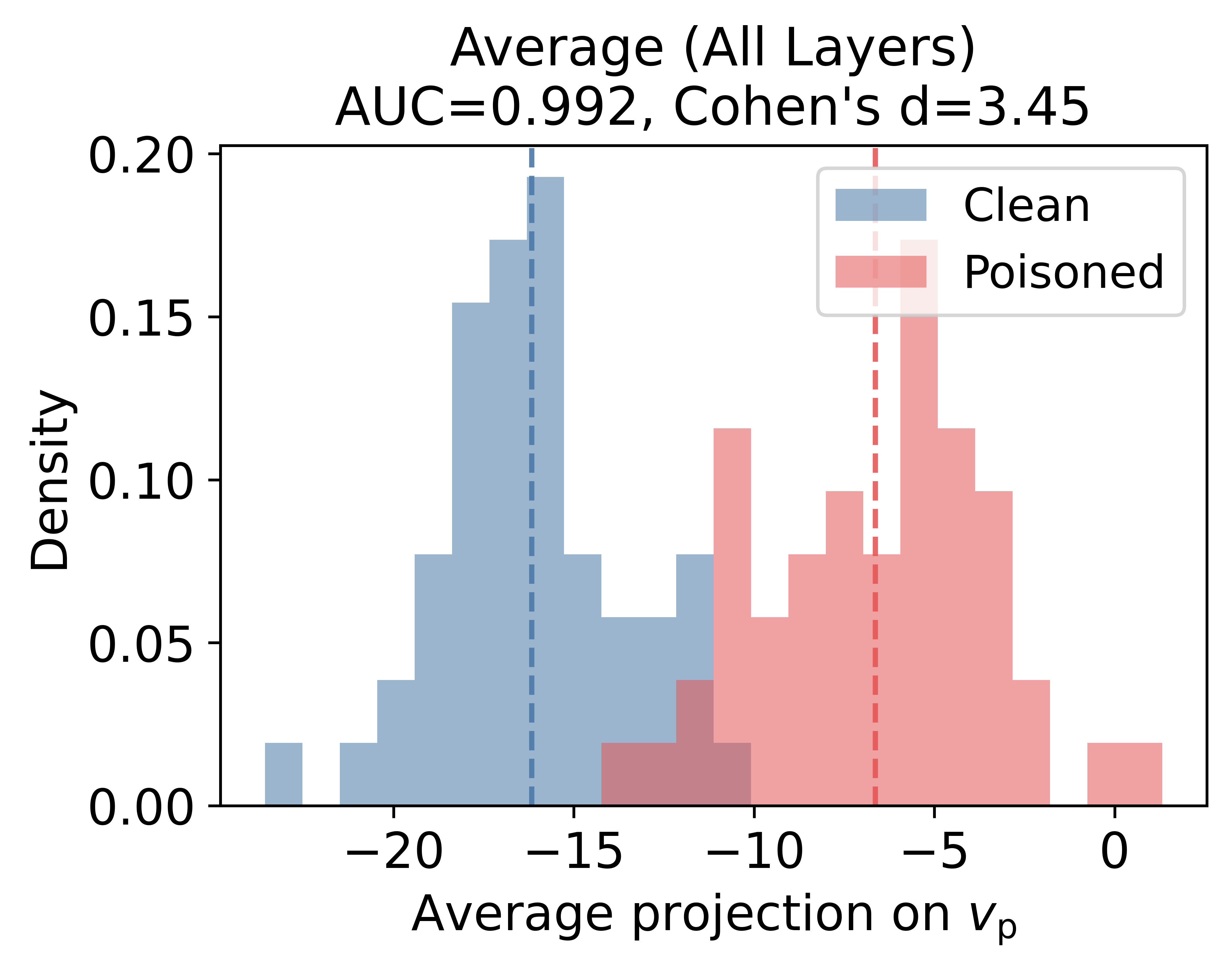}
        \caption{MS-MARCO}
        \label{fig:qwen_msmarco_pr}
    \end{subfigure}
    \hfill
    \begin{subfigure}{0.28\textwidth}
        \centering
        \includegraphics[width=\linewidth]{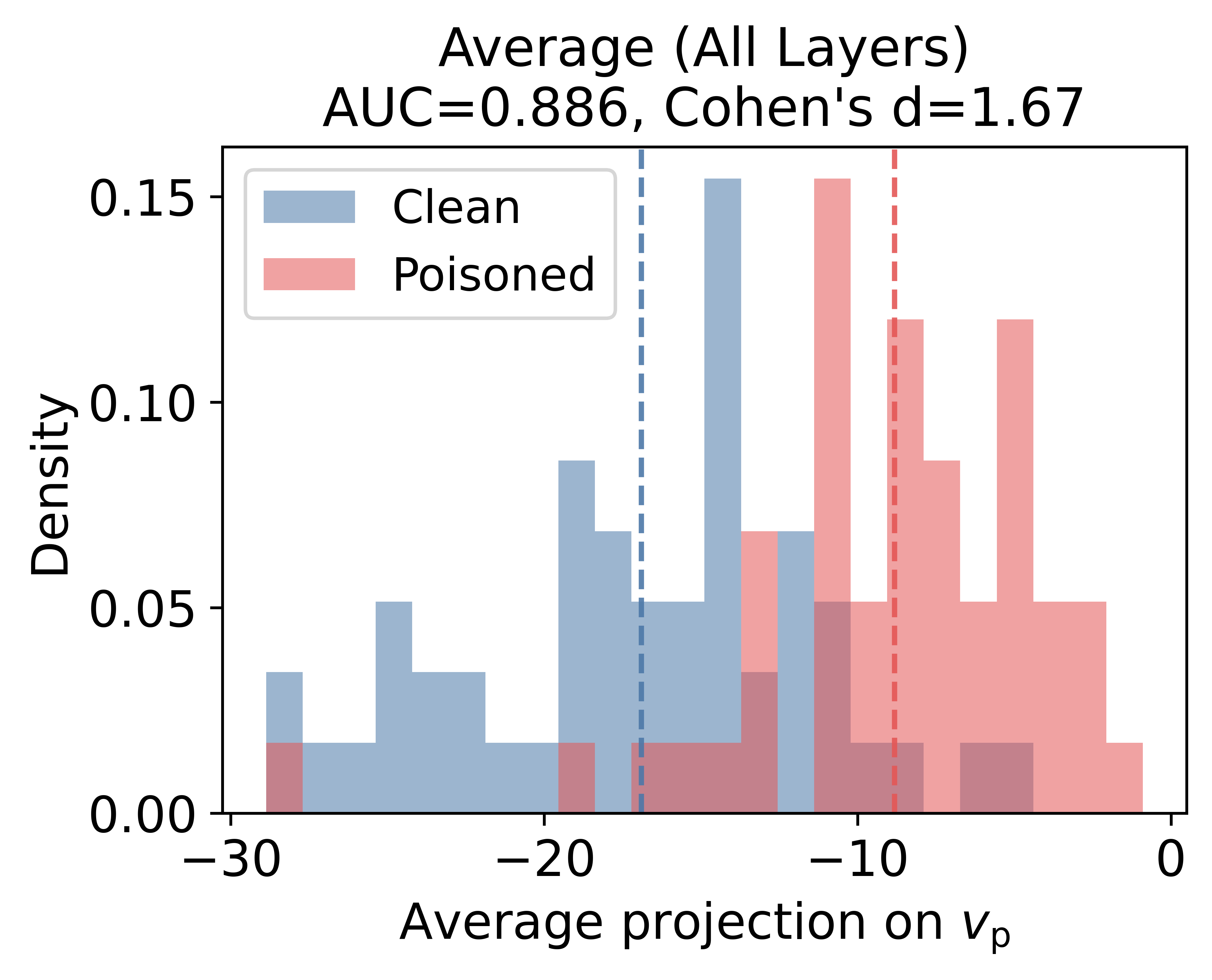}
        \caption{Amazon Reviews}
        \label{fig:qwen_reviews_pr}
    \end{subfigure}

    \caption{Projection distributions of clean and poisoned activations along the poison direction. Results are shown for four different LLM backbones across three datasets under the PoisonedRAG attack.}
    \label{fig:activation_projections_poisonedrag}
\end{figure*}

\end{document}